\documentclass[sigplan,10pt,nonacm]{acmart}
\renewcommand\footnotetextcopyrightpermission[1]{}
\AtBeginDocument{%
  }

\setcopyright{acmlicensed}
\copyrightyear{2025}
\acmYear{2025}
\acmDOI{xxxxxxx.xxxxxxx}
\acmISBN{978-1-4503-xxxx-x/xxxx/xx}

\usepackage[dvipsnames,table]{xcolor}

\usepackage{xspace}
\usepackage{amsmath}
\usepackage{bold-extra}
\usepackage{graphbox}

\usepackage[export]{adjustbox}
\usepackage{setspace}

\usepackage{etoolbox}
\makeatletter
\preto{\@verbatim}{\topsep=1pt \partopsep=1pt}
\makeatother

\usepackage[compact]{titlesec}
\titlespacing*{\section}{0pt}{5pt}{2pt}
\titlespacing*{\subsection}{0pt}{4pt}{1pt}
\titlespacing*{\subsubsection}{0pt}{4pt}{1pt}

\usepackage[ruled,linesnumbered]{algorithm2e}
\SetArgSty{textup}

\usepackage{float}
\usepackage{enumitem}
\SetLabelAlign{center}{\strut\smash{\parbox[t]\labelwidth{\centering#1}}}

\usepackage{caption}
\setlist{leftmargin=9pt,itemsep=2pt,topsep=3pt}

\usepackage[labelformat=simple]{subcaption}

\usepackage{xurl}
\usepackage{hyperref}
\hypersetup{
    colorlinks,
    linkcolor={green!70!black},
    citecolor={red!70!black},
    urlcolor={blue!70!black}
}

\usepackage{tikz}
\usetikzlibrary{shapes.geometric}
\usetikzlibrary{arrows.meta}

\usepackage{bm}
\usepackage{wasysym}
\usepackage{amsthm}

\usepackage{array}
\usepackage{makecell}
\usepackage{multirow}

\usepackage{lipsum}
\usepackage{duckuments}

\usepackage{listings}
\usepackage{pdfpages}

\usepackage[framemethod=TikZ]{mdframed}

\newcommand{\revise}[1]{\textcolor{black}{#1}\xspace}
\newcommand{\revisetwo}[1]{\textcolor{black}{#1}\xspace}
\newcommand{\revisethree}[1]{\textcolor{black}{#1}\xspace}

\newcommand{\bodega}{\textsc{Bodega}\xspace}
\newcommand{\summerset}{Vineyard\xspace}
\newcommand{\tlaplus}{$\textrm{TLA}^{+}$\xspace}

\newcommand{\paratitle}[1]{\vspace{2pt}\noindent\textbf{#1}.}
\newcommand{\paratitlenodot}[1]{\vspace{2pt}\noindent\textbf{#1}}

\newcommand*\circleb[1]{\tikz[baseline=(char.base)]{
    \node[shape=circle,fill,inner sep=.5pt](char){\textcolor{white}{\small #1}};}}

\newcommand{\floatcap}[2]{\caption[#1]{\textbf{#1.} \textit{#2}}}
\newcommand{\floatcapnodot}[2]{\caption[#1]{\textbf{#1} \textit{#2}}}
\newcommand{\floatcapoffig}[2]{\captionof{figure}{\textbf{#1.} \textit{#2}}}
\newcommand{\floatcapoffignodot}[2]{\captionof{figure}{\textbf{#1} \textit{#2}}}

\newmdenv[%
  linecolor=gray,
  linewidth=1pt,
  backgroundcolor=gray!8,
  roundcorner=5pt,
  frametitlealignment=\centering,
  frametitlerule=true,
  frametitlerulewidth=0.5pt,
  frametitleaboveskip=3pt,
  frametitlebelowskip=3pt,
  innertopmargin=3pt,
  innerbottommargin=3pt,
  innerleftmargin=3pt,
  innerrightmargin=3pt,
  font=\small,
  fontcolor=black!80,
  frametitlefont=\normalfont\bfseries\color{black!90}
]{algoblock}
\newcommand{\acblue}{RoyalBlue!20}
\newcommand{\acgold}{Goldenrod!25}
\newcommand{\acred}{Maroon!30}
\newcommand{\acorange}{YellowOrange!30}
\newcommand{\acgreen}{ForestGreen!25}
\newcommand{\acviolet}{Orchid!25}

\newcommand{\acbox}[1]{\colorbox{#1}{\phantom{\small i}}}

\newlist{algolist}{enumerate}{2}
\setlist[algolist,1]{
    label*=\arabic*:,font=\scriptsize\color{NavyBlue!70},before=\small,
    itemsep=0.5pt,topsep=-3pt,labelsep=1pt,leftmargin=8pt
}
\setlist[algolist,2]{
    label*=\arabic*:,font=\scriptsize\color{NavyBlue!70},before=\small,
    itemsep=0.5pt,topsep=-3pt,labelsep=1pt,leftmargin=12pt
}

\newlist{algolistnonum}{enumerate}{2}
\setlist[algolistnonum,1]{
    label*=,
    itemsep=0.5pt,topsep=-3pt,labelsep=0pt,leftmargin=0pt
}
\setlist[algolistnonum,2]{
    label*=,
    itemsep=0.5pt,topsep=-3pt,labelsep=0pt,leftmargin=8pt
}

\newcommand{\avar}[1]{\textnormal{\textit{#1}}}
\newcommand{\avpm}[1]{\textnormal{\textit{#1}}$'$}
\newcommand{\afun}[2]{\texttt{\textbf{#1}(}\textnormal{#2}\texttt{)}}
\newcommand{\amsg}[2]{\texttt{#1}$\langle$\textnormal{#2}$\rangle$}
\newcommand{\atimer}[1]{$T_{\text{#1}}$}
\newcommand{\atmdur}[1]{$t_{\text{#1}}$}
\newcommand{\aset}[1]{\{\}$_{\text{#1}}$}

\newcommand{\tporange}[1]{\textcolor{Orange}{#1}}
\newcommand{\tpgreen}[1]{\textcolor{ForestGreen}{#1}}
\newcommand{\tpblue}[1]{\textcolor{NavyBlue}{#1}}

\begin{document}

\title{\vspace*{-18pt}\LARGE \bodega: Serving Linearizable Reads Locally from Anywhere\\at Anytime via \revise{Roster} Leases}


\author{Guanzhou Hu}
\affiliation{%
  \institution{University of Wisconsin--Madison}
  \city{}
  \country{}}

\author{Andrea C. Arpaci-Dusseau}
\affiliation{%
  \institution{University of Wisconsin--Madison}
  \city{}
  \country{}}

\author{Remzi H. Arpaci-Dusseau}
\affiliation{%
  \institution{University of Wisconsin--Madison}
  \city{}
  \country{}}


\begin{abstract}
  \revise{We present \bodega, \revisethree{the first} consensus protocol that \revisethree{serves linearizable reads locally} \revisethree{from any desired node, regardless of interfering writes}. \revisethree{\bodega achieves this via a novel \textit{roster leases} algorithm that safeguards the \textit{roster},} a new \revisetwo{notion} of cluster metadata. The roster \revisethree{is a generalization of leadership; it tracks} arbitrary subsets of replicas as \textit{responder} nodes \revisethree{for local reads}. \revisethree{A consistent agreement on the roster is established through roster leases}, an all-to-all leasing mechanism that \revisethree{generalizes} existing \revisetwo{all-to-one leasing} approaches (Leader Leases, Quorum Leases), \revisetwo{unlocking a new point in the protocol design space}. \bodega further employs \textit{optimistic holding} and \textit{early accept notifications} to minimize interruption from interfering writes, and incorporates \textit{smart roster coverage} and \textit{lightweight heartbeats} to maximize practicality.} \bodega is a non-intrusive extension to classic consensus; it imposes no special requirements on writes other than \revisetwo{a responder-covering quorum.}

We implement \bodega and related works in \summerset, a protocol-generic replicated key-value store written in async Rust. We compare it to previous protocols (Leader Leases, EPaxos, PQR, and Quorum Leases) and two production coordination services (etcd and ZooKeeper). \revisetwo{\bodega speeds up average client read requests by 5.6x$\sim$13.1x on real WAN clusters versus previous approaches under moderate write interference, delivers comparable write performance, supports fast proactive roster changes as well as fault tolerance via leases, and} closely matches the performance of sequentially-consistent etcd and ZooKeeper deployments across all YCSB workloads. We will open-source \summerset upon publication.



\vspace{-5pt}

\end{abstract}

\maketitle
\pagestyle{plain}

\vspace{-0.02in}
\section{Introduction}
\label{sec:introduction}

\revisetwo{Paxos-based consensus~\cite{paxos-parliament, paxos-made-simple, abcd-of-paxos} (and protocols alike~\cite{raft, viewstamped-replication}) serves as a critical foundation for modern distributed systems infrastructure. Originally used in limited contexts, e.g., for bespoke configuration information as in Petal~\cite{petal}, for critical metadata stores such as etcd~\cite{etcd, kubernetes} and FireScroll~\cite{firescroll, redpanda}, and for lock services such as Chubby~\cite{chubby}, consensus now forms the foundation of widely used cloud-native databases such as Spanner~\cite{spanner}, CockroachDB~\cite{cockroachdb}, TiDB~\cite{tidb}, ScyllaDB~\cite{scylladb}, and Physalia~\cite{physalia}.}





\revisetwo{In these systems, simple access semantics are critical, enabling scalable services to be readily built atop them~\cite{zookeeper-consistency-question-1, s3-strong-consistency}. In particular, \textit{linearizability} is a strong consistency level they strive to provide: for interrelated requests, clients observe a real-time serial ordering, as if talking to a single node~\cite{linearizability, practical-consistency-summary}.}



\paratitle{\revisethree{Local Linearizable Reads}}
\revisetwo{Delivering high performance in linearizable systems remains a daunting challenge. In the cloud era, systems replicate critical data across multiple geographically-distinct availability zones~\cite{aws-global-infra, gcp-global-infra, azure-global-infra}, to guard against correlated failures caused by power outage, fire, natural disaster, or operator error~\cite{alibaba-singapore-fire, maelstrom, what-bugs-in-the-cloud, google-unisuper-delete}. By spreading replicas globally, robustness is achieved, but at the cost of performance \revisethree{due to quorum round trips}~\cite{rifl}.}


\revisetwo{The physical distribution of replicas yields an opportunity to serve \revisethree{read} requests locally from a client's nearest replica. \revisethree{Reads comprise a majority of the workloads~\cite{ycsb, benchmarking-spanner, live-optics-rw-ratio, aws-prescriptive-workload-characteristics}}; achieving local reads can greatly reduce read latency and drastically increase overall throughput.}




\paratitle{\revisethree{Existing Solutions Fall Short}}
Existing \revisetwo{consensus} protocols have demonstrated effective wide-area optimizations, but none, \revisetwo{to our knowledge}, supports coherently fast linearizable reads \revisethree{for workloads containing even small amounts of interfering writes}. Leaderless protocols~\cite{generalized-paxos, mencius, epaxos, hermes, paxos-quorum-read, swiftpaxos} allow near quorums but not local reads. Others explore flexible quorums~\cite{fpaxos, edge-pqr, tradeoffs-geo-distributed, dynamic-quorum, vertical-paxos}, utilize special hardware or client validation~\cite{nezha, glean-consensus, nopaxos, speculative-paxos, clock-rsm, flair, caesar-consensus}, or exploit API semantics favoring writes~\cite{consistency-aware-durability, skyros-nil-externality, lazylog, curp-commutativity, raft-thesis, chain-replication, chain-paxos, sdpaxos}. Read leases \revisethree{(covered later)}~\cite{megastore, paxos-made-live, quorum-leases, quorum-leases-report, leases-mechanism, read-leases-theory, read-leases-theory-parameterized} are so far the most compelling, but are only effective at the leader or during quiescent periods without interfering writes.


\begin{figure}[!t]
    \centering
    
    \begin{subfigure}[t]{0.27\linewidth}
        \centering
        \raisebox{8.5pt}{\includegraphics[width=\columnwidth]{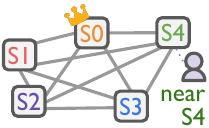}}
    \end{subfigure}
    \hfill
    \vrule
    \hfill
    \begin{subfigure}[t]{0.70\linewidth}
        \centering
        \includegraphics[width=\columnwidth]{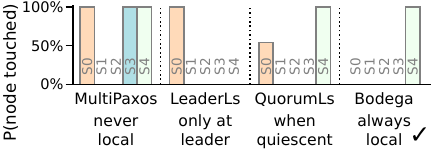}
    \end{subfigure}

    \addtocounter{figure}{-1}  
    \captionsetup{justification=raggedright}
    \vspace{-8pt}
    \floatcapoffignodot{Frequency of touching a node on the critical path of reads}{by a client near S4, in a cluster where S0 is the leader, with infrequent writes. \revisethree{Ideal outcome is 100\% at S4.}}
    \label{fig:intro-access_cnts}
    \vspace{-0.14in}
\end{figure}


\paratitle{\revisethree{Self-Containment Necessitates Leases}}
\revise{A primary \revisethree{challenge of designing consensus protocols for critical workloads} \revisetwo{is that the protocols must be} \textit{self-contained}, i.e., they cannot depend on external \revisethree{services to deliver essential information about the current role of nodes. Self-containment is needed for several reasons.}} First, to provide local linearizable reads, the consensus protocol must know whether or not the local data is the most recent; contacting an external service to obtain this information would defeat the purpose of local reads. Second, given that fault tolerance must be provided, having external dependencies would reduce its guarantees to those of the external services~\cite{kraft, redpanda, cockroachdb}.




\revise{Within the design space of self-contained protocols, we observe that} \textit{leases} are a \revisetwo{vital} and powerful primitive. They carry timed promises that naturally tolerate faults through expiration~\cite{leases-mechanism}, while only requiring bounded clock drifts (typical in today's cloud environments~\cite{rfc-ntp, aws-time-sync-service, nezha, dtp-clock-sync}). \revisethree{This opens the gateway to local linearizable read protocols.}

\paratitle{\revisethree{Leases Were Not Fully Exploited}}
Existing lease-infused protocols, however, do not employ the most suitable types of promises for local reads, and thus cannot fully realize their potential. As a motivating example, Figure~\ref{fig:intro-access_cnts} shows \revisethree{a 5-site cluster where S0 is the leader (and S4 is a local-read-enabled replica, if eligible), and reports} the frequency of servers being touched by read requests from a client near S4. The workload contains 99\% reads and only 1\% writes, which favors existing approaches. Classic consensus (MultiPaxos) requires a majority accept quorum around S0. Leader Leases \revisethree{only protect stable leadership, so} S0 can reply to reads directly, yet the delay between client and S0 persists. Quorum Leases \revisethree{allow granting leases to followers but use leases to guard against individual writes, rendering them vulnerable to even small amounts of concurrent writes to the key. As a result,} a significant portion are redirected to the leader S0.





\paratitle{Our Approach}
\revise{We introduce the notion of a \textit{roster}: a \revisetwo{generalized} cluster metadata that \revisethree{dictates} not only leadership but also an assignment of local-read-enabled replicas (called \textit{responders}) for arbitrary keys. \revisethree{Accordingly}, we introduce \textit{roster leases}, a novel all-to-all \revisetwo{generalization of leader leases}, deployed off the critical path to protect the agreement on the roster with no observable overhead. Roster leases stay valid in the absence of failures or proactive changes.}


\revise{\revisetwo{We present} \bodega, \revisetwo{a consensus protocol that uses roster leases to empower local linearizable reads}. \bodega assures that writes never commit before reaching all of the key's active responders. A responder that holds a \revise{majority} of leases can thus serve reads directly if it knows the latest value will commit. When unsure, it \textit{optimistically holds} the read locally until enough information is gathered, optionally utilizing \textit{early accept notifications} to accelerate the hold. The roster may be changed manually by users, automatically according to statistics, or in reaction to failures. In Figure~\ref{fig:intro-access_cnts}, \bodega is able to handle all reads by the client locally at S4.}


\revisethree{Our evaluation shows that \revisetwo{\bodega speeds up client read requests by 5.6x$\sim$13.1x versus previous approaches under slight write interference, delivers comparable write performance, supports proactive roster changes in two message rounds as well as self-contained fault tolerance via leases, and} matches the performance of sequentially-consistent etcd~\cite{etcd} and ZooKeeper~\cite{zookeeper} deployments across all YCSB variants.}

\paratitle{Summary of Contributions}
\revise{\circleb{1} We introduce the notion of roster and propose \bodega, a consensus protocol equipped with a novel all-to-all roster leases algorithm plus various optimizations, enabling local linearizable reads anywhere in a replicated cluster at anytime.} \circleb{2} We provide a thorough comparison across wide-area linearizable read approaches. \circleb{3} We implement \bodega and related protocols in \summerset, a protocol-generic replicated key-value store, with 25.6k lines of async Rust. \circleb{4} We evaluate \bodega comprehensively against previous works and two production coordination services, etcd and ZooKeeper, on 5-site CloudLab clusters, \revisethree{delivering aforementioned evaluation results}. \circleb{5} We provide a formal \tlaplus specification in the appendix.

The paper organization: \S\ref{sec:background} provides background and discusses existing solutions; \S\ref{sec:design} presents the \bodega design; \revisetwo{\S\ref{sec:comparison-proof} gives a formal comparison and proof}; \S\ref{sec:implementation} covers the implementation of \bodega in \summerset; \S\ref{sec:evaluation} presents our evaluation; \S\ref{sec:discussion}-\ref{sec:related-work} add discussions and related work; \S\ref{sec:conclusion} concludes.

\section{Background and Motivation}
\label{sec:background}

We provide background on consensus and linearizable reads, discuss existing solutions, and derive our goals for \bodega.

\subsection{Consensus \& Linearizable Reads}
\label{sec:bg-linearizable-reads}

We consider the typical consensus problem of reaching agreement across message-passing server nodes, where nodes can be fail-slow/stop and the network is asynchronous~\cite{paxos-parliament}.

Following well-established practice, nodes agree upon an ordered \textit{log} of commands to behave as a \revisethree{replicated state machine (RSM)}~\cite{smr-approach, paxos-made-simple}. For clarity, \revisethree{we use key-value \texttt{Put}/\texttt{Get} commands}. In practice, \texttt{Get}s map to read-only requests that only query but do not update state (hereby referred to as \textit{reads}), and \texttt{Put}s map to all other requests (\textit{writes}). We restrict our discussion to non-transactional commands and assume no blind writes~\cite{ansi-sql-isolation}; \revisethree{these are out of the scope of this paper}.


\paratitle{Linearizable Reads}
The consistency level dictates what results are allowed to be observed by clients~\cite{practical-consistency-summary}. \textit{Linearizability} is the strongest non-transactional consistency level, where clients expect a serial ordering of commands with the real-time property: a command on key $k$ issued at physical time $t$ must be ordered after all the interfering writes on $k$ acknowledged before $t$, and observe its latest value~\cite{linearizability, sequential-vs-linearizability, epaxos}. This semantic is mandatory for critical use cases such as metadata storage and coordination~\cite{etcd, firescroll, chubby, physalia}, and is generally desirable as clients would otherwise get stale reads from the past (\textit{sequential consistency})~\cite{sequential-vs-linearizability, sequential-consistency} or weaker guarantees~\cite{session-guarantees, eventual-consistency}, complicating development.


\paratitle{Availability Requirements}
A practical protocol must also offer \textit{availability} for fault tolerance, allowing \revisethree{client requests to proceed under any minority number of node/network faults, and retaining consistency in all circumstances}~\cite{practical-consistency-summary, paxos-made-simple}.


\subsection{Distributed Lease}
\label{sec:bg-distributed-leases}

\textit{Leases} are a common distributed system technique~\cite{leases-mechanism}. They may be deployed as user-facing APIs through locks~\cite{chubby} and TTL-tagged objects~\cite{etcd}, or as protocol-internal optimizations; we focus on the latter.


A lease is, conceptually, a directional limited-time \textit{promise} that a \textit{grantor} node makes to a \textit{grantee}. It relies on bounded clock speed drift between the two ends, that is, over a given physical expiration time \atmdur{lease} elapsed, the two nodes' clocks do not deviate more than a small \atmdur{$\Delta$}. This is typically true in today's cloud environments~\cite{rfc-ntp, aws-time-sync-service, nezha, dtp-clock-sync}; note that it does not assume synchronized clock timestamps~\cite{spanner, megastore, sundial}.

\begin{figure}[t!]
    \centering

    \includegraphics[height=60pt]{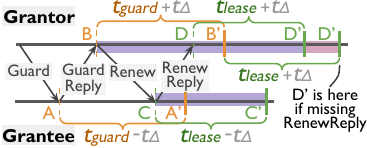}
    \hfill
    \vrule
    \hfill
    \hspace{1pt}
    \raisebox{1.5pt}{\includegraphics[height=58pt]{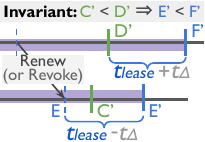}}
    \captionsetup{justification=raggedright}
    \vspace{-6pt}
    \floatcapoffig{\revisetwo{Demonstration of standard leasing}}{\revisetwo{Left: the guard phase establishes the first iteration of promise coverage; grantee welcomes the first \texttt{Renew} only if it is received within the guarded period (\tpgreen{C} $<$ \tporange{A'}). This allows the grantor to derive a safe \tpgreen{D'} $=$ \tporange{B'}$+$\atmdur{lease}$+$\atmdur{$\Delta$} even if the \texttt{RenewReply} is lost, such that \tpgreen{C'} $<$ \tpgreen{D'}. Right: the grantor attempts to extend the promise with a \texttt{Renew} (or to actively revoke it with a \texttt{Revoke}), but has not received the grantee's reply. The leasing logic assures that \tpblue{E'} $<$ \tpblue{F'} holds; therefore, if the grantee indeed failed, after \tpblue{F'} the grantor can assert the promise is no longer believed by the grantee. Optimizations exist when replies are successful~\cite{quorum-leases-report}.}}
    \label{fig:lease-renew-example}
\end{figure}

\paratitle{\revisetwo{Standard One-to-One Leasing}}
The procedure of activating one lease between two nodes consists of an initial \textit{guard} phase and repeated \textit{endow} (\revise{i.e., \textit{renew}}) phases, \revisetwo{depicted in Figure~\ref{fig:lease-renew-example}. The guard phase (left half) establishes the first iteration, and the endow phases (right half) keep it refreshed periodically~\cite{leases-mechanism, quorum-leases, quorum-leases-report}. The goal is to maintain this invariant: the grantor-side expiration time is \textit{never earlier than} the grantee-side.} A lease is considered held by the grantee when its clock has not surpassed \atmdur{lease}$-$\atmdur{$\Delta$} after the last endowment received. The grantor can proactively deactivate the lease with a \texttt{Revoke} or, in the case of unresponsiveness, wait for \atmdur{lease}$+$\atmdur{$\Delta$} without endowing to let it safely expire.

\subsection{Previous Work on Read Optimizations}
\label{sec:bg-previous-approaches}

\begin{figure}[!t]
    \centering

    \includegraphics[width=0.999\linewidth]{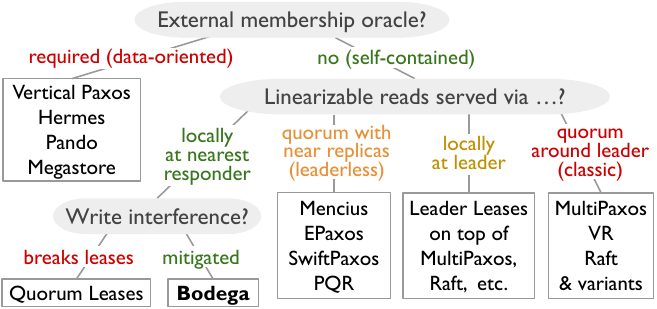}
    \captionsetup{justification=raggedright}
    \vspace{-18pt}
    \floatcapoffig{\revise{Categorization chart of relevant protocols}}{\revisethree{Ideal properties for local reads are marked in \textcolor{ForestGreen!70!black}{green}. See \S\ref{sec:bg-previous-approaches}.}}
    \label{fig:categorization-chart}
\end{figure}

\revise{Figure~\ref{fig:categorization-chart} presents a coarse-grained categorization of previous approaches to linearizable reads. The following sections discuss them in the general order from right to left.}


\subsubsection{Classic Protocols \& Leader Leases}
\label{sec:bg-previous-classic}

Protocols such as \textbf{MultiPaxos}~\cite{paxos-made-simple}, \textbf{VR}~\cite{viewstamped-replication}, and \textbf{Raft}~\cite{raft} are the de-facto standards implemented in the wild~\cite{etcd, spanner, cockroachdb, tidb, tigerbeetle}. While stale read options exist~\cite{paxos-made-live, raft-thesis}, normal reads are treated obliviously just like other commands.

With MultiPaxos terminology, a typical protocol is as follows. A node S first makes a ``covering-all'' \texttt{Prepare} phase to settle a unique, higher \textit{ballot} number for all non-committed \textit{instances} (i.e., \textit{slots}) in the log, effectively stepping up as a leader. Without competing leaders, S takes a client command, assigns the next vacant log slot, broadcasts \texttt{Accept} messages, and waits for $\geqslant m = \lceil\frac{n}{2}\rceil$ \texttt{AcceptReply}s with matching ballot including self (where $n$ is an odd cluster size), after which the slot is marked committed and \texttt{Commit} notifications are broadcast asynchronously as announcement. S executes the commands in contiguously-committed slots in order and replies to their clients.


\paratitlenodot{Leader Leases}~\cite{paxos-made-live} are a commonly deployed optimization to establish \textit{stable leadership}. \revisetwo{All} nodes grant lease to the most recent leader they are aware of (including self) after invalidating any old lease given out. If a leader S is holding $\geqslant m$ leases, it can safely assert that it is the only such leader in the cluster, i.e., the stable leader. Therefore, S (and only S) can reply to read requests locally using the latest committed value, knowing that no newer values could have committed.

\subsubsection{Leaderless Approaches}
\label{sec:bg-previous-leaderless}

Leaderless (or multi-leader) protocols distribute the responsibilities of a leader onto all nodes, improving scalability and latency under wide-area settings by allowing a fast-path quorum nearer to the clients. However, they are sensitive to command interference and often make local reads infeasible without degrading back to a leader-based protocol.

\paratitlenodot{Mencius}~\cite{mencius} assigns the leader role Round-Robin across nodes based on slot index. This mainly benefits scalability.

\paratitlenodot{EPaxos}~\cite{epaxos, epaxos-revisited} absorbs the idea of inter-command dependencies from \textbf{Generalized Paxos}~\cite{generalized-paxos} and lets any node to act as the \textit{command leader} for nearby clients. Nodes attach to each command its dependency set; without concurrent conflicting proposals, consensus can be reached on the fast path of \texttt{PreAccept}s by a (super-)majority quorum. Conflicts in dependencies require a second phase to resolve. Local reads are inherently hard to achieve in such a protocol without degrading to a leader-based protocol on keys of interest~\cite{epaxos-revisited}.

\paratitlenodot{SwiftPaxos}~\cite{swiftpaxos} improves EPaxos slow path to 1.5 RTTs (vs. 2) by re-introducing a leader in the slow phase.

\paratitlenodot{PQR}~\cite{paxos-quorum-read, edge-pqr} applies EPaxos-like leaderless optimization to only reads and not writes. Clients broadcast read requests directly to the nearest majority of servers. If all replies contain the same latest-seen value, all in committed status, then this value must be a valid linearizable read result. Otherwise, the client starts a repeated \textit{rinse} phase, retrying on arbitrary servers until the value becomes committed.

\subsubsection{Enhanced Read Leases}
\label{sec:bg-previous-extended-leases}

Several works explored enhancements to \textit{read leases} beyond stable leadership, enabling broader local reads.

\paratitlenodot{Megastore}~\cite{megastore} grants read leases to all replicas by a standalone coordinator. These leases carry the promise of not permitting any writes to covered keys. When writes arrive, leases are actively revoked (requiring an extra round-trip to all replicas) and local reads at followers are disabled until leases are re-granted. Megastore leases cover either all replicas or none; they also require external coordination and experience long downtimes during concurrent writes.


\paratitlenodot{Quorum Leases}~\cite{quorum-leases, quorum-leases-report} extend leases to configurable subsets of replicas. Leases are granted by replicas themselves, removing the need for an external coordinator. Upon writes, revocation actions are merged with the natural \texttt{Accept} messages and their replies, avoiding extra round-trips for writes in failure-free cases. Quorum Leases improve the configurability and write performance aspects of Megastore, but insufficiencies with reads remain. \circleb{1} Lease actions remain on the critical path, leading to frequent interruptions from writes. \circleb{2} When fast-path local reads fail during lease downtimes, they are redirected to the leader or retried indefinitely by clients, leading to suboptimal slow-path latency. \circleb{3} Assignment of grantees is configurable but only through normal consensus commands, making failure cases hard to reason about and implement.

\subsubsection{With External Coordination}
\label{sec:bg-previous-data-oriented}

\revise{Protocols below assuming external coordination are notable.}

\paratitlenodot{Hermes}~\cite{hermes} is a primary-backup replication protocol inspired by cache coherence protocols. It allows reads to be completed by individual nodes assuming that writes reach all nodes and are resolved synchronously with respect to each other (similar to CPU shared cache invalidation). \revise{Hermes inherits its architectural assumption from \textbf{Vertical Paxos}~\cite{vertical-paxos}, requiring an external membership manager for reconfigurations upon failures.}

\paratitlenodot{Pando}~\cite{tradeoffs-geo-distributed} is a WAN-aware, erasure-coded protocol that emphasizes cost efficiency. It allows \revise{statically} tunable read-write quorums, which \revise{are settable ahead-of-time before deployment}, and assumes a network topology with frontends and an external service for membership management.

\subsection{Summary of Goals}
\label{sec:bg-summary-of-goals}

After reviewing existing solutions, we summarize the desired properties of a linearizable read protocol as our design goals:

\begin{itemize}
    \item \revise{\textit{Self-contained}: no external metadata oracle dependencies.}
    \item \revise{\textit{Local reads anywhere}: enable local linearizable reads at arbitrary subsets of replicas as appropriate.}
    \item \revise{\textit{Local reads at anytime}: keep reads localized during concurrent interfering writes, minimizing degradation time and maintaining good slow-case latency.}
    \item \revise{\textit{Configurable}: tunable against arbitrary ranges of keys.}
    \item \revise{\textit{Non-intrusive}: designed atop classic consensus, introducing marginal performance impacts on writes and retaining availability under any minority number of failures.}
\end{itemize}

\vspace{2pt}
\noindent\revisethree{Via these goals, \bodega delivers superior performance characteristics compared to aforementioned approaches. We will show them both theoretically (\S\ref{sec:comparison-proof}) and experimentally (\S\ref{sec:evaluation}).}

\section{Design}
\label{sec:design}

In this section, we present the core design of \bodega, an always-local linearizable read protocol.

\paratitle{Design Outline}
We derive the complete design in three steps: \circleb{1} define \revise{the roster}, \circleb{2} design optimal normal case operations assuming replicas agree on the same stable \revise{roster}, and \circleb{3} introduce \revisetwo{all-to-all} \revise{roster} leases, the enabler behind the fault-resilient agreement on the \revise{roster}.

For clarity, we adhere to Paxos-style terminology throughout this paper. All the optimizations are applicable to Raft-style protocols due to their fundamental duality~\cite{parallel-paxos-raft}.

\subsection{\revise{The Roster}}
\label{sec:design-roster}

\revise{We start by introducing the core concepts behind \bodega: responder status and the roster.} A node is a \textit{responder} for a key if it is expected to serve read requests on that key locally without actively contacting other nodes. A \textit{\revise{roster}} is the collection of each node's desired capabilities at a certain time; specifically, it dictates:

\begin{itemize}[itemsep=1pt]
    \item The node ID of the current leader node.
    \item For each (range of) key(s): the node IDs of its responders.
\end{itemize}

\noindent The \revise{roster} is a generalization of leadership from classic protocols: besides the one special leader role, we now have special responder roles for selected keys. The leader can be implicitly treated as a responder for all keys, \revise{and different keys can additionally mark different nodes as responders.}

\revise{The optimal choice of responders for each key depends on various factors: \circleb{1} client locations and proximity, \circleb{2} workload read-heaviness and skewness, and \circleb{3} cluster topology and status. This paper focuses on the mechanisms supporting the roster rather than the policies for tuning it; we recognize that the latter could be an intriguing study on its own.}


The system starts from an empty \revise{roster} with a null leader ID and an empty responder set for the entire keyspace. Every newly-proposed \revise{roster} is associated with (and identified by) a unique ballot number, forming a $\langle$\avar{bal}, \avar{ros}$\rangle$ pair, where \avar{bal} is the ballot number formed by concatenating a monotonically increasing integer $b$ with the proposing node's ID $r$ to ensure uniqueness. \revise{Rosters} of different ballots may contain the same content but are still considered different. \revise{Roster changes may happen due to explicit requests by users, automatic tuning from statistics, or mandatorily in reaction to failures.}

\subsection{Normal Case Operations}
\label{sec:design-normal-case}

We first describe normal case operations, using Figure~\ref{fig:design-agreed-roster} as a demonstrative example. In a 5-node cluster, S0 is the leader (depicted by the crown) and S2,3,4 are additional responders (depicted by the red star symbols) for a \revise{specific} key $k$. Assume, in this section, that this is the latest \revise{roster} all nodes know and consider stable according to leases. \revise{Nodes use their known roster to assure that writes to $k$ would never commit before reaching all of its active responders. A responder can therefore serve reads directly if it knows the latest value of $k$ will commit; when unsure, optimizations exist.}



\subsubsection{Writes}
\label{sec:design-normal-case-writes}

\begin{figure}[!t]
    \centering
    
    \includegraphics[width=0.915\linewidth]{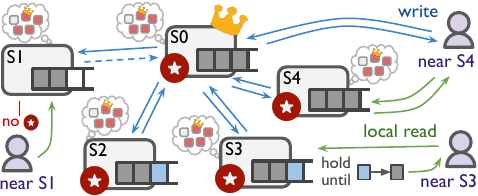}
    \captionsetup{justification=raggedright}
    \vspace{-10pt}
    \floatcap{Normal case operations of \bodega}{Assume all nodes agree on the same example \revise{roster}: S0 is the leader and S0,2-4 are responders for a key while S1 is not. See \S\ref{sec:design-normal-case}.}
    \label{fig:design-agreed-roster}
    \vspace{-3pt}
\end{figure}

Writes follow the same leader-based process as in MultiPaxos \revise{(Figure~\ref{fig:design-agreed-roster} blue arrows)}, except for an updated commit condition. Normally, a write to key $k$ can be marked as committed and acknowledged once $\geqslant m$ \texttt{AcceptReply}s are received. We impose an additional constraint that it must also have received replies from all the responders for $k$, according to the leader's current \revise{roster}.

Requiring a write quorum that covers all responders is an unavoidable penalty that any local linearizable read algorithm must pay. Luckily, without far-off responders, this penalty is marginal as the system usually picks a leader with relatively uniform distances to other replicas, and the write must anyway reach a majority. This aligns with previous observations~\cite{quorum-leases} and our evaluation results (\S\ref{sec:evaluation}). Distant responders could still be appropriate for certain workloads.

\subsubsection{Reads}
\label{sec:design-normal-case-reads}

Clients send read requests on key $k$ to the closest responder server for $k$ \revise{(Figure 4 green arrows)}. It is common for clients of wide-area systems to be co-located with some replica; for example, consensus is usually part of an outer system, e.g., a database, where requests come directly from participating sites, but this is not a requirement.


When a server S (with a stable \revise{roster}) takes a read, there are three cases. \circleb{1} S is the leader, \revisetwo{in which case} S simply finds the highest committed slot in its log that contains a write to $k$ and returns the value. The leader does not need to worry about in-progress writes~\cite{paxos-made-live}. \circleb{2} S is neither a leader nor a responder for $k$ (e.g., S1 in Fig.~\ref{fig:design-agreed-roster}), \revisetwo{in which case} S rejects the read and promptly redirects the client to another server, preferably a close-by responder or the leader. \circleb{3} S is a non-leader responder. \revisetwo{In this case}, S looks up the highest slot in its log that contains any write to $k$. If the slot is in committed status, S immediately replies with the value (e.g., the read at S4 in Fig.~\ref{fig:design-agreed-roster}); otherwise, S cannot yet determine whether that value will surely commit or will be overwritten due to \revisetwo{impending} failures (e.g., the read at S3), \revisetwo{in which case S optimistically} \textit{holds} the read.


\paratitle{Optimistic Holding}
\revisetwo{The holding mechanism is optimistic in that it expects S and its connection with the leader to remain healthy and for the slot to be committed soon. In failure-free cases, S may need to wait up to one RTT to be notified that an interfering write has committed (from when S receives the \texttt{Accept} from leader and replies to it, to when S receives the \texttt{Commit} notification). Note that even with a constant stream of writes, held reads will not be blocked indefinitely: they are released as soon as their associated slot turns committed.}

A responder S optimistically holds a local read by adding it to a \textit{pending set} attached to the corresponding slot. Upon receiving the commit notification for a slot, S releases the pending reads and replies with the committed value. \revisetwo{To handle cases where the leader fails to notify S promptly}, clients start an \textit{unhold} timeout when sending local read requests; if the timeout is reached, clients proactively issue the same request to another responder or the leader (with the same req ID, which is safe since reads are idempotent) and use the earliest reply. A good timeout length is longer than the usual RTT between S and the current leader.




\paratitle{Early Accept Notifications}
While optimistic holding already delivers outstanding local read performance, \revisetwo{\bodega} introduces a further optimization \revise{that reduces average holding time}: followers not only reply to \texttt{Accept}s on key $k$ to the leader, but also broadcast notifications to $k$'s responders. \revisetwo{Once a responder S has received $m$ \revisethree{notifications} (counting self), it} can assert that a pending slot will surely commit even across minority failures. \revisetwo{A similar optimization exists for BFT writes~\cite{pbft-recovery}.} On average, this halves the expected holding time for interfered local reads.


\subsection{\revise{Roster} Leases}
\label{sec:design-roster-leases}

So far, we have \revisetwo{assumed} a consistent \revise{roster} across the cluster without showing how it is achieved. \revisetwo{The idea is to exchange roster leases in an all-to-all manner, between at least a majority of nodes and all nodes that may be responders for some key. When holding a majority number of leases, responders know that the roster is stable, and the leader (as an implicit responder) also knows the identity of all responders and will not commit writes without notifying them.}

\begin{center}
    \vspace{-2pt}
    \small
    \setlength\tabcolsep{3pt}
    \def\arraystretch{0.95}
    \begin{tabular}{c|c|c|c}
         & Lease Primitive & Leader Leases & Roster Leases \\ \hline
        Pattern & one-to-one & all-to-one & all-to-all \\
        Safe \#grantors & $1$ & $m$ & $m$ \\
        Must to whom & - & leader & responders \\
    \end{tabular}
\end{center}

We present how \bodega deploys off-the-critical-path \textit{roster leases} to establish \revisetwo{a stable} \revise{roster} elegantly and efficiently. We use Figure~\ref{fig:design-roster-leases} as an illustration when needed.

\paratitle{Lease-related States}
Besides the SMR log and the $\langle$\avar{bal}, \avar{ros}$\rangle$ pair, we let every node S \revise{\revisetwo{act as both} a lease grantor and a grantee} \revisetwo{(recall \S\ref{sec:bg-distributed-leases} for how a standard lease grant primitive works)}. This means S maintains the following additional data structures: \circleb{1} two lists of grantor-side timers \atimer{guarding, $p$} and \atimer{endowing, $p$} per peer node P; \circleb{2} two corresponding sets \aset{guarding} and \aset{endowing} for tracking which peers are S currently guarding/endowing to; \circleb{3} two lists of grantee-side timers \atimer{guarded, $p$} and \atimer{endowed, $p$} per peer P; \circleb{4} two sets \aset{guarded} and \aset{endowed} for tracking the guards/endowments S currently holds; \circleb{5} a list of safety slot numbers \avar{thresh$_p$}, \revisethree{that specifies the highest slot S has accepted from each peer P}.


\subsubsection{\revise{Roster} Leases Activation}
\label{sec:design-roster-leases-activation}

We first describe how \revise{roster} leases are activated. Consider node \revisetwo{X} wants to announce a new \revise{roster} \avpm{ros}; this could be due to, e.g., stepping up as new leader (by setting \revisetwo{X} as the leader in \avpm{ros}) or other reasons covered in \S\ref{sec:design-roster-leases-expiration}. \revisetwo{X} composes a unique, higher ballot \avpm{bal} by concatenating $(b+1)$ with its node ID, where $b$ is the higher part of the current \avar{bal}. \revisetwo{X} then broadcasts the $\langle$\avpm{bal}, \avpm{ros}$\rangle$ pair to all nodes including self.

For any node S upon receiving a ballot \avpm{bal} higher than ever seen, it first ensures all old leases are safely revoked \revisetwo{or expired} (discussed later in \S\ref{sec:design-roster-leases-expiration}). Then, it moves on to $\langle$\avpm{bal}, \avpm{ros}$\rangle$ and starts a \afun{initiate\_leases}{\avpm{bal}} procedure, where it begins granting leases for the new roster to all peers asynchronously in parallel.


\revisetwo{To each peer P, the procedure obeys standard lease granting: S and P first complete the guard phase, exchanging a sequence of \texttt{Guard}, \texttt{GuardReply}, \texttt{Renew}, and \texttt{RenewReply}, and utilizing proper timers along the way. If all goes well, S should have P in its \aset{endowing} and have \atimer{endowing} properly extended; it repeats renewals periodically to keep the S-to-P lease refreshed. Similarly, P should have S in its \aset{endowed} and have \atimer{endowed} kicked off properly.} Whenever a \atimer{intent, $p$} times out for any intent among the four, the peer is removed from the corresponding set \aset{intent}, \revisetwo{leading to a retry of the guard phase or a proposal of a new roster}.


After transitioning to \avpm{ros}, if S sees itself being the leader of \avpm{ros}, it does the usual step-up routine of redoing the \texttt{Prepare} phase for non-committed slots of its log.


\paratitle{Stable Condition \& Safety Thresholds}
As shown in \S\ref{sec:bg-distributed-leases} and above, a node P is considered granted a lease by S when S $\in$ P's \aset{endowed} set. Assume P itself is always in the set. The size of this set, $|$\aset{endowed}$|$, indicates the number of lease grants P currently holds. When $|$\aset{endowed}$|$ $\geqslant m$, then P knows at least a majority number of nodes in the cluster has the same latest $\langle$\avar{bal}, \avar{ros}$\rangle$ as P and that at most one such \revise{roster} exists; this is called the \textit{stable} \revise{roster} of the cluster and is a necessary precondition for all optimizations described in \S\ref{sec:design-normal-case}. For example, in Figure~\ref{fig:design-roster-leases}, the local reads at S1 and S2 are rejected due to an insufficient lease count.

This condition alone is not enough, though. When a node directly inspects its log and uses the highest slot index for local reads, it is assuming that its log is up-to-date and contains all the recently accepted instances; this is normally true, but could be violated when a fell-behind node joins a new \revise{roster}. To address this, \revisetwo{a node should be informed of other peer's acceptance progress when transitioning to a new roster}.

We let \texttt{Guard} messages from S to P carry an extra number, which is the highest slot number that S has ever accepted. P stores the number in its \avar{thresh} list. A node permits local reads only if it has committed all the slots up to the $m$-th smallest slot number in its \avar{thresh} list; \revisetwo{otherwise, it might not have observed the latest committed writes yet}. S4 in Figure~\ref{fig:design-roster-leases}, for example, has not reached this condition.

In summary, all the stable leader and local read operations of \S\ref{sec:design-normal-case} are preceded by the following stable condition check:
\begin{align} \label{eqn:stable-condition}
    & |\text{\{\}}_{\text{endowed}}| \geqslant m \\
    & \wedge \,\, \exists \text{ size-}m \text{ subset } E \subseteq \text{\{\}}_{\text{endowed}}: \nonumber \\[-2pt]
    & \quad\,\,\,\,\,\,\, \text{ committed all slots up to \textit{thresh}}_p, \forall p \in E \nonumber
\end{align}
\noindent If the check fails, the operation falls back to classic consensus as if it is a write, which does not require this check.

\begin{figure}[!t]
    \centering

    \includegraphics[width=0.99\linewidth]{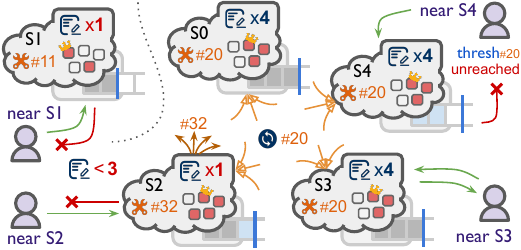}
    \captionsetup{justification=raggedright}
    \vspace{-6pt}
    \floatcap{All-to-all \revise{roster} leases demonstrated}{S0,3,4 are each holding $\ge$ majority grants of \revise{roster} \#20; S4 has not seen all slots up to \#20's safety threshold. S1 is stuck with an older \revise{roster} of \#11. S2 is initiating a new \revise{roster} of \#32. See \S\ref{sec:design-roster-leases}.}
    \label{fig:design-roster-leases}
\end{figure}


\begin{figure*}[t]
    \centering
    
    \vspace*{-2pt}
    \begin{subfigure}[t]{0.145\linewidth}
        \centering
        \includegraphics[width=0.99\columnwidth]{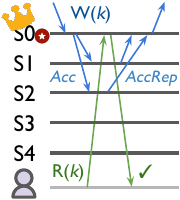}
        \captionsetup{justification=centering}
        \caption{Leader Leases}
        \label{fig:timeline-leader-leases}
    \end{subfigure}\hspace{4.2pt}%
    \begin{subfigure}[t]{0.193\linewidth}
        \centering
        \includegraphics[width=0.99\columnwidth]{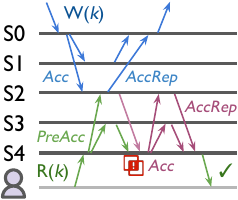}
        \captionsetup{justification=centering}
        \caption{EPaxos}
        \label{fig:timeline-epaxos}
    \end{subfigure}\hspace{2.5pt}%
    \begin{subfigure}[t]{0.224\linewidth}
        \centering
        \includegraphics[width=0.99\columnwidth]{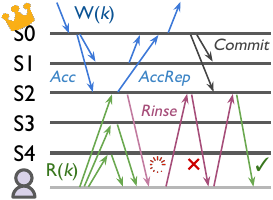}
        \captionsetup{justification=centering}
        \caption{PQR}
        \label{fig:timeline-pqr}
    \end{subfigure}\hspace{2.5pt}%
    \begin{subfigure}[t]{0.247\linewidth}
        \centering
        \includegraphics[width=0.99\columnwidth]{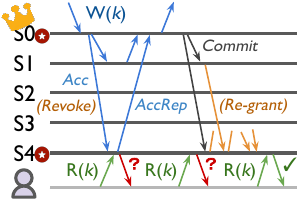}
        \captionsetup{justification=centering}
        \caption{Quorum Leases}
        \label{fig:timeline-quorum-leases}
    \end{subfigure}\hspace{2.5pt}%
    \begin{subfigure}[t]{0.167\linewidth}
        \centering
        \includegraphics[width=0.99\columnwidth]{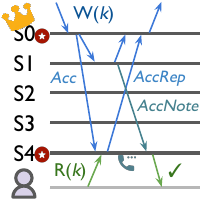}
        \captionsetup{justification=centering}
        \caption{Bodega}
        \label{fig:timeline-bodega}
    \end{subfigure}
    
    \addtocounter{figure}{-1}  
    \captionsetup{justification=raggedright}
    \vspace{-2pt}
    \floatcapoffignodot{Timeline comparison across protocols of linearizable reads}{in the presence of an interfering write. See \S\ref{sec:bg-previous-approaches}, \ref{sec:formal-comparison-across}.}
    \label{fig:timeline-comparison}
\end{figure*}

\subsubsection{Revocation \& Expiration}
\label{sec:design-roster-leases-expiration}

Most \revise{roster} lease activations happen when there are ongoing old leases in the system. Broadly speaking, a \revise{roster} change may be triggered by one of the following reasons.

\begin{itemize}
    \item Node initiates a new \revise{roster} in reaction to suspected failures, removing failed nodes from special responder roles.
    \item Node autonomously proposes a new, more optimized \revise{roster} according to collected workload statistics.
    \item Node receives an explicit \revise{roster} change request from user.
\end{itemize}

\noindent In either case, before \afun{initiate\_leases}{}, a node S always invokes the \afun{revoke\_leases}{\avar{bal}} procedure synchronously to ensure that all the leases it is granting or holding with the older ballot \avar{bal} are safely revoked and removed. To do so, S clears its \aset{guarding} set and broadcasts \texttt{Revoke} messages carrying the old ballot. Whenever a node P receives a \texttt{Revoke} with matching ballot from S, it removes S from the \aset{guarded} and \aset{endowed} sets and replies with \texttt{RevokeReply}.

S either receives a \texttt{RevokeReply} from P promptly (common fast case) or has to wait for expiration timeout (failure case), after which it removes P from \aset{endowing}. \revise{Note that, unless failures occur and force a wait on expiration, a roster change completes swiftly within two message rounds: one for the revocation and the other for the initiation guards.}

\subsubsection{Piggybacking on Heartbeats}
\label{sec:design-roster-leases-piggyback}

Heartbeats are ubiquitous in modern distributed systems; many systems already deploy all-to-all heartbeats for tasks such as health tracking~\cite{etcd, zookeeper, epaxos, rabbitmq}. This opens the opportunity to enable \revise{roster} leases without any common-case overheads, by piggybacking lease messages onto existing periodic heartbeats. \bodega piggybacks all the \texttt{Renew} and \texttt{RenewReply} messages onto heartbeats, and uses a proper heartbeat interval \atmdur{hb\_send} such that leases are refreshed in time. Heartbeat messages also carry the sender's $\langle$\avar{bal}, \avar{ros}$\rangle$ pair to let receivers discover \revise{roster} changes.

Each node has per-peer timers \atimer{heartbeat, $p$} which are used for detecting failures from peers; a peer is considered down if no heartbeats were received from it for \atmdur{hb\_fail}. A rule of thumb for choosing good timeout lengths for a cluster is:
\begin{align}
    \text{avg. RTT} < \text{\atmdur{hb\_send}} \ll \text{\atmdur{hb\_fail}} < \text{\atmdur{guard}} = \text{\atmdur{lease}}
\end{align}
\noindent \bodega uses the following defaults for wide-area replication: \atmdur{hb\_send} $=$ 120ms, \atmdur{hb\_fail} $\approx$ 1200ms, \atmdur{guard} $=$ \atmdur{lease} $=$ 2500ms.




\revise{We provide a complete presentation of the \bodega protocol in Figure~\ref{fig:algo-blocks-full} of Appendix~\ref{sec:appendix-full-algo} as an implementation guide.}

\section{Formal Comparison and Proof}
\label{sec:comparison-proof}

\revise{For completeness, we give a qualitative comparison across all the notable linearizable read optimizations (Figure~\ref{fig:timeline-comparison}, Table~\ref{tab:comparison}), and provide a concise proof of correctness.}

\subsection{Comparison Across Protocols}
\label{sec:formal-comparison-across}

In Table~\ref{tab:comparison}, we model the normal-case write and read latency, degraded read latency under write interference, and degradation period length of related protocols. \revisetwo{Cells are shaded according to example values from the Figure~\ref{fig:eval-setting-geo} GEO setting (lighter is better).} We also indicate whether a protocol retains the fault tolerance of classic protocols and whether it allows tunable \revise{rosters}. If tunable, we use its most read-optimized \revise{roster} that tolerates $f = \lfloor\frac{n}{2}\rfloor$ faults. Assume only one interfering write.

The following symbols are used to model the performance metrics. $\underline{l}$: RTT between client and the leader, $\underline{c}$: RTT between client and its nearest server, $\underline{m}$: time to establish a simple majority quorum (i.e., to reach majority nodes from some server and receive replies), $\underline{M}$: time to establish a super majority quorum (as in EPaxos~\cite{epaxos}), $\underline{N}$: time to form a quorum composed of all nodes. For an average client in typical WAN-scale settings, one should expect $c \ll l \approx m < M < N$.

Most results are derived naturally from Figure~\ref{fig:timeline-comparison}, \S\ref{sec:bg-previous-approaches}, and \S\ref{sec:design-roster}-\ref{sec:design-roster-leases}. We provide supplementary explanations. \textbf{PQR (+ Ldr Ls)} is a straightforward variant of PQR combined with Leader Leases; if a near quorum read attempt fails, the client contacts the stable leader directly, bounding slow-path latency by $c+m+l$. We assume Quorum Leases always incorporate Leader Leases. \textbf{Qrm Ls (passive)} is a variant of Quorum Leases where we deliberately let grantees keep the leases upon accept to show the upper bound of Quorum Leases performance, saving one re-granting RTT from the degradation time. Doing so risks blocking fault-induced \revise{roster} change commands as described in \S\ref{sec:design-roster-leases}. Hermes uses primary-backup broadcast and thus requires external coordination for fault tolerance; Megastore is similar. Pando uses a pre-deployment planner to dictate erasure coding and quorum composition. \bodega achieves the best across all metrics and retains fault tolerance and configurability.

\begin{table}[t]
    \renewcommand{\arraystretch}{1.1}
    \setlength\tabcolsep{2.56pt}
    \small
    \begin{tabular}{c|c|c|c|c|c|c}
        \hline \hline
            \textbf{Protocol} &
            $\pmb{W}$ &
            $\pmb{R}$ &
            $\pmb{R^*}$ &
            $\pmb{D^*}$ &
            \raisebox{-0.23\height}{\includegraphics[height=12.6pt]{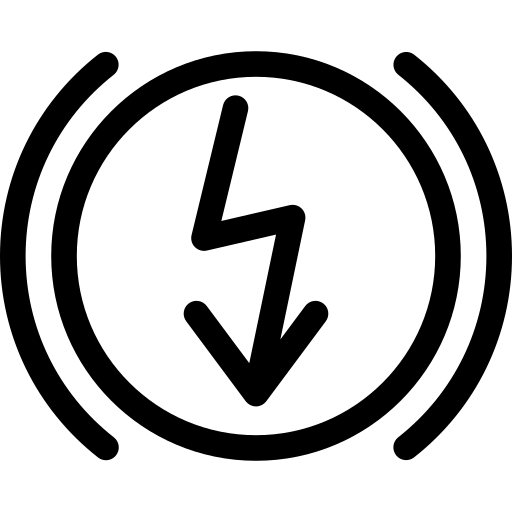}} &
            \raisebox{-0.17\height}{\includegraphics[height=10.3pt]{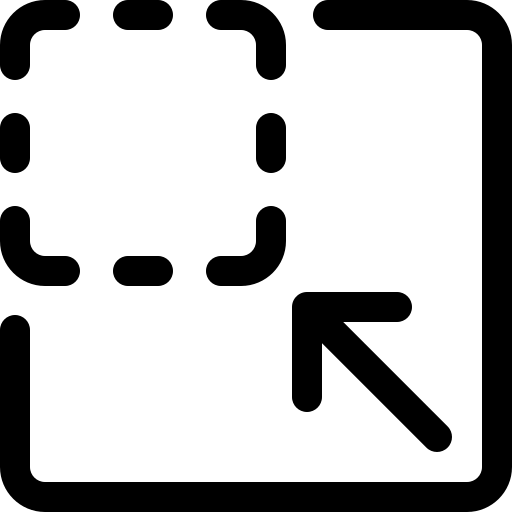}} \\
        \hline \hline
            Leader Ls &
            \cellcolor{black!8}$l+m$ &
            \cellcolor{black!22}$l$ &
            \cellcolor{black!9}$l$ &
            - &
            \CIRCLE &
            \Circle \\
        \hline
            EPaxos &
            \cellcolor{black!4}$c+M$ &
            \cellcolor{black!27}$c+M$ &
            \cellcolor{black!25}$c+M+m$ &
            \cellcolor{black!7}$M$ &
            \CIRCLE &
            \Circle \\
        \hline
            Hermes &
            \cellcolor{black!9}$c+N$ &
            \cellcolor{black!3}$c$ &
            \cellcolor{black!11}$c \sim c+\frac{N}{2}$ &
            \cellcolor{black!6}$\frac{N}{2}$ &
            \LEFTcircle &
            \Circle \\
        \hline
            PQR &
            \cellcolor{black!8}$l+m$ &
            \cellcolor{black!22}$c+m$ &
            \cellcolor{black!23}$c+m * \text{rinses}$ &
            \cellcolor{black!5}$m$ &
            \CIRCLE &
            \Circle \\
        \hline
            PQR (+ Ldr Ls) &
            \cellcolor{black!8}$l+m$ &
            \cellcolor{black!22}$c+m$ &
            \cellcolor{black!23}$c+m+l$ &
            \cellcolor{black!5}$m$ &
            \CIRCLE &
            \Circle \\
        \hline
            Pando &
            \cellcolor{black!3}$c+m$ &
            \cellcolor{black!22}$c+m$ &
            \cellcolor{black!27}$c+N$ &
            \cellcolor{black!13}$N$ &
            \LEFTcircle &
            \LEFTcircle \\
        \hline
            Megastore &
            \cellcolor{black!27}$l+2N$ &
            \cellcolor{black!3}$c$ &
            \cellcolor{black!9}$c+l$ &
            \cellcolor{black!27}$2N$ &
            \LEFTcircle &
            \Circle \\
        \hline
            Quorum Ls &
            \cellcolor{black!14}$l+N$ &
            \cellcolor{black!3}$c$ &
            \cellcolor{black!9}$c+l$ &
            \cellcolor{black!27}$2N$ &
            \CIRCLE &
            \CIRCLE \\
        \hline
            Qrm Ls (passive) &
            \cellcolor{black!14}$l+N$ &
            \cellcolor{black!3}$c$ &
            \cellcolor{black!9}$c+l$ &
            \cellcolor{black!13}$N$ &
            \LEFTcircle &
            \LEFTcircle \\
        \hline
            \bodega &
            \cellcolor{black!14}$l+N$ &
            \cellcolor{black!3}$c$ &
            \cellcolor{black!3}$c \sim c+\frac{m}{2}$ &
            \cellcolor{black!3}$\frac{m}{2}$ &
            \CIRCLE &
            \CIRCLE \\
        \hline \hline
    \end{tabular}

    \vspace{11pt}
    \floatcapnodot{Qualitative comparison across protocols}{assuming the most read-optimized \revise{roster} of each protocol. $\pmb{W}$: write latency; $\pmb{R}$: read latency if quiescent; $\pmb{R^*}$: read latency if there is an interfering write; $\pmb{D^*}$: read performance degradation period length. \raisebox{-0.23\height}{\includegraphics[height=12.6pt]{figures/icon-fault-tolerance.png}}: \revise{fault tolerance (without external oracle)}. \raisebox{-0.17\height}{\includegraphics[height=10.3pt]{figures/icon-tunable-config.png}}: allows tunable \revise{rosters}. See \S\ref{sec:formal-comparison-across} for the use of metric symbols. \revisetwo{Cells shaded darker if value higher using Fig.\ref{fig:eval-setting-geo} GEO as example.}}
    \label{tab:comparison}
    \vspace*{-25pt}
\end{table}

\subsection{Proof}
\label{sec:formal-proof}

We provide a proof of \bodega's local read linearizability and write liveness, assuming well-established results of the safety and liveness of leases~\cite{leases-mechanism}. \revise{For linearizability, only locally-served reads need proof, as \bodega behaves the same as classic consensus otherwise.}

\paratitle{Linearizability} A local read $R$ served by server S observes any write $W$ acknowledged cluster-wise before $R$ was issued.

\vspace{-5pt}
\begin{proof}[\textsc{Proof}]
$R$ is served locally only if S is a responder that passes the stable \revise{roster} check (\ref{eqn:stable-condition}). Let the stable \revise{roster} be $\langle$\avar{bal}, \avar{ros}$\rangle$.

\vspace{2pt}
\noindent \underline{Case \#1}: $W$ was committed on a ballot $>$ \avar{bal}. It is impossible because the latest ballot on at least a majority is \avar{bal}.

\vspace{2pt}
\noindent \underline{Case \#2}: $W$ was committed on a ballot $=$ \avar{bal}. By the injective ballot-\revise{roster} mapping and the commit condition of writes, S must be in $W$'s write quorum and have $W$ in its log.

\vspace{2pt}
\noindent \underline{Case \#3}: $W$ was committed on a ballot $<$ \avar{bal}. By majority intersection, for any size-$m$ subset $E \in$ S's \aset{endowed}, at least one of the lease grantors P $\in E$ accepted $W$ at its committed slot $x$ before granting to S. This implies \avar{thresh}$_p$ $\geqslant x$.
\end{proof}
\vspace{-6pt}

\paratitle{Liveness of Writes}
A write $W$ can always eventually make progress if retried on a majority group $G$ of healthy servers.

\vspace{-5pt}
\begin{proof}[\textsc{Proof}]
By the property of leases, after old leases expire, a \revise{roster} change can eventually be made on all servers $\in G$ to restrict the leader and all the responders to be contained in $G$. Then, normal consensus applies.
\end{proof}
\vspace{-6pt}




\section{Implementation}
\label{sec:implementation}

We present details of our practical \bodega implementation.

\paratitle{The \summerset KV-store} We develop \summerset, a distributed, replicated, protocol-generic key-value store. \summerset is written in async Rust/\texttt{tokio} using a lock-less architecture and serves as a fair codebase for evaluating consensus and replication protocols. We do not stack our implementation on top of previous research codebases~\cite{epaxos, chain-paxos} due to their lack of extensibility and noticeable language runtime overheads.

The codebase has 13.6k lines of infrastructure code and includes five protocols of interest: MultiPaxos w/ Leader Leases (2.5k), EPaxos (3.1k), PQR \& variant (2.8k), Quorum Leases \& variant (3.2k), and \bodega (3.0k). All protocol implementations have passed extensive tests. We will open-source our complete codebase upon publication.

\subsection{Smart \revise{Roster} Coverage}
\label{sec:impl-roster-coverage}

In cases where users desire local reads but cannot observe workload patterns externally, \bodega servers can collect statistics and automatically propose \revise{roster} changes to mark servers as responders for proper keys. Our default implementation traces per-key read/write request counts grouped by clients' preferred nearby server IDs. For a key, if $>$ 95\% requests are reads at a periodic check, then servers near $>$ 20\% of the reads are added as responders. More sophisticated strategies exist; for example, straggler detection can help remove fail-slow nodes from responders promptly~\cite{gray-failures, canopy}.

\subsection{Lightweight Heartbeats}
\label{sec:impl-lightweight-heartbeats}

In \S\ref{sec:design-roster-leases}, we described \revise{roster} leases as if all heartbeats carry the complete \revise{roster} data structure. In practice, \revise{rosters} with fine-grained key ranges can get large (tens of KBs). Luckily, most heartbeats in \bodega are \textit{lightweight heartbeats}: the sender puts in only the ballot number to indicate that the \revise{roster} has not changed from previous heartbeats. Full-sized heartbeats are sent when changes occur.

Similarly, clients may request a server to send the \revise{roster} along with a command reply, and then cache this \revise{roster} as a heuristic for choosing the best responder for local reads.

\subsection{Other Practical Details}
\label{sec:impl-other-practical}


\paratitle{Request Batching}
As is common practice, \bodega deploys request batching at servers (at 1 ms intervals) for non-local-read commands. Each log slot contains a batch of requests and the commit condition is checked for all writes contained.

\paratitle{Snapshots}
\bodega servers take periodic snapshots of the executed prefix of the log~\cite{raft}. Local reads past the beginning of truncated log looks up the latest snapshot directly.

\paratitle{Membership Management}
Membership changes are handled identically to \textit{reconfigurations} in other protocols~\cite{paxos-made-live, viewstamped-replication, epaxos}, just with an extra step of proposing and stabilizing an empty \revise{roster} with no responders ahead of the change.

\section{Evaluation}
\label{sec:evaluation}

\begin{figure}[t]
    \centering
    
    \begin{subfigure}[t]{0.46\linewidth}
        \centering
        \includegraphics[width=\columnwidth]{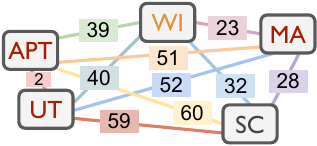}
        \captionsetup{justification=centering}
        \caption{WAN (CloudLab sites)}
        \label{fig:eval-setting-wan}
    \end{subfigure}
    \hfill
    \vrule
    \hfill
    \begin{subfigure}[t]{0.52\linewidth}
        \centering
        \includegraphics[width=\columnwidth]{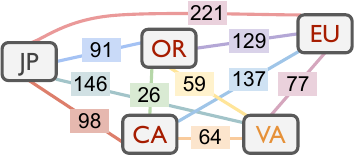}
        \captionsetup{justification=centering}
        \caption{GEO (GCP~\cite{epaxos-revisited}, emulated)}
        \label{fig:eval-setting-geo}
    \end{subfigure}

    \addtocounter{figure}{-1}  
    \captionsetup{justification=raggedright}
    \vspace{-3pt}
    \floatcapoffig{Evaluation settings}{\textcolor{BurntOrange}{Orange} denotes designated leader node and \textcolor{Maroon}{Red} denotes other responders, if relevant. \revisethree{The edges mark per-pair RTT values in milliseconds. See \S\ref{sec:evaluation}.}}
    \label{fig:eval-settings}
    \vspace*{-5pt}
\end{figure}

\begin{figure*}[t]
    \centering
    
    \vspace*{-3pt}
    \begin{minipage}{0.03\linewidth}
        \begin{subfigure}[t]{\columnwidth}
            \centering
            \includegraphics[width=\columnwidth]{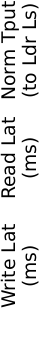}
            \captionsetup{justification=centering}
        \end{subfigure}
    \end{minipage}
    \hfill
    \begin{minipage}{0.349\linewidth}
        \begin{subfigure}[t]{\columnwidth}
            \centering
            \includegraphics[width=\columnwidth]{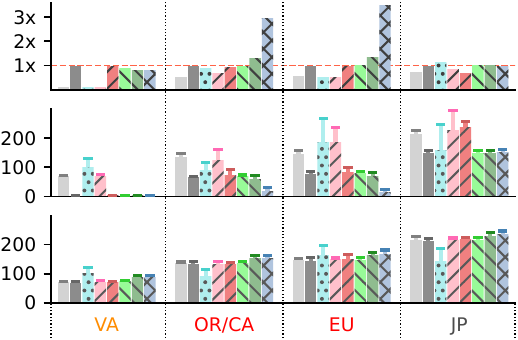}
            \captionsetup{justification=centering}
            \caption{GEO, 10\% writes}
            \label{fig:exper-loc_grid_geo-w10}
        \end{subfigure}
    \end{minipage}
    \hfill
    \begin{minipage}{0.349\linewidth}
        \begin{subfigure}[t]{\columnwidth}
            \centering
            \includegraphics[width=\columnwidth]{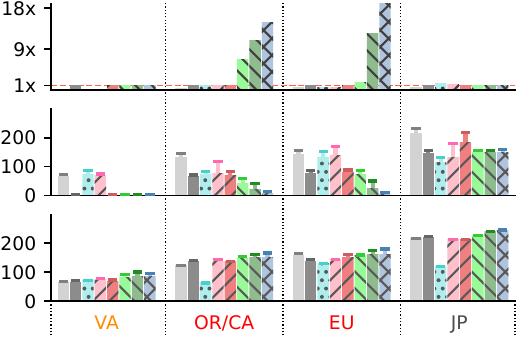}
            \captionsetup{justification=centering}
            \caption{GEO, 1\% writes}
            \label{fig:exper-loc_grid_geo-w1}
        \end{subfigure}
    \end{minipage}
    \hfill
    \begin{minipage}{0.256\linewidth}
        \vspace{-1pt}
        \begin{subfigure}[t]{\columnwidth}
            \centering
            \includegraphics[width=\columnwidth]{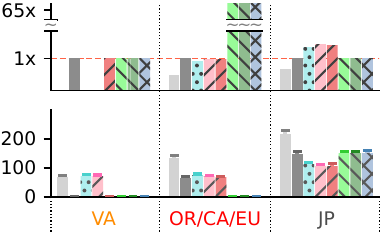}
        \end{subfigure}
        \begin{subfigure}[t]{\columnwidth}
            \centering
            \vspace{-12pt}
            \includegraphics[width=\columnwidth]{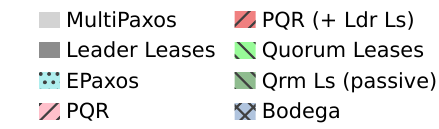}
            \captionsetup{justification=centering}
            \vspace{-16pt}
            \caption{GEO, 0\% writes}
            \label{fig:exper-loc_grid_geo-w0}
        \end{subfigure}
    \end{minipage}

    \vspace*{10pt}
    \begin{minipage}{0.03\linewidth}
        \begin{subfigure}[t]{\columnwidth}
            \centering
            \includegraphics[width=\columnwidth]{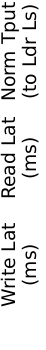}
            \captionsetup{justification=centering}
        \end{subfigure}
    \end{minipage}
    \hfill
    \begin{minipage}{0.349\linewidth}
        \begin{subfigure}[t]{\columnwidth}
            \centering
            \includegraphics[width=\columnwidth]{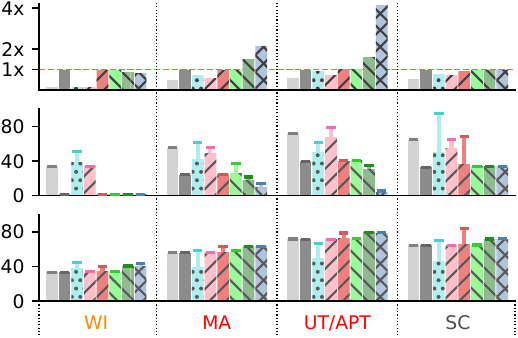}
            \captionsetup{justification=centering}
            \caption{WAN, 10\% writes}
            \label{fig:exper-loc_grid_wan-w10}
        \end{subfigure}
    \end{minipage}
    \hfill
    \begin{minipage}{0.349\linewidth}
        \begin{subfigure}[t]{\columnwidth}
            \centering
            \includegraphics[width=\columnwidth]{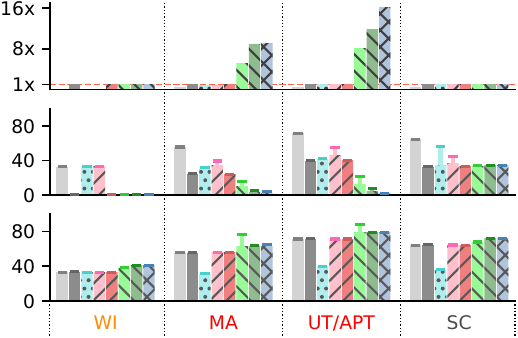}
            \captionsetup{justification=centering}
            \caption{WAN, 1\% writes}
            \label{fig:exper-loc_grid_wan-w1}
        \end{subfigure}
    \end{minipage}
    \hfill
    \begin{minipage}{0.256\linewidth}
        \vspace{-1pt}
        \begin{subfigure}[t]{\columnwidth}
            \centering
            \includegraphics[width=\columnwidth]{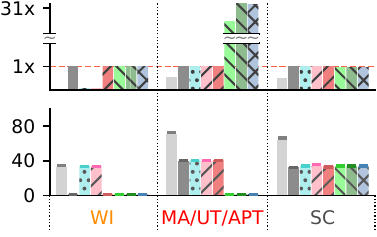}
        \end{subfigure}
        \begin{subfigure}[t]{\columnwidth}
            \centering
            \vspace{-12pt}
            \includegraphics[width=\columnwidth]{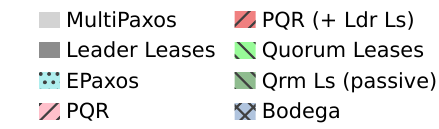}
            \captionsetup{justification=centering}
            \vspace{-16pt}
            \caption{WAN, 0\% writes}
            \label{fig:exper-loc_grid_wan-w0}
        \end{subfigure}
    \end{minipage}
    
    \addtocounter{figure}{-1}  
    \captionsetup{justification=raggedright}
    \floatcapoffignodot{Normalized throughput and latency at different client locations}{across different write intensities. See \S\ref{sec:eval-normal-case}.}
    \label{fig:exper-loc_grid}
\end{figure*}

We do comprehensive evaluations to answer these questions:

\begin{itemize}[itemsep=1pt,topsep=2pt]
    \item How does \bodega perform compared to other protocols under microbenchmarks of various write intensities? (\S\ref{sec:eval-normal-case})
    \item What do the request latency distributions look like? (\S\ref{sec:eval-detailed-anatomy})
    \item What are the impacts of write interference, and how does performance change with write ratios \& value sizes? (\S\ref{sec:eval-detailed-anatomy})
    \item How do \revise{roster} changes impact performance? (\S\ref{sec:eval-roster-change-compo})
    \item How does \bodega behave with different choices of responders and different coverages of keys? (\S\ref{sec:eval-roster-change-compo})
    \item How does \bodega compare with production coordination services, etcd \& ZooKeeper, under YCSB? (\S\ref{sec:eval-ycsb-zk-etcd})
\end{itemize}

\paratitle{Experimental Setup}
All experiments are run on two CloudLab~\cite{cloudlab} clusters, hereafter called WAN and GEO, shown in Figure~\ref{fig:eval-settings}. WAN is a wide-area cluster spanning five CloudLab sites with nodes of similar hardware types: WI-\texttt{c220g5}, UT-\texttt{xl170}, SC-\texttt{c6320}, MA-\texttt{rs620}, and APT-\texttt{r320}. \S\ref{sec:eval-normal-case} also includes a GEO cluster of five \texttt{c220g5} nodes emulated with Google Cloud RTTs reported in previous work~\cite{epaxos-revisited} using Linux kernel \texttt{netem}~\cite{tc-netem}. All nodes' public NICs have 1Gbps bandwidth. In experiments where the responder roles are controlled, the orange-colored site in Figure~\ref{fig:eval-settings} denotes the leader and the red-colored ones denote responders.


Clients are launched on machines evenly distributed across all datacenters, each marking the nearby server as their preferred server for local reads when eligible. All machines run Linux kernel v6.1.64 and pin processes to disjoint cores. All protocols use 120 ms heartbeat interval, 1200$\pm$300 ms randomized heartbeat timeout, and 2500$\pm$100 ms lease expiration (if applicable). All protocols send immediate \texttt{Commit} notifications: whenever a commit decision is made, \texttt{Commit}s are broadcast to other servers promptly. Non-self-contained protocols are not considered in the experiments.


\begin{figure*}[t]
    \centering
    
    \vspace*{-3pt}
    \begin{minipage}{\linewidth}
        \centering
        \includegraphics[width=0.95\columnwidth]{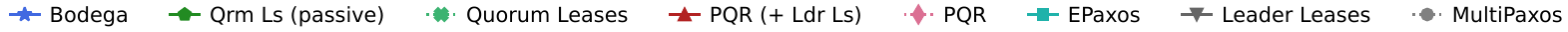}
        \captionsetup{justification=centering}
    \end{minipage}

    \vspace*{4pt}
    \begin{minipage}{0.015\linewidth}
        \begin{subfigure}[t]{\columnwidth}
            \centering
            \vspace{20pt}
            \includegraphics[width=\columnwidth]{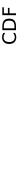}
            \captionsetup{justification=centering}
        \end{subfigure}
    \end{minipage}
    \hfill
    \begin{minipage}{0.24\linewidth}
        \begin{subfigure}[t]{\columnwidth}
            \centering
            \includegraphics[width=\columnwidth]{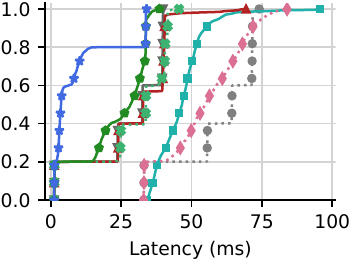}
            \captionsetup{justification=centering}
            \caption{Reads w/ 10\% writes}
            \label{fig:exper-latency_cdfs-w10-rd}
        \end{subfigure}
    \end{minipage}
    \hfill
    \begin{minipage}{0.24\linewidth}
        \begin{subfigure}[t]{\columnwidth}
            \centering
            \includegraphics[width=\columnwidth]{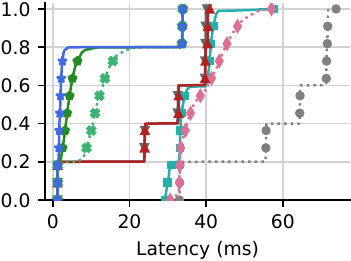}
            \captionsetup{justification=centering}
            \caption{Reads w/ 1\% writes}
            \label{fig:exper-latency_cdfs-w1-rd}
        \end{subfigure}
    \end{minipage}
    \hfill
    \begin{minipage}{0.24\linewidth}
        \begin{subfigure}[t]{\columnwidth}
            \centering
            \includegraphics[width=\columnwidth]{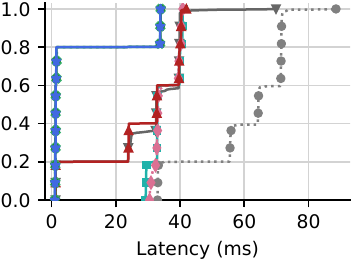}
            \captionsetup{justification=centering}
            \caption{Reads w/ 0\% writes}
            \label{fig:exper-latency_cdfs-w0-rd}
        \end{subfigure}
    \end{minipage}
    \hfill
    \begin{minipage}{0.24\linewidth}
        \begin{subfigure}[t]{\columnwidth}
            \centering
            \includegraphics[width=\columnwidth]{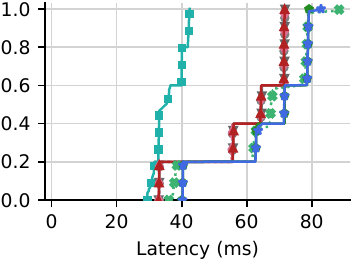}
            \captionsetup{justification=centering}
            \caption{Writes}
            \label{fig:exper-latency_cdfs-w1-wr}
        \end{subfigure}
    \end{minipage}
    
    \addtocounter{figure}{-1}  
    \captionsetup{justification=raggedright}
    \vspace{-3pt}
    \floatcapoffignodot{Latency CDFs of requests}{in the WAN setting across different write intensities, focusing on one specific key. See \S\ref{sec:eval-detailed-anatomy}.}
    \label{fig:exper-latency_cdfs}
\end{figure*}

\begin{figure*}[t]
    \centering

    \vspace{6pt}
    \begin{minipage}{0.436\textwidth}
        \includegraphics[width=\linewidth]{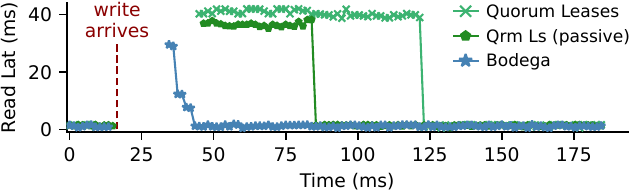}
        \captionsetup{justification=raggedright}
        \vspace{-21pt}
        \floatcapoffig{Read latency after an interfering write}{The x-axis is time at which the reads finish. See \S\ref{sec:eval-detailed-anatomy}.}
        \label{fig:exper-rlats_on_write}
    \end{minipage}
    \hfill
    \begin{minipage}{0.325\textwidth}
        \includegraphics[width=\linewidth]{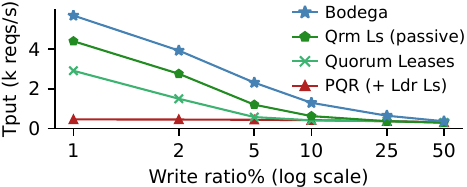}
        \captionsetup{justification=raggedright}
        \vspace{-20pt}
        \floatcapoffig{Throughput vs. write ratio}{The x-axis is log-scale (same for Fig.~\ref{fig:exper-writes_sizes-size}).}
        \label{fig:exper-writes_sizes-puts}
    \end{minipage}
    \hfill
    \begin{minipage}{0.212\textwidth}
        \includegraphics[width=\linewidth]{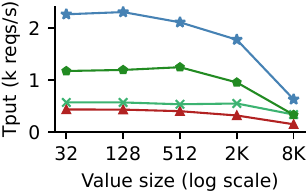}
        \captionsetup{justification=raggedright}
        \vspace{-21pt}
        \floatcapoffig{Throughput vs. value size}{See \S\ref{sec:eval-detailed-anatomy}.}
        \label{fig:exper-writes_sizes-size}
    \end{minipage}
\end{figure*}

\subsection{Normal Case Performance}
\label{sec:eval-normal-case}

We run microbenchmarks on both cluster settings and compare the following linearizable read protocols: ordinary MultiPaxos, Leader Leases, EPaxos, PQR, PQR (+ Leader Leases) variant, Quorum Leases, Quorum Leases (passive) variant, and \bodega. We spawn 50 closed-loop clients with 10 near each server and let all clients run a microbenchmark with 1k 8B-size keys and 128B values; keys are chosen uniformly. We test three write percentages in the workload mix: 0\%, 1\%, and 10\%. Figure~\ref{fig:exper-loc_grid} shows the normalized throughput (w.r.t. Leader Leases), avg. read latency, and avg. write latency perceived by clients at different locations. Leader and responders (of the full key range) are set as depicted in Figure~\ref{fig:eval-settings}. The red dashed lines indicate baseline Leader Leases throughput, and the top Ts on latency bars indicate P99 latency.

The results yield the following observations. First, except for a few datapoints (which we soon discuss), both GEO and WAN clusters exhibit similar performance patterns, just with different absolute values due to RTT differences.

\revise{Second}, for writes, all protocols except EPaxos exhibit similar performance. Quorum Leases and \bodega have slightly higher write latency due to the requirement of writes reaching responders; this explains the small throughput gap between them and Leader Leases for near-leader clients with 10\% writes. EPaxos delivers better average (but not P99) write latency due to its leaderless write protocol design.

\revise{Third}, we observe coherent patterns for read performance. \circleb{1} Compared to ordinary MultiPaxos, Leader Leases cut read latency for near-leader clients to nearly zero, but do not help other clients as much; they still pay an RTT to the leader for reads. \circleb{2} In nearly all cases, PQR (and its Leader Leases variant) show worse (or identical) performance compared to Leader Leases. The only exception is in the GEO, 0\% writes setting for the JP clients; they are so far away from the leader that a nearer majority quorum actually helps, letting them outperform local read protocols (since JP is not marked as a responder). \circleb{3} EPaxos has similar read performance as PQR but with higher P99 latency when there are writes. \circleb{4} Both Quorum Leases variants and \bodega show the same performance as Leader Leases for clients near the leader or a non-responder. \circleb{5} Quorum Leases and \bodega both deliver extraordinary read performance for clients near responders when with 0\% writes. \circleb{6} \revise{\bodega sustains this read performance advantage and keeps read latency close to zero for higher write intensities. In contrast, Quorum Leases performance quickly drops and almost degrades back to Leader Leases for 10\% writes. This shows the \bodega's resilience to write interference, which is a crucial advantage over previous approaches under practical workloads.}

\subsection{Detailed Performance Anatomy}
\label{sec:eval-detailed-anatomy}

\revise{We conduct a closer study across various dimensions.}

\paratitle{Latency CDFs}
We collect request latency CDFs across all 50 clients of the WAN setting (Fig.\ref{fig:exper-loc_grid_wan-w10}-\ref{fig:exper-loc_grid_wan-w0}) and plot them in Figure~\ref{fig:exper-latency_cdfs}. Results are filtered to show a single key for a clean pattern. Each site contributes an equal 20\% of datapoints.

We make four observations. \circleb{1} Write latencies across all workloads are similar and are presented as one Figure~\ref{fig:exper-latency_cdfs-w1-wr}. Quorum Leases and \bodega show slightly higher write latencies in favor of responder local reads, while EPaxos delivers the same level of latencies as its reads due to its leaderless design. These results align with \S\ref{sec:eval-normal-case}. \circleb{2} At 0\% writes, all protocols deliver a read performance close to their theoretical best, though a few outlier datapoints remain. \circleb{3} \revise{At 1\% writes, slight write interference occurs. Quorum Leases reads deviate from \bodega, with the passive variant delivering roughly half the latency of the original variant.} \circleb{4} At 10\% writes, differences in read latency distributions are the most obvious. MultiPaxos clients' read latency clearly correlates with their distance to the leader; Leader Leases are similar but with a majority-quorum latency subtracted. PQR and EPaxos exhibit suboptimal latency and have high tail latency of up to 100 ms for a read; this is due to the need for conflict resolution. \revise{Quorum Leases variants both degrade to Leader Leases. \bodega delivers outstanding local read performance as expected (except for the 20\% non-local SC clients).}


\begin{figure*}[t]
    \centering

    \begin{minipage}{0.47\textwidth}
        \includegraphics[width=\linewidth]{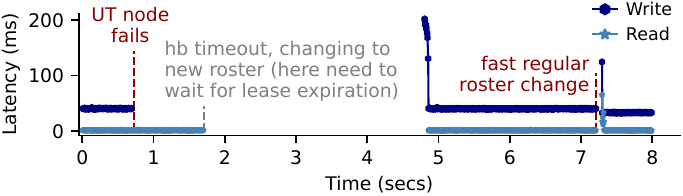}
        \captionsetup{justification=raggedright}
        \vspace{-21pt}
        \floatcapoffignodot{Failure-triggered vs. regular \revise{roster} changes}{in \bodega. The x-axis is time at which reqs finish. See \S\ref{sec:eval-roster-change-compo}.}
        \label{fig:exper-wlats_on_conf}
    \end{minipage}
    \hfill
    \begin{minipage}{0.236\textwidth}
        \includegraphics[width=\linewidth]{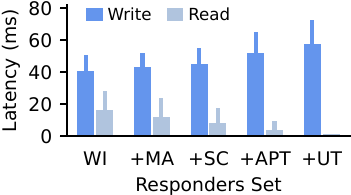}
        \captionsetup{justification=raggedright}
        \vspace{-19pt}
        \floatcapoffig{Latency vs. the set of responders}{See \S\ref{sec:eval-roster-change-compo}.}
        \label{fig:exper-conf_coverage-site}
    \end{minipage}
    \hfill
    \begin{minipage}{0.236\textwidth}
        \includegraphics[width=\linewidth]{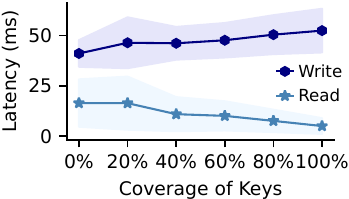}
        \captionsetup{justification=raggedright}
        \vspace{-20pt}
        \floatcapoffig{Latency vs. keys covered by \revise{roster}}{See \S\ref{sec:eval-roster-change-compo}.}
        \label{fig:exper-conf_coverage-keys}
    \end{minipage}
\end{figure*}

\begin{figure*}[t]
    \centering
    
    \vspace*{5pt}
    \begin{minipage}{0.026\linewidth}
        \begin{subfigure}[t]{\columnwidth}
            \centering
            \raisebox{4pt}{\includegraphics[width=\columnwidth]{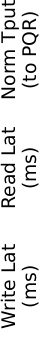}}
            \captionsetup{justification=centering}
        \end{subfigure}
    \end{minipage}
    \hfill
    \begin{minipage}{0.425\linewidth}
        \begin{subfigure}[t]{\columnwidth}
            \centering
            \includegraphics[width=\columnwidth]{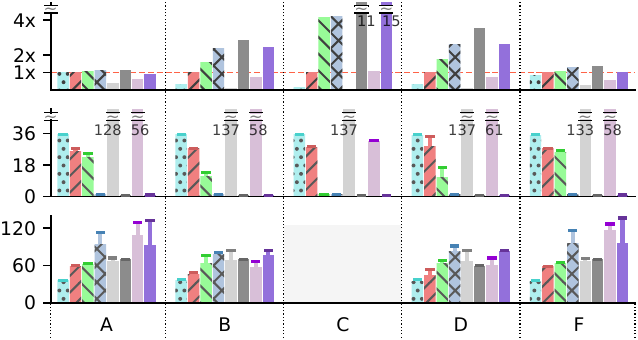}
            \captionsetup{justification=centering}
            \caption{Uniform, full-coverage \revise{roster}}
            \label{fig:exper-ycsb_zk_etcd-uniform}
        \end{subfigure}
    \end{minipage}
    \hfill
    \begin{minipage}{0.425\linewidth}
        \begin{subfigure}[t]{\columnwidth}
            \centering
            \includegraphics[width=\columnwidth]{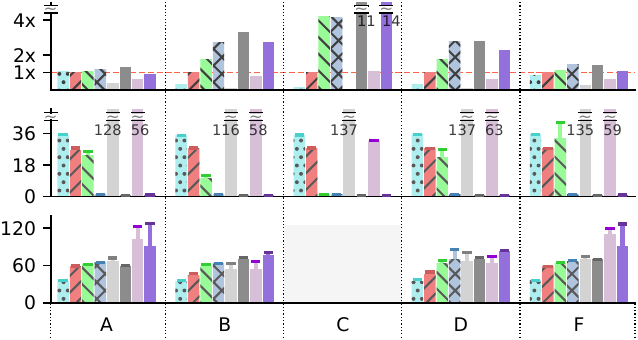}
            \captionsetup{justification=centering}
            \caption{Zipfian, top-20\%-coverage \revise{roster}}
            \label{fig:exper-ycsb_zk_etcd-zipfian}
        \end{subfigure}
    \end{minipage}
    \hfill
    \begin{minipage}{0.11\linewidth}
        \begin{subfigure}[t]{\columnwidth}
            \centering
            \includegraphics[width=\columnwidth]{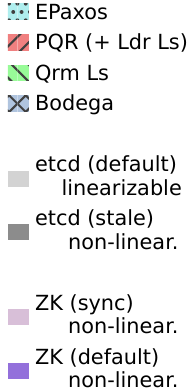}
            \captionsetup{justification=centering}
        \end{subfigure}
    \end{minipage}
    
    \addtocounter{figure}{-1}  
    \captionsetup{justification=raggedright}
    \vspace{-2pt}
    \floatcapoffig{YCSB workloads on \summerset, etcd, \& ZooKeeper}{etcd (stale) \& ZK (both modes) are non-linearizable. See \S\ref{sec:eval-ycsb-zk-etcd}.}
    \label{fig:exper-ycsb_zk_etcd}
\end{figure*}



\paratitle{Visualizing \revise{Write Interference}}
We use a similar setting to the read-only workload in \S\ref{sec:eval-normal-case} on the WAN cluster, but this time with open-loop clients, each sending reads at a rate of 400 reqs/sec to a key. At \textasciitilde15 secs, we let one client issue a write command to the key. We monitor the average read latency across the three non-leader responders for the local read protocol variants, and plot them over a time axis in Figure~\ref{fig:exper-rlats_on_write}. We see that the write introduces an interruption to local reads for all three protocols. Both Quorum Leases variants degrade to 40 ms read latency, which is the average RTT to the stable leader; the passive variant shows a shorter degradation duration. \bodega \revise{\circleb{1} shortens the degradation time to \textasciitilde25 ms; recall Table~\ref{tab:comparison}, this is {\textasciitilde}$\frac{m}{2}$, and \circleb{2} allows all reads to be held locally and released at the end of the degradation, leading to better latencies also for disrupted reads.}

\paratitle{Varying Write Ratios}
We take the same setup as \S\ref{sec:eval-normal-case} on the WAN cluster and vary the write ratios of the workload mix from 1\% to 50\% while fixing value size to 128B. We report the aggregate throughput in Figure~\ref{fig:exper-writes_sizes-puts}. All protocols except the PQR (+ Ldr Ls) baseline show a trend of lower throughput with higher write interference as local reads become less profitable. The results match Figure~\ref{fig:exper-loc_grid_wan-w10}-\ref{fig:exper-loc_grid_wan-w1}.

\paratitle{Varying Value Sizes}
We repeat the same setup as above and vary the value size while fixing the write ratio at 5\%. As expected, Figure~\ref{fig:exper-writes_sizes-size} shows that smaller values have little impact on performance, but throughput drops with larger values due to slower writes and larger read results to transfer.

\subsection{\revise{Roster Changes \& Composition}}
\label{sec:eval-roster-change-compo}

\revise{We evaluate roster changes and the impact of their coverage.}

\paratitle{Roster Change Duration}
We compare the duration of two different types of \revise{roster} changes: failure-induced changes (where waiting for lease expiration is necessary) vs. regular (where revocations complete quickly). \revise{Regular roster changes finish in just two message rounds, because it is no more than an all-to-all lease revocation followed by the initiation of new leases.} We create an open-loop client near the WI server and let it issue a 50\%-read workload at a rate of 400 reqs/sec to 1k keys. We plot real-time latencies in Figure~\ref{fig:exper-wlats_on_conf}.

At \textasciitilde800 ms, we crash the UT node, which is one of the responders for the full key range. Writes are immediately blocked since UT as a responder is unreachable. Reads, however, can still be served locally without interruption until \textasciitilde1.1 secs later, when some healthy server in the cluster raises a heartbeat timeout and initiates a change to a new \revise{roster} where UT is removed from all responder roles. Since UT is unresponsive to lease revocations, waiting 2.6 secs for expiration is required, after which normal operations continue.

At \textasciitilde7.2 secs, we make an explicit \revise{roster} change request to a server. In contrast to the failure case, this \revise{roster} change completes in just \revise{\textasciitilde75 ms, which is \textasciitilde2x cluster-wise RTTs as expected}. Impacts on client requests are minor.




\paratitle{Choice of Responders}
We run a 10\%-write workload on all clients in the WAN setting, while using an increasing set of responders for all keys; WI node is still the leader. We report the cluster-wide average read/write latency and their standard deviation in Figure~\ref{fig:exper-conf_coverage-site}. With more nodes added as responders, read latency tends to zero while write latency increases, revealing the expected tradeoff. This demonstrates the importance of allowing adjustable \revise{rosters} to help avoid unnecessary taxes on writes.

\paratitle{Coverage of Keys}
We repeat the same experiment, but vary the percentage of local-read-enabled keys while fixing the choice of responders to Figure~\ref{fig:eval-setting-wan}. The cluster-wide average latencies and their standard deviation across the coverage spectrum are plotted in Figure~\ref{fig:exper-conf_coverage-keys}. Results show an expected decrease in read latency and a corresponding increase in write latency as local reads are enabled on more keys. This implies the general strategy of enabling local reads for read-heavy keys while avoiding local reads for write-heavy keys.

\subsection{Macrobenchmark vs. etcd \& ZooKeeper}
\label{sec:eval-ycsb-zk-etcd}

To evaluate the protocols in a more realistic setup, we compare \summerset protocols with two widely-used coordination services, etcd~\cite{etcd} and ZooKeeper~\cite{zookeeper}, on the WAN cluster. We drive all systems with YCSB~\cite{ycsb}, the standard KV macrobenchmark. Workloads have the following approximate write ratios (we treat insertions as updates). \underline{A}: 50\% w, \underline{B}: 5\% w, \underline{C}: 0\% w, \underline{D}: 5\% w, \underline{F}: 25\% w.

\paratitle{YCSB Request Distributions}
We use 10k keys and construct two scenarios corresponding to two request distributions. \circleb{1} For the Uniform distribution, clients at all locations choose keys uniformly randomly across the key space. Since there are no site-specific preferences for keys, all sites are added as responders for all keys to secure local reads. \circleb{2} For Zipfian, clients at each location choose keys according to a Zipfian-0.99 distribution, skewed towards different sets of keys at different sites; this creates per-site preferences for keys. We then add each site as a responder only for its top-20\% accessed keys to derive an asymmetric \revise{roster} \revise{that imposes unnoticeable impacts on write performance}.

\paratitle{etcd Modes}
We deploy etcd in two modes, both with 120 ms heartbeat intervals. The default mode showcases a standard implementation of vanilla Raft~\cite{raft}. The \textit{stale} mode turns on the serializable member-local read option for all read requests, delivering sequential consistency by \revise{always} serving reads locally with past committed values at any server\revise{; this represents the ideal upper bound for \bodega}.

\paratitle{ZooKeeper Modes}
Similarly, we deploy ZooKeeper in two modes, though both are non-linearizable. The default mode is a standard implementation of the ZAB primary-backup protocol~\cite{zookeeper} that pushes writes to all servers and serves reads locally from anywhere. The \textit{sync} mode is the closest mode to linearizable reads that ZooKeeper clients can get: every read request is preceded by a \texttt{sync} API call to force flush all the in-progress writes from the leader to its endpoint server, but all writes that may have completed after the start of the flush are not guaranteed to be seen by the read.

\paratitle{Results}
We present the performance results in Figure~\ref{fig:exper-ycsb_zk_etcd}, grouped by workload type and with PQR (+ Ldr Ls) as the normalized throughput baseline. We make the following observations. \circleb{1} \revise{\bodega matches (and sometimes surpasses) the performance of sequentially-consistent default ZooKeeper, and is able to keep up with stale etcd across all workloads. This illustrates \bodega's powerful local linearizable read capabilities. The advantage over ZK is due to avoiding Java runtime overheads.} \circleb{2} \revise{In workload C, both non-linearizable services deliver \textasciitilde0.3 ms read latency and over 10x throughput gain, while \bodega and Quorum Leases deliver \textasciitilde1.2 ms latency due to the 1 ms request batching applied.} \circleb{3} EPaxos, PQR (+ Ldr Ls), Quorum Leases, and \bodega all show similar patterns coherent to \S\ref{sec:eval-normal-case}. With no writes (C), both local read protocols deliver excellent performance. With higher write ratios, \bodega sustains this advantage better than Quorum Leases. \circleb{4} Default etcd and sync ZK have high read latencies of $>$50 ms because they are classic consensus without leases. \circleb{5} Comparing the Uniform scenario with Zipfian, the only notable difference is that \bodega exhibits higher write latencies close to ZK in Uniform. This is expected because \bodega writes need to reach all nodes as they are all responders.


\section{Discussion}
\label{sec:discussion}

We discuss topics that are out of the scope of this paper but are interesting directions for future work.

\paratitle{Partial Network Partitioning}
Using heartbeat timeout-based failure detection for leader step-up is known to \revise{risk liveness} under partial network partitioning~\cite{raft-thesis}, and the same holds true for \revise{roster} lease activations. \revise{Common} techniques such as pre-votes~\cite{raft-thesis} and transparent re-routing~\cite{nifty} can be deployed to easily eliminate this issue.

\paratitle{Generalization of \revise{Roster} Leases}
We observe that the activation procedure of \revise{roster} leases shares similarities with \textit{broadcast-based} (randomized) consensus~\cite{rabia, ben-or-algorithm, randomized-consensus-coins}. It is a practical application of all-to-all broadcast in a non-adversarial setting for one-off agreements on the \revise{roster} ballot. Combined with leases, this technique can be used to establish fault-tolerant agreement on any general ``\revise{roster} metadata''  that change infrequently, not limited to leadership and assignment of responders as in \bodega. Possible extensions may include cluster membership, asymmetric quorum sizes, and node-specific performance and reliability hints.

\paratitle{Bounded Staleness Support}
With simple modifications, \bodega can extend beyond linearizability and support fast local reads that can tolerate (but require) bounded staleness measured in the maximum version difference with the latest committed write. When a non-leader responder receives a read that allows up to $x$ versions stale, it can traverse the tail of its log in reverse and search for at most $x$ occurrences of the key, returning the latest committed version if found.

\section{Related Work}
\label{sec:related-work}

We list \revise{additional} notable related work in this section.

\paratitle{Consensus \& Read Optimizations}
\S\ref{sec:bg-previous-approaches} and \S\ref{sec:formal-comparison-across} have covered in detail the most essential related work, including classic consensus algorithms~\cite{paxos-parliament, paxos-made-simple, raft, raft-thesis, viewstamped-replication, abcd-of-paxos}, leaderless or multi-leader approaches~\cite{mencius, epaxos, epaxos-revisited, generalized-paxos, fast-paxos, swiftpaxos, hermes, paxos-quorum-read, tradeoffs-geo-distributed, wpaxos, atlas}, and read leases~\cite{paxos-made-live, leases-mechanism, quorum-leases, quorum-leases-report, megastore, t-lease}. Flexible quorum sizes are discussed in classic literature~\cite{dynamic-quorum} as well as recent proposals such as Flexible Paxos~\cite{fpaxos, dpaxos, edge-pqr, quorums-made-practical}. \revise{Optimistic holding shares similarity to wait-vs.-abort in database concurrency control~\cite{evaluation-of-cc, occ-methods}.}

\paratitle{Shared Logs \& Lazy Ordering for Writes}
Shared logs are a common abstraction found in cloud systems and are usually backed by primary-backup-style protocols~\cite{scalog, corfu, tango, fuzzylog, gryff, delos-virtual-consensus, delos-log-structured}. CAD~\cite{consistency-aware-durability}, Skyros~\cite{skyros-nil-externality}, and LazyLog~\cite{lazylog} are a series of work on a \textit{lazy ordering} optimization for writes and shared log appends. It hides a significant portion of write latency but could hurt read performance in contended cases.

\paratitle{Synchronized Clocks}
Recent works demonstrate production-ready implementations of synchronized clocks~\cite{spanner, sundial} and designs that take advantage of them through timestamp heuristics~\cite{nezha, clock-rsm}. Chandra et al. presented a formal, optimal lease algorithm that assumes synchronized clocks~\cite{read-leases-theory, read-leases-theory-parameterized}.

\paratitle{Orthogonal Designs}
Other works address orthogonal problems in consensus: programmability~\cite{paxos-made-transparent, depfast, pacifica}, membership~\cite{viewstamped-replication, raft, chain-paxos, vertical-paxos, reconfigurable-linearizable-reads, ukharon}, fail-slow mitigation~\cite{copilots, occult, gray-failures}, utilizing specialized hardware and/or eBPF~\cite{derecho, swarm-disaggregated-memory, aceso-disaggregated-memory, p4xos, nopaxos, netpaxos, off-path-smartnic, fisslock, apus-rdma-paxos, mu-consensus, hydra-network-ordering, speculative-paxos, electrode}, scalability~\cite{scalable-smr, compartmentalization, insanely-scalable-smr, dynastar, consul, chain-replication, chain-paxos, craq, chainreaction, pull-based-consensus-mongodb, pigpaxos, sdpaxos, canopus}, weaker consistency~\cite{sequential-consistency, regular-sequential-consistency, occult, eventual-consistency, noctua, cops, bayou}, formal methods~\cite{I4, sift-veri, ironfleet, verus, anvil}, and BFT~\cite{pbft, hotstuff, ubft, basil, autobahn, bidl-blockchain, pbft-read-only-liveness, byzantine-ordered-consensus}.

\section{Conclusion}
\label{sec:conclusion}

We present \bodega, a wide-area consensus protocol that enables always-local linearizable reads \revisetwo{anywhere (i.e., at arbitrary responder replicas)} at \revisetwo{anytime (i.e., remains local under interfering writes, with minimal disruption)}. \bodega \revisetwo{achieves this via introducing the notion of a roster}, and deploying novel all-to-all \revise{roster} leases off the critical path to establish responders assignment without compromising fault tolerance. \bodega combines optimistic holding with early accept notifications in the normal case to keep reads localized, and employs smart roster coverage and lightweight heartbeats, delivering extreme read performance comparable to sequentially-consistent production systems with negligible overhead. We believe \bodega is a valuable step towards performance-optimal wide-area replication for critical workloads of the modern cloud.



\clearpage
\bibliographystyle{ACM-Reference-Format}
\bibliography{references}


\begin{thebibliography}{150}


\ifx \showCODEN    \undefined \def \showCODEN     #1{\unskip}     \fi
\ifx \showISBNx    \undefined \def \showISBNx     #1{\unskip}     \fi
\ifx \showISBNxiii \undefined \def \showISBNxiii  #1{\unskip}     \fi
\ifx \showISSN     \undefined \def \showISSN      #1{\unskip}     \fi
\ifx \showLCCN     \undefined \def \showLCCN      #1{\unskip}     \fi
\ifx \shownote     \undefined \def \shownote      #1{#1}          \fi
\ifx \showarticletitle \undefined \def \showarticletitle #1{#1}   \fi
\ifx \showURL      \undefined \def \showURL       {\relax}        \fi
\providecommand\bibfield[2]{#2}
\providecommand\bibinfo[2]{#2}
\providecommand\natexlab[1]{#1}
\providecommand\showeprint[2][]{arXiv:#2}

\bibitem[Aguilera et~al\mbox{.}(2020)]%
        {mu-consensus}
\bibfield{author}{\bibinfo{person}{Marcos~K. Aguilera}, \bibinfo{person}{Naama Ben-David}, \bibinfo{person}{Rachid Guerraoui}, \bibinfo{person}{Virendra~J. Marathe}, \bibinfo{person}{Athanasios Xygkis}, {and} \bibinfo{person}{Igor Zablotchi}.} \bibinfo{year}{2020}\natexlab{}.
\newblock \showarticletitle{Microsecond Consensus for Microsecond Applications}. In \bibinfo{booktitle}{\emph{14th USENIX Symposium on Operating Systems Design and Implementation (OSDI 20)}}. \bibinfo{publisher}{USENIX Association}, \bibinfo{pages}{599--616}.
\newblock
\showISBNx{978-1-939133-19-9}
\urldef\tempurl%
\url{https://www.usenix.org/conference/osdi20/presentation/aguilera}
\showURL{%
\tempurl}


\bibitem[Aguilera et~al\mbox{.}(2023)]%
        {ubft}
\bibfield{author}{\bibinfo{person}{Marcos~K. Aguilera}, \bibinfo{person}{Naama Ben-David}, \bibinfo{person}{Rachid Guerraoui}, \bibinfo{person}{Antoine Murat}, \bibinfo{person}{Athanasios Xygkis}, {and} \bibinfo{person}{Igor Zablotchi}.} \bibinfo{year}{2023}\natexlab{}.
\newblock \showarticletitle{UBFT: Microsecond-Scale BFT Using Disaggregated Memory}. In \bibinfo{booktitle}{\emph{Proceedings of the 28th ACM International Conference on Architectural Support for Programming Languages and Operating Systems, Volume 2}} (Vancouver, BC, Canada) \emph{(\bibinfo{series}{ASPLOS 2023})}. \bibinfo{publisher}{Association for Computing Machinery}, \bibinfo{address}{New York, NY, USA}, \bibinfo{pages}{862–877}.
\newblock
\showISBNx{9781450399166}
\href{https://doi.org/10.1145/3575693.3575732}{doi:\nolinkurl{10.1145/3575693.3575732}}


\bibitem[Ailijiang et~al\mbox{.}(2020)]%
        {wpaxos}
\bibfield{author}{\bibinfo{person}{Ailidani Ailijiang}, \bibinfo{person}{Aleksey Charapko}, \bibinfo{person}{Murat Demirbas}, {and} \bibinfo{person}{Tevfik Kosar}.} \bibinfo{year}{2020}\natexlab{}.
\newblock \showarticletitle{WPaxos: Wide Area Network Flexible Consensus}.
\newblock \bibinfo{journal}{\emph{IEEE Trans. Parallel Distrib. Syst.}} \bibinfo{volume}{31}, \bibinfo{number}{1} (\bibinfo{date}{Jan.} \bibinfo{year}{2020}), \bibinfo{pages}{211–223}.
\newblock
\showISSN{1045-9219}
\href{https://doi.org/10.1109/TPDS.2019.2929793}{doi:\nolinkurl{10.1109/TPDS.2019.2929793}}


\bibitem[Alfatafta et~al\mbox{.}(2020)]%
        {nifty}
\bibfield{author}{\bibinfo{person}{Mohammed Alfatafta}, \bibinfo{person}{Basil Alkhatib}, \bibinfo{person}{Ahmed Alquraan}, {and} \bibinfo{person}{Samer Al-Kiswany}.} \bibinfo{year}{2020}\natexlab{}.
\newblock \showarticletitle{Toward a Generic Fault Tolerance Technique for Partial Network Partitioning}. In \bibinfo{booktitle}{\emph{14th USENIX Symposium on Operating Systems Design and Implementation (OSDI 20)}}. \bibinfo{publisher}{USENIX Association}, \bibinfo{pages}{351--368}.
\newblock
\showISBNx{978-1-939133-19-9}
\urldef\tempurl%
\url{https://www.usenix.org/conference/osdi20/presentation/alfatafta}
\showURL{%
\tempurl}


\bibitem[Almeida et~al\mbox{.}(2013)]%
        {chainreaction}
\bibfield{author}{\bibinfo{person}{S\'{e}rgio Almeida}, \bibinfo{person}{Jo\~{a}o Leit\~{a}o}, {and} \bibinfo{person}{Lu\'{\i}s Rodrigues}.} \bibinfo{year}{2013}\natexlab{}.
\newblock \showarticletitle{ChainReaction: a causal+ consistent datastore based on chain replication}. In \bibinfo{booktitle}{\emph{Proceedings of the 8th ACM European Conference on Computer Systems}} (Prague, Czech Republic) \emph{(\bibinfo{series}{EuroSys '13})}. \bibinfo{publisher}{Association for Computing Machinery}, \bibinfo{address}{New York, NY, USA}, \bibinfo{pages}{85–98}.
\newblock
\showISBNx{9781450319942}
\href{https://doi.org/10.1145/2465351.2465361}{doi:\nolinkurl{10.1145/2465351.2465361}}


\bibitem[{Artem on StackOverflow}(2017)]%
        {zookeeper-consistency-question-1}
\bibfield{author}{\bibinfo{person}{{Artem on StackOverflow}}.} \bibinfo{year}{2017}\natexlab{}.
\newblock \bibinfo{title}{Is ZooKeeper always consistent in terms of CAP theorem?}
\newblock \bibinfo{howpublished}{\url{https://stackoverflow.com/questions/35387774}}.
\newblock
\newblock
\shownote{Accessed: 2024-12-01}.


\bibitem[Arun et~al\mbox{.}(2017)]%
        {caesar-consensus}
\bibfield{author}{\bibinfo{person}{Balaji Arun}, \bibinfo{person}{Sebastiano Peluso}, \bibinfo{person}{Roberto Palmieri}, \bibinfo{person}{Giuliano Losa}, {and} \bibinfo{person}{Binoy Ravindran}.} \bibinfo{year}{2017}\natexlab{}.
\newblock \showarticletitle{Speeding up Consensus by Chasing Fast Decisions}. In \bibinfo{booktitle}{\emph{47th IEEE/IFIP International Conference on Dependable Systems and Networks (DSN)}}. \bibinfo{pages}{49--60}.
\newblock
\href{https://doi.org/10.1109/DSN.2017.35}{doi:\nolinkurl{10.1109/DSN.2017.35}}


\bibitem[Attiya and Welch(1994)]%
        {sequential-vs-linearizability}
\bibfield{author}{\bibinfo{person}{Hagit Attiya} {and} \bibinfo{person}{Jennifer~L. Welch}.} \bibinfo{year}{1994}\natexlab{}.
\newblock \showarticletitle{Sequential Consistency versus Linearizability}.
\newblock \bibinfo{journal}{\emph{ACM Trans. Comput. Syst.}} \bibinfo{volume}{12}, \bibinfo{number}{2} (\bibinfo{date}{may} \bibinfo{year}{1994}), \bibinfo{pages}{91–122}.
\newblock
\showISSN{0734-2071}
\href{https://doi.org/10.1145/176575.176576}{doi:\nolinkurl{10.1145/176575.176576}}


\bibitem[AWS(2024a)]%
        {aws-global-infra}
\bibfield{author}{\bibinfo{person}{AWS}.} \bibinfo{year}{2024}\natexlab{a}.
\newblock \bibinfo{title}{AWS Global Infrastructure}.
\newblock
\newblock
\shownote{\url{https://aws.amazon.com/about-aws/global-infrastructure/}, Last accessed on 2024-04-28}.


\bibitem[AWS(2024b)]%
        {aws-prescriptive-workload-characteristics}
\bibfield{author}{\bibinfo{person}{AWS}.} \bibinfo{year}{2024}\natexlab{b}.
\newblock \bibinfo{title}{Workload Characteristics}.
\newblock \bibinfo{howpublished}{\url{https://docs.aws.amazon.com/prescriptive-guidance/latest/oracle-exadata-blueprint/workload-characteristics.html}}.
\newblock
\newblock
\shownote{Accessed: 2024-12-01}.


\bibitem[Baker et~al\mbox{.}(2011)]%
        {megastore}
\bibfield{author}{\bibinfo{person}{Jason Baker}, \bibinfo{person}{Chris Bond}, \bibinfo{person}{James~C. Corbett}, \bibinfo{person}{JJ Furman}, \bibinfo{person}{Andrey Khorlin}, \bibinfo{person}{James Larson}, \bibinfo{person}{Jean-Michel Leon}, \bibinfo{person}{Yawei Li}, \bibinfo{person}{Alexander Lloyd}, {and} \bibinfo{person}{Vadim Yushprakh}.} \bibinfo{year}{2011}\natexlab{}.
\newblock \showarticletitle{Megastore: Providing Scalable, Highly Available Storage for Interactive Services}. In \bibinfo{booktitle}{\emph{Proceedings of the Conference on Innovative Data system Research (CIDR)}}. \bibinfo{pages}{223--234}.
\newblock
\urldef\tempurl%
\url{http://www.cidrdb.org/cidr2011/Papers/CIDR11_Paper32.pdf}
\showURL{%
\tempurl}


\bibitem[Balakrishnan et~al\mbox{.}(2020)]%
        {delos-virtual-consensus}
\bibfield{author}{\bibinfo{person}{Mahesh Balakrishnan}, \bibinfo{person}{Jason Flinn}, \bibinfo{person}{Chen Shen}, \bibinfo{person}{Mihir Dharamshi}, \bibinfo{person}{Ahmed Jafri}, \bibinfo{person}{Xiao Shi}, \bibinfo{person}{Santosh Ghosh}, \bibinfo{person}{Hazem Hassan}, \bibinfo{person}{Aaryaman Sagar}, \bibinfo{person}{Rhed Shi}, \bibinfo{person}{Jingming Liu}, \bibinfo{person}{Filip Gruszczynski}, \bibinfo{person}{Xianan Zhang}, \bibinfo{person}{Huy Hoang}, \bibinfo{person}{Ahmed Yossef}, \bibinfo{person}{Francois Richard}, {and} \bibinfo{person}{Yee~Jiun Song}.} \bibinfo{year}{2020}\natexlab{}.
\newblock \showarticletitle{Virtual Consensus in Delos}. In \bibinfo{booktitle}{\emph{14th USENIX Symposium on Operating Systems Design and Implementation (OSDI 20)}}. \bibinfo{publisher}{USENIX Association}, \bibinfo{pages}{617--632}.
\newblock
\showISBNx{978-1-939133-19-9}
\urldef\tempurl%
\url{https://www.usenix.org/conference/osdi20/presentation/balakrishnan}
\showURL{%
\tempurl}


\bibitem[Balakrishnan et~al\mbox{.}(2013a)]%
        {corfu}
\bibfield{author}{\bibinfo{person}{Mahesh Balakrishnan}, \bibinfo{person}{Dahlia Malkhi}, \bibinfo{person}{John~D. Davis}, \bibinfo{person}{Vijayan Prabhakaran}, \bibinfo{person}{Michael Wei}, {and} \bibinfo{person}{Ted Wobber}.} \bibinfo{year}{2013}\natexlab{a}.
\newblock \showarticletitle{CORFU: A distributed shared log}.
\newblock \bibinfo{journal}{\emph{ACM Trans. Comput. Syst.}} \bibinfo{volume}{31}, \bibinfo{number}{4}, Article \bibinfo{articleno}{10} (\bibinfo{date}{Dec.} \bibinfo{year}{2013}), \bibinfo{numpages}{24}~pages.
\newblock
\showISSN{0734-2071}
\href{https://doi.org/10.1145/2535930}{doi:\nolinkurl{10.1145/2535930}}


\bibitem[Balakrishnan et~al\mbox{.}(2013b)]%
        {tango}
\bibfield{author}{\bibinfo{person}{Mahesh Balakrishnan}, \bibinfo{person}{Dahlia Malkhi}, \bibinfo{person}{Ted Wobber}, \bibinfo{person}{Ming Wu}, \bibinfo{person}{Vijayan Prabhakaran}, \bibinfo{person}{Michael Wei}, \bibinfo{person}{John~D. Davis}, \bibinfo{person}{Sriram Rao}, \bibinfo{person}{Tao Zou}, {and} \bibinfo{person}{Aviad Zuck}.} \bibinfo{year}{2013}\natexlab{b}.
\newblock \showarticletitle{Tango: Distributed Data Structures over a Shared Log}. In \bibinfo{booktitle}{\emph{Proceedings of the Twenty-Fourth ACM Symposium on Operating Systems Principles}} (Farminton, Pennsylvania) \emph{(\bibinfo{series}{SOSP '13})}. \bibinfo{publisher}{Association for Computing Machinery}, \bibinfo{address}{New York, NY, USA}, \bibinfo{pages}{325–340}.
\newblock
\showISBNx{9781450323888}
\href{https://doi.org/10.1145/2517349.2522732}{doi:\nolinkurl{10.1145/2517349.2522732}}


\bibitem[Balakrishnan et~al\mbox{.}(2021)]%
        {delos-log-structured}
\bibfield{author}{\bibinfo{person}{Mahesh Balakrishnan}, \bibinfo{person}{Chen Shen}, \bibinfo{person}{Ahmed Jafri}, \bibinfo{person}{Suyog Mapara}, \bibinfo{person}{David Geraghty}, \bibinfo{person}{Jason Flinn}, \bibinfo{person}{Vidhya Venkat}, \bibinfo{person}{Ivailo Nedelchev}, \bibinfo{person}{Santosh Ghosh}, \bibinfo{person}{Mihir Dharamshi}, \bibinfo{person}{Jingming Liu}, \bibinfo{person}{Filip Gruszczynski}, \bibinfo{person}{Jun Li}, \bibinfo{person}{Rounak Tibrewal}, \bibinfo{person}{Ali Zaveri}, \bibinfo{person}{Rajeev Nagar}, \bibinfo{person}{Ahmed Yossef}, \bibinfo{person}{Francois Richard}, {and} \bibinfo{person}{Yee~Jiun Song}.} \bibinfo{year}{2021}\natexlab{}.
\newblock \showarticletitle{Log-structured Protocols in Delos}. In \bibinfo{booktitle}{\emph{Proceedings of the ACM 28th Symposium on Operating Systems Principles}} (Germany) \emph{(\bibinfo{series}{SOSP '21})}. \bibinfo{publisher}{Association for Computing Machinery}, \bibinfo{address}{New York, NY, USA}, \bibinfo{pages}{538–552}.
\newblock
\showISBNx{9781450387095}
\href{https://doi.org/10.1145/3477132.3483544}{doi:\nolinkurl{10.1145/3477132.3483544}}


\bibitem[Barr(2023)]%
        {s3-strong-consistency}
\bibfield{author}{\bibinfo{person}{Jeff Barr}.} \bibinfo{year}{2023}\natexlab{}.
\newblock \bibinfo{title}{Amazon S3 Update – Strong Read-After-Write Consistency}.
\newblock
\newblock
\shownote{\url{https://aws.amazon.com/blogs/aws/amazon-s3-update-strong-read-after-write-consistency/}, Last accessed on 2023-11-19}.


\bibitem[Ben-Or(1983)]%
        {ben-or-algorithm}
\bibfield{author}{\bibinfo{person}{Michael Ben-Or}.} \bibinfo{year}{1983}\natexlab{}.
\newblock \showarticletitle{Another advantage of free choice (Extended Abstract): Completely asynchronous agreement protocols}. In \bibinfo{booktitle}{\emph{Proceedings of the Second Annual ACM Symposium on Principles of Distributed Computing}} (Montreal, Quebec, Canada) \emph{(\bibinfo{series}{PODC '83})}. \bibinfo{publisher}{Association for Computing Machinery}, \bibinfo{address}{New York, NY, USA}, \bibinfo{pages}{27–30}.
\newblock
\showISBNx{0897911105}
\href{https://doi.org/10.1145/800221.806707}{doi:\nolinkurl{10.1145/800221.806707}}


\bibitem[Berenson et~al\mbox{.}(1995)]%
        {ansi-sql-isolation}
\bibfield{author}{\bibinfo{person}{Hal Berenson}, \bibinfo{person}{Phil Bernstein}, \bibinfo{person}{Jim Gray}, \bibinfo{person}{Jim Melton}, \bibinfo{person}{Elizabeth O'Neil}, {and} \bibinfo{person}{Patrick O'Neil}.} \bibinfo{year}{1995}\natexlab{}.
\newblock \showarticletitle{A critique of ANSI SQL isolation levels}. In \bibinfo{booktitle}{\emph{Proceedings of the 1995 ACM SIGMOD International Conference on Management of Data}} (San Jose, California, USA) \emph{(\bibinfo{series}{SIGMOD '95})}. \bibinfo{publisher}{Association for Computing Machinery}, \bibinfo{address}{New York, NY, USA}, \bibinfo{pages}{1–10}.
\newblock
\showISBNx{0897917316}
\href{https://doi.org/10.1145/223784.223785}{doi:\nolinkurl{10.1145/223784.223785}}


\bibitem[Berger et~al\mbox{.}(2021)]%
        {pbft-read-only-liveness}
\bibfield{author}{\bibinfo{person}{Christian Berger}, \bibinfo{person}{Hans~P. Reiser}, {and} \bibinfo{person}{Alysson Bessani}.} \bibinfo{year}{2021}\natexlab{}.
\newblock \showarticletitle{Making Reads in BFT State Machine Replication Fast, Linearizable, and Live}. In \bibinfo{booktitle}{\emph{2021 40th International Symposium on Reliable Distributed Systems (SRDS)}}. \bibinfo{pages}{1--12}.
\newblock
\href{https://doi.org/10.1109/SRDS53918.2021.00010}{doi:\nolinkurl{10.1109/SRDS53918.2021.00010}}


\bibitem[Bezerra et~al\mbox{.}(2014)]%
        {scalable-smr}
\bibfield{author}{\bibinfo{person}{Carlos~Eduardo Bezerra}, \bibinfo{person}{Fernando Pedone}, {and} \bibinfo{person}{Robbert Van~Renesse}.} \bibinfo{year}{2014}\natexlab{}.
\newblock \showarticletitle{Scalable State-Machine Replication}. In \bibinfo{booktitle}{\emph{2014 44th Annual IEEE/IFIP International Conference on Dependable Systems and Networks}}. \bibinfo{pages}{331--342}.
\newblock
\href{https://doi.org/10.1109/DSN.2014.41}{doi:\nolinkurl{10.1109/DSN.2014.41}}


\bibitem[Bi et~al\mbox{.}(2022)]%
        {read-leases-theory-parameterized}
\bibfield{author}{\bibinfo{person}{Changyu Bi}, \bibinfo{person}{Vassos Hadzilacos}, {and} \bibinfo{person}{Sam Toueg}.} \bibinfo{year}{2022}\natexlab{}.
\newblock \bibinfo{title}{Parameterized algorithm for replicated objects with local reads}.
\newblock
\showeprint[arxiv]{2204.01228}~[cs.DC]
\urldef\tempurl%
\url{https://arxiv.org/abs/2204.01228}
\showURL{%
\tempurl}


\bibitem[Brooker et~al\mbox{.}(2020)]%
        {physalia}
\bibfield{author}{\bibinfo{person}{Marc Brooker}, \bibinfo{person}{Tao Chen}, {and} \bibinfo{person}{Fan Ping}.} \bibinfo{year}{2020}\natexlab{}.
\newblock \showarticletitle{Millions of tiny databases}. In \bibinfo{booktitle}{\emph{NSDI 2020}}.
\newblock
\urldef\tempurl%
\url{https://www.amazon.science/publications/millions-of-tiny-databases}
\showURL{%
\tempurl}


\bibitem[Burke et~al\mbox{.}(2020)]%
        {gryff}
\bibfield{author}{\bibinfo{person}{Matthew Burke}, \bibinfo{person}{Audrey Cheng}, {and} \bibinfo{person}{Wyatt Lloyd}.} \bibinfo{year}{2020}\natexlab{}.
\newblock \showarticletitle{Gryff: Unifying Consensus and Shared Registers}. In \bibinfo{booktitle}{\emph{17th USENIX Symposium on Networked Systems Design and Implementation (NSDI 20)}}. \bibinfo{publisher}{USENIX Association}, \bibinfo{address}{Santa Clara, CA}, \bibinfo{pages}{591--617}.
\newblock
\showISBNx{978-1-939133-13-7}
\urldef\tempurl%
\url{https://www.usenix.org/conference/nsdi20/presentation/burke}
\showURL{%
\tempurl}


\bibitem[Burrows(2006)]%
        {chubby}
\bibfield{author}{\bibinfo{person}{Mike Burrows}.} \bibinfo{year}{2006}\natexlab{}.
\newblock \showarticletitle{The Chubby Lock Service for Loosely-Coupled Distributed Systems}. In \bibinfo{booktitle}{\emph{Proceedings of the 7th Symposium on Operating Systems Design and Implementation}} (Seattle, Washington) \emph{(\bibinfo{series}{OSDI '06})}. \bibinfo{publisher}{USENIX Association}, \bibinfo{address}{USA}, \bibinfo{pages}{335–350}.
\newblock
\showISBNx{1931971471}


\bibitem[Butler(2024)]%
        {google-unisuper-delete}
\bibfield{author}{\bibinfo{person}{Georgia Butler}.} \bibinfo{year}{2024}\natexlab{}.
\newblock \showarticletitle{Google Cloud accidentally deleted UniSuper's Private Cloud Subscription}.
\newblock \bibinfo{journal}{\emph{Data Center Dynamics}} (\bibinfo{year}{2024}).
\newblock
\urldef\tempurl%
\url{https://www.datacenterdynamics.com/en/news/google-cloud-accidentally-deleted-unisupers-private-cloud-subscription/}
\showURL{%
\tempurl}


\bibitem[Castro and Liskov(1999)]%
        {pbft}
\bibfield{author}{\bibinfo{person}{Miguel Castro} {and} \bibinfo{person}{Barbara Liskov}.} \bibinfo{year}{1999}\natexlab{}.
\newblock \showarticletitle{Practical Byzantine Fault Tolerance}. In \bibinfo{booktitle}{\emph{Third Symposium on Operating Systems Design and Implementation (OSDI)}}. \bibinfo{publisher}{USENIX Association, Co-sponsored by IEEE TCOS and ACM SIGOPS}, \bibinfo{address}{New Orleans, Louisiana}.
\newblock


\bibitem[Castro and Liskov(2002)]%
        {pbft-recovery}
\bibfield{author}{\bibinfo{person}{Miguel Castro} {and} \bibinfo{person}{Barbara Liskov}.} \bibinfo{year}{2002}\natexlab{}.
\newblock \showarticletitle{Practical byzantine fault tolerance and proactive recovery}.
\newblock \bibinfo{journal}{\emph{ACM Trans. Comput. Syst.}} \bibinfo{volume}{20}, \bibinfo{number}{4} (\bibinfo{date}{Nov.} \bibinfo{year}{2002}), \bibinfo{pages}{398–461}.
\newblock
\showISSN{0734-2071}
\href{https://doi.org/10.1145/571637.571640}{doi:\nolinkurl{10.1145/571637.571640}}


\bibitem[Centre(2024)]%
        {alibaba-singapore-fire}
\bibfield{author}{\bibinfo{person}{Data Centre}.} \bibinfo{year}{2024}\natexlab{}.
\newblock \showarticletitle{Alibaba Cloud hit by Digital Realty fire in Singapore}.
\newblock \bibinfo{journal}{\emph{Frontier Enterprise}} (\bibinfo{year}{2024}).
\newblock
\urldef\tempurl%
\url{https://www.frontier-enterprise.com/alibaba-cloud-hit-by-digital-realty-fire-in-singapore/}
\showURL{%
\tempurl}


\bibitem[Chandra et~al\mbox{.}(2007)]%
        {paxos-made-live}
\bibfield{author}{\bibinfo{person}{Tushar~D. Chandra}, \bibinfo{person}{Robert Griesemer}, {and} \bibinfo{person}{Joshua Redstone}.} \bibinfo{year}{2007}\natexlab{}.
\newblock \showarticletitle{Paxos Made Live: An Engineering Perspective}. In \bibinfo{booktitle}{\emph{Proceedings of the 26th ACM Symposium on Principles of Distributed Computing}} (Portland, OR, USA) \emph{(\bibinfo{series}{PODC '07})}. \bibinfo{publisher}{Association for Computing Machinery}, \bibinfo{address}{New York, NY, USA}, \bibinfo{pages}{398–407}.
\newblock
\showISBNx{9781595936165}
\href{https://doi.org/10.1145/1281100.1281103}{doi:\nolinkurl{10.1145/1281100.1281103}}


\bibitem[Chandra et~al\mbox{.}(2016)]%
        {read-leases-theory}
\bibfield{author}{\bibinfo{person}{Tushar~D. Chandra}, \bibinfo{person}{Vassos Hadzilacos}, {and} \bibinfo{person}{Sam Toueg}.} \bibinfo{year}{2016}\natexlab{}.
\newblock \showarticletitle{An Algorithm for Replicated Objects with Efficient Reads}. In \bibinfo{booktitle}{\emph{Proceedings of the 2016 ACM Symposium on Principles of Distributed Computing}} (Chicago, Illinois, USA) \emph{(\bibinfo{series}{PODC '16})}. \bibinfo{publisher}{Association for Computing Machinery}, \bibinfo{address}{New York, NY, USA}, \bibinfo{pages}{325–334}.
\newblock
\showISBNx{9781450339643}
\href{https://doi.org/10.1145/2933057.2933111}{doi:\nolinkurl{10.1145/2933057.2933111}}


\bibitem[Charapko et~al\mbox{.}(2019)]%
        {paxos-quorum-read}
\bibfield{author}{\bibinfo{person}{Aleksey Charapko}, \bibinfo{person}{Ailidani Ailijiang}, {and} \bibinfo{person}{Murat Demirbas}.} \bibinfo{year}{2019}\natexlab{}.
\newblock \showarticletitle{Linearizable Quorum Reads in Paxos}. In \bibinfo{booktitle}{\emph{11th USENIX Workshop on Hot Topics in Storage and File Systems (HotStorage 19)}}. \bibinfo{publisher}{USENIX Association}, \bibinfo{address}{Renton, WA}.
\newblock
\urldef\tempurl%
\url{https://www.usenix.org/conference/hotstorage19/presentation/charapko}
\showURL{%
\tempurl}


\bibitem[Charapko et~al\mbox{.}(2021)]%
        {pigpaxos}
\bibfield{author}{\bibinfo{person}{Aleksey Charapko}, \bibinfo{person}{Ailidani Ailijiang}, {and} \bibinfo{person}{Murat Demirbas}.} \bibinfo{year}{2021}\natexlab{}.
\newblock \showarticletitle{PigPaxos: Devouring the Communication Bottlenecks in Distributed Consensus}. In \bibinfo{booktitle}{\emph{Proceedings of the 2021 International Conference on Management of Data}} (Virtual Event, China) \emph{(\bibinfo{series}{SIGMOD '21})}. \bibinfo{publisher}{Association for Computing Machinery}, \bibinfo{address}{New York, NY, USA}, \bibinfo{pages}{235–247}.
\newblock
\showISBNx{9781450383431}
\href{https://doi.org/10.1145/3448016.3452834}{doi:\nolinkurl{10.1145/3448016.3452834}}


\bibitem[Choi et~al\mbox{.}(2023)]%
        {hydra-network-ordering}
\bibfield{author}{\bibinfo{person}{Inho Choi}, \bibinfo{person}{Ellis Michael}, \bibinfo{person}{Yunfan Li}, \bibinfo{person}{Dan R.~K. Ports}, {and} \bibinfo{person}{Jialin Li}.} \bibinfo{year}{2023}\natexlab{}.
\newblock \showarticletitle{Hydra: {Serialization-Free} Network Ordering for Strongly Consistent Distributed Applications}. In \bibinfo{booktitle}{\emph{20th USENIX Symposium on Networked Systems Design and Implementation (NSDI 23)}}. \bibinfo{publisher}{USENIX Association}, \bibinfo{address}{Boston, MA}, \bibinfo{pages}{293--320}.
\newblock
\showISBNx{978-1-939133-33-5}
\urldef\tempurl%
\url{https://www.usenix.org/conference/nsdi23/presentation/choi}
\showURL{%
\tempurl}


\bibitem[Cloud(2024)]%
        {gcp-global-infra}
\bibfield{author}{\bibinfo{person}{Google Cloud}.} \bibinfo{year}{2024}\natexlab{}.
\newblock \bibinfo{title}{Google Cloud locations}.
\newblock
\newblock
\shownote{\url{https://cloud.google.com/about/locations/}, Last accessed on 2024-11-30}.


\bibitem[Cooper et~al\mbox{.}(2010)]%
        {ycsb}
\bibfield{author}{\bibinfo{person}{Brian~F. Cooper}, \bibinfo{person}{Adam Silberstein}, \bibinfo{person}{Erwin Tam}, \bibinfo{person}{Raghu Ramakrishnan}, {and} \bibinfo{person}{Russell Sears}.} \bibinfo{year}{2010}\natexlab{}.
\newblock \showarticletitle{Benchmarking Cloud Serving Systems with YCSB}. In \bibinfo{booktitle}{\emph{Proceedings of the 1st ACM Symposium on Cloud Computing}} (Indianapolis, IN, USA) \emph{(\bibinfo{series}{SoCC '10})}. \bibinfo{publisher}{Association for Computing Machinery}, \bibinfo{address}{New York, NY, USA}, \bibinfo{pages}{143–154}.
\newblock
\showISBNx{9781450300360}
\href{https://doi.org/10.1145/1807128.1807152}{doi:\nolinkurl{10.1145/1807128.1807152}}


\bibitem[Corbett et~al\mbox{.}(2013)]%
        {spanner}
\bibfield{author}{\bibinfo{person}{James~C. Corbett}, \bibinfo{person}{Jeffrey Dean}, \bibinfo{person}{Michael Epstein}, \bibinfo{person}{Andrew Fikes}, \bibinfo{person}{Christopher Frost}, \bibinfo{person}{J.~J. Furman}, \bibinfo{person}{Sanjay Ghemawat}, \bibinfo{person}{Andrey Gubarev}, \bibinfo{person}{Christopher Heiser}, \bibinfo{person}{Peter Hochschild}, \bibinfo{person}{Wilson Hsieh}, \bibinfo{person}{Sebastian Kanthak}, \bibinfo{person}{Eugene Kogan}, \bibinfo{person}{Hongyi Li}, \bibinfo{person}{Alexander Lloyd}, \bibinfo{person}{Sergey Melnik}, \bibinfo{person}{David Mwaura}, \bibinfo{person}{David Nagle}, \bibinfo{person}{Sean Quinlan}, \bibinfo{person}{Rajesh Rao}, \bibinfo{person}{Lindsay Rolig}, \bibinfo{person}{Yasushi Saito}, \bibinfo{person}{Michal Szymaniak}, \bibinfo{person}{Christopher Taylor}, \bibinfo{person}{Ruth Wang}, {and} \bibinfo{person}{Dale Woodford}.} \bibinfo{year}{2013}\natexlab{}.
\newblock \showarticletitle{Spanner: Google’s Globally Distributed Database}.
\newblock \bibinfo{journal}{\emph{ACM Trans. Comput. Syst.}} \bibinfo{volume}{31}, \bibinfo{number}{3}, Article \bibinfo{articleno}{8} (\bibinfo{date}{aug} \bibinfo{year}{2013}), \bibinfo{numpages}{22}~pages.
\newblock
\showISSN{0734-2071}
\href{https://doi.org/10.1145/2491245}{doi:\nolinkurl{10.1145/2491245}}


\bibitem[Cui et~al\mbox{.}(2015)]%
        {paxos-made-transparent}
\bibfield{author}{\bibinfo{person}{Heming Cui}, \bibinfo{person}{Rui Gu}, \bibinfo{person}{Cheng Liu}, \bibinfo{person}{Tianyu Chen}, {and} \bibinfo{person}{Junfeng Yang}.} \bibinfo{year}{2015}\natexlab{}.
\newblock \showarticletitle{Paxos made transparent}. In \bibinfo{booktitle}{\emph{Proceedings of the 25th Symposium on Operating Systems Principles}} (Monterey, California) \emph{(\bibinfo{series}{SOSP '15})}. \bibinfo{publisher}{Association for Computing Machinery}, \bibinfo{address}{New York, NY, USA}, \bibinfo{pages}{105–120}.
\newblock
\showISBNx{9781450338349}
\href{https://doi.org/10.1145/2815400.2815427}{doi:\nolinkurl{10.1145/2815400.2815427}}


\bibitem[Dang et~al\mbox{.}(2020)]%
        {p4xos}
\bibfield{author}{\bibinfo{person}{Huynh~Tu Dang}, \bibinfo{person}{Pietro Bressana}, \bibinfo{person}{Han Wang}, \bibinfo{person}{Ki~Suh Lee}, \bibinfo{person}{Noa Zilberman}, \bibinfo{person}{Hakim Weatherspoon}, \bibinfo{person}{Marco Canini}, \bibinfo{person}{Fernando Pedone}, {and} \bibinfo{person}{Robert Soul\'{e}}.} \bibinfo{year}{2020}\natexlab{}.
\newblock \showarticletitle{P4xos: Consensus as a Network Service}.
\newblock \bibinfo{journal}{\emph{IEEE/ACM Trans. Netw.}} \bibinfo{volume}{28}, \bibinfo{number}{4} (\bibinfo{date}{Aug.} \bibinfo{year}{2020}), \bibinfo{pages}{1726–1738}.
\newblock
\showISSN{1063-6692}
\href{https://doi.org/10.1109/TNET.2020.2992106}{doi:\nolinkurl{10.1109/TNET.2020.2992106}}


\bibitem[Dang et~al\mbox{.}(2015)]%
        {netpaxos}
\bibfield{author}{\bibinfo{person}{Huynh~Tu Dang}, \bibinfo{person}{Daniele Sciascia}, \bibinfo{person}{Marco Canini}, \bibinfo{person}{Fernando Pedone}, {and} \bibinfo{person}{Robert Soul\'{e}}.} \bibinfo{year}{2015}\natexlab{}.
\newblock \showarticletitle{NetPaxos: consensus at network speed}. In \bibinfo{booktitle}{\emph{Proceedings of the 1st ACM SIGCOMM Symposium on Software Defined Networking Research}} (Santa Clara, California) \emph{(\bibinfo{series}{SOSR '15})}. \bibinfo{publisher}{Association for Computing Machinery}, \bibinfo{address}{New York, NY, USA}, Article \bibinfo{articleno}{5}, \bibinfo{numpages}{7}~pages.
\newblock
\showISBNx{9781450334518}
\href{https://doi.org/10.1145/2774993.2774999}{doi:\nolinkurl{10.1145/2774993.2774999}}


\bibitem[Ding et~al\mbox{.}(2020)]%
        {scalog}
\bibfield{author}{\bibinfo{person}{Cong Ding}, \bibinfo{person}{David Chu}, \bibinfo{person}{Evan Zhao}, \bibinfo{person}{Xiang Li}, \bibinfo{person}{Lorenzo Alvisi}, {and} \bibinfo{person}{Robbert~Van Renesse}.} \bibinfo{year}{2020}\natexlab{}.
\newblock \showarticletitle{Scalog: Seamless Reconfiguration and Total Order in a Scalable Shared Log}. In \bibinfo{booktitle}{\emph{17th USENIX Symposium on Networked Systems Design and Implementation (NSDI 20)}}. \bibinfo{publisher}{USENIX Association}, \bibinfo{address}{Santa Clara, CA}, \bibinfo{pages}{325--338}.
\newblock
\showISBNx{978-1-939133-13-7}
\urldef\tempurl%
\url{https://www.usenix.org/conference/nsdi20/presentation/ding}
\showURL{%
\tempurl}


\bibitem[Du et~al\mbox{.}(2014)]%
        {clock-rsm}
\bibfield{author}{\bibinfo{person}{Jiaqing Du}, \bibinfo{person}{Daniele Sciascia}, \bibinfo{person}{Sameh Elnikety}, \bibinfo{person}{Willy Zwaenepoel}, {and} \bibinfo{person}{Fernando Pedone}.} \bibinfo{year}{2014}\natexlab{}.
\newblock \showarticletitle{Clock-RSM: Low-Latency Inter-datacenter State Machine Replication Using Loosely Synchronized Physical Clocks}. In \bibinfo{booktitle}{\emph{2014 44th Annual IEEE/IFIP International Conference on Dependable Systems and Networks}}. \bibinfo{pages}{343--354}.
\newblock
\href{https://doi.org/10.1109/DSN.2014.42}{doi:\nolinkurl{10.1109/DSN.2014.42}}


\bibitem[Duplyakin et~al\mbox{.}(2019)]%
        {cloudlab}
\bibfield{author}{\bibinfo{person}{Dmitry Duplyakin}, \bibinfo{person}{Robert Ricci}, \bibinfo{person}{Aleksander Maricq}, \bibinfo{person}{Gary Wong}, \bibinfo{person}{Jonathon Duerig}, \bibinfo{person}{Eric Eide}, \bibinfo{person}{Leigh Stoller}, \bibinfo{person}{Mike Hibler}, \bibinfo{person}{David Johnson}, \bibinfo{person}{Kirk Webb}, \bibinfo{person}{Aditya Akella}, \bibinfo{person}{Kuangching Wang}, \bibinfo{person}{Glenn Ricart}, \bibinfo{person}{Larry Landweber}, \bibinfo{person}{Chip Elliott}, \bibinfo{person}{Michael Zink}, \bibinfo{person}{Emmanuel Cecchet}, \bibinfo{person}{Snigdhaswin Kar}, {and} \bibinfo{person}{Prabodh Mishra}.} \bibinfo{year}{2019}\natexlab{}.
\newblock \showarticletitle{The Design and Operation of {CloudLab}}. In \bibinfo{booktitle}{\emph{Proceedings of the {USENIX} Annual Technical Conference}}. \bibinfo{pages}{1--14}.
\newblock
\urldef\tempurl%
\url{https://www.flux.utah.edu/paper/duplyakin-atc19}
\showURL{%
\tempurl}


\bibitem[Enes et~al\mbox{.}(2020)]%
        {atlas}
\bibfield{author}{\bibinfo{person}{Vitor Enes}, \bibinfo{person}{Carlos Baquero}, \bibinfo{person}{Tuanir~Fran\c{c}a Rezende}, \bibinfo{person}{Alexey Gotsman}, \bibinfo{person}{Matthieu Perrin}, {and} \bibinfo{person}{Pierre Sutra}.} \bibinfo{year}{2020}\natexlab{}.
\newblock \showarticletitle{State-Machine Replication for Planet-Scale Systems}. In \bibinfo{booktitle}{\emph{Proceedings of the Fifteenth European Conference on Computer Systems}} (Heraklion, Greece) \emph{(\bibinfo{series}{EuroSys '20})}. \bibinfo{publisher}{Association for Computing Machinery}, \bibinfo{address}{New York, NY, USA}, Article \bibinfo{articleno}{24}, \bibinfo{numpages}{15}~pages.
\newblock
\showISBNx{9781450368827}
\href{https://doi.org/10.1145/3342195.3387543}{doi:\nolinkurl{10.1145/3342195.3387543}}


\bibitem[etcd(2023)]%
        {etcd}
\bibfield{author}{\bibinfo{person}{etcd}.} \bibinfo{year}{2023}\natexlab{}.
\newblock \bibinfo{title}{etcd: A distributed, reliable key-value store for the most critical data}.
\newblock
\newblock
\shownote{\url{https://etcd.io/}, Last accessed on 2023-11-13}.


\bibitem[FireScroll(2023)]%
        {firescroll}
\bibfield{author}{\bibinfo{person}{FireScroll}.} \bibinfo{year}{2023}\natexlab{}.
\newblock \bibinfo{title}{FireScroll: The config database to deploy everywhere}.
\newblock
\newblock
\shownote{\url{https://github.com/FireScroll/FireScroll}, Last accessed on 2024-09-05}.


\bibitem[Fouto et~al\mbox{.}(2022)]%
        {chain-paxos}
\bibfield{author}{\bibinfo{person}{Pedro Fouto}, \bibinfo{person}{Nuno Pregui{\c c}a}, {and} \bibinfo{person}{Joao Leit{\~a}o}.} \bibinfo{year}{2022}\natexlab{}.
\newblock \showarticletitle{High Throughput Replication with Integrated Membership Management}. In \bibinfo{booktitle}{\emph{2022 USENIX Annual Technical Conference (USENIX ATC 22)}}. \bibinfo{publisher}{USENIX Association}, \bibinfo{address}{Carlsbad, CA}, \bibinfo{pages}{575--592}.
\newblock
\showISBNx{978-1-939133-29-67}
\urldef\tempurl%
\url{https://www.usenix.org/conference/atc22/presentation/fouto}
\showURL{%
\tempurl}


\bibitem[Ganesan et~al\mbox{.}(2020)]%
        {consistency-aware-durability}
\bibfield{author}{\bibinfo{person}{Aishwarya Ganesan}, \bibinfo{person}{Ramnatthan Alagappan}, \bibinfo{person}{Andrea Arpaci-Dusseau}, {and} \bibinfo{person}{Remzi Arpaci-Dusseau}.} \bibinfo{year}{2020}\natexlab{}.
\newblock \showarticletitle{Strong and Efficient Consistency with {Consistency-Aware} Durability}. In \bibinfo{booktitle}{\emph{18th USENIX Conference on File and Storage Technologies (FAST 20)}}. \bibinfo{publisher}{USENIX Association}, \bibinfo{address}{Santa Clara, CA}, \bibinfo{pages}{323--337}.
\newblock
\showISBNx{978-1-939133-12-0}
\urldef\tempurl%
\url{https://www.usenix.org/conference/fast20/presentation/ganesan}
\showURL{%
\tempurl}


\bibitem[Ganesan et~al\mbox{.}(2021)]%
        {skyros-nil-externality}
\bibfield{author}{\bibinfo{person}{Aishwarya Ganesan}, \bibinfo{person}{Ramnatthan Alagappan}, \bibinfo{person}{Andrea~C. Arpaci-Dusseau}, {and} \bibinfo{person}{Remzi~H. Arpaci-Dusseau}.} \bibinfo{year}{2021}\natexlab{}.
\newblock \showarticletitle{Exploiting Nil-Externality for Fast Replicated Storage}. In \bibinfo{booktitle}{\emph{Proceedings of the ACM SIGOPS 28th Symposium on Operating Systems Principles}} (Virtual Event, Germany) \emph{(\bibinfo{series}{SOSP '21})}. \bibinfo{publisher}{Association for Computing Machinery}, \bibinfo{address}{New York, NY, USA}, \bibinfo{pages}{440–456}.
\newblock
\showISBNx{9781450387095}
\href{https://doi.org/10.1145/3477132.3483543}{doi:\nolinkurl{10.1145/3477132.3483543}}


\bibitem[Geng et~al\mbox{.}(2022)]%
        {nezha}
\bibfield{author}{\bibinfo{person}{Jinkun Geng}, \bibinfo{person}{Anirudh Sivaraman}, \bibinfo{person}{Balaji Prabhakar}, {and} \bibinfo{person}{Mendel Rosenblum}.} \bibinfo{year}{2022}\natexlab{}.
\newblock \showarticletitle{Nezha: Deployable and High-Performance Consensus Using Synchronized Clocks}.
\newblock \bibinfo{journal}{\emph{Proc. VLDB Endow.}} \bibinfo{volume}{16}, \bibinfo{number}{4} (\bibinfo{date}{dec} \bibinfo{year}{2022}), \bibinfo{pages}{629–642}.
\newblock
\showISSN{2150-8097}
\href{https://doi.org/10.14778/3574245.3574250}{doi:\nolinkurl{10.14778/3574245.3574250}}


\bibitem[Giridharan et~al\mbox{.}(2024)]%
        {autobahn}
\bibfield{author}{\bibinfo{person}{Neil Giridharan}, \bibinfo{person}{Florian Suri-Payer}, \bibinfo{person}{Ittai Abraham}, \bibinfo{person}{Lorenzo Alvisi}, {and} \bibinfo{person}{Natacha Crooks}.} \bibinfo{year}{2024}\natexlab{}.
\newblock \showarticletitle{Autobahn: Seamless high speed BFT}. In \bibinfo{booktitle}{\emph{Proceedings of the ACM SIGOPS 30th Symposium on Operating Systems Principles}} (Austin, TX, USA) \emph{(\bibinfo{series}{SOSP '24})}. \bibinfo{publisher}{Association for Computing Machinery}, \bibinfo{address}{New York, NY, USA}, \bibinfo{pages}{1–23}.
\newblock
\showISBNx{9798400712517}
\href{https://doi.org/10.1145/3694715.3695942}{doi:\nolinkurl{10.1145/3694715.3695942}}


\bibitem[Gray and Cheriton(1989)]%
        {leases-mechanism}
\bibfield{author}{\bibinfo{person}{C. Gray} {and} \bibinfo{person}{D. Cheriton}.} \bibinfo{year}{1989}\natexlab{}.
\newblock \showarticletitle{Leases: an efficient fault-tolerant mechanism for distributed file cache consistency}. In \bibinfo{booktitle}{\emph{Proceedings of the Twelfth ACM Symposium on Operating Systems Principles}} \emph{(\bibinfo{series}{SOSP '89})}. \bibinfo{publisher}{Association for Computing Machinery}, \bibinfo{address}{New York, NY, USA}, \bibinfo{pages}{202–210}.
\newblock
\showISBNx{0897913388}
\href{https://doi.org/10.1145/74850.74870}{doi:\nolinkurl{10.1145/74850.74870}}


\bibitem[Guarnieri and Charapko(2023)]%
        {edge-pqr}
\bibfield{author}{\bibinfo{person}{Joshua Guarnieri} {and} \bibinfo{person}{Aleksey Charapko}.} \bibinfo{year}{2023}\natexlab{}.
\newblock \showarticletitle{Linearizable Low-latency Reads at the Edge}. In \bibinfo{booktitle}{\emph{Proceedings of the 10th Workshop on Principles and Practice of Consistency for Distributed Data}} (Rome, Italy) \emph{(\bibinfo{series}{PaPoC '23})}. \bibinfo{publisher}{Association for Computing Machinery}, \bibinfo{address}{New York, NY, USA}, \bibinfo{pages}{77–83}.
\newblock
\showISBNx{9798400700866}
\href{https://doi.org/10.1145/3578358.3591327}{doi:\nolinkurl{10.1145/3578358.3591327}}


\bibitem[Guerraoui et~al\mbox{.}(2022)]%
        {ukharon}
\bibfield{author}{\bibinfo{person}{Rachid Guerraoui}, \bibinfo{person}{Antoine Murat}, \bibinfo{person}{Javier Picorel}, \bibinfo{person}{Athanasios Xygkis}, \bibinfo{person}{Huabing Yan}, {and} \bibinfo{person}{Pengfei Zuo}.} \bibinfo{year}{2022}\natexlab{}.
\newblock \showarticletitle{{uKharon}: A Membership Service for Microsecond Applications}. In \bibinfo{booktitle}{\emph{2022 USENIX Annual Technical Conference (ATC 22)}}. \bibinfo{publisher}{USENIX Association}, \bibinfo{address}{Carlsbad, CA}, \bibinfo{pages}{101--120}.
\newblock
\showISBNx{978-1-939133-29-24}
\urldef\tempurl%
\url{https://www.usenix.org/conference/atc22/presentation/guerraoui}
\showURL{%
\tempurl}


\bibitem[Gunawi et~al\mbox{.}(2014)]%
        {what-bugs-in-the-cloud}
\bibfield{author}{\bibinfo{person}{Haryadi~S. Gunawi}, \bibinfo{person}{Mingzhe Hao}, \bibinfo{person}{Tanakorn Leesatapornwongsa}, \bibinfo{person}{Tiratat Patana-anake}, \bibinfo{person}{Thanh Do}, \bibinfo{person}{Jeffry Adityatama}, \bibinfo{person}{Kurnia~J. Eliazar}, \bibinfo{person}{Agung Laksono}, \bibinfo{person}{Jeffrey~F. Lukman}, \bibinfo{person}{Vincentius Martin}, {and} \bibinfo{person}{Anang~D. Satria}.} \bibinfo{year}{2014}\natexlab{}.
\newblock \showarticletitle{What Bugs Live in the Cloud? A Study of 3000+ Issues in Cloud Systems}. In \bibinfo{booktitle}{\emph{Proceedings of the ACM Symposium on Cloud Computing}} (Seattle, WA, USA) \emph{(\bibinfo{series}{SOCC '14})}. \bibinfo{publisher}{Association for Computing Machinery}, \bibinfo{address}{New York, NY, USA}, \bibinfo{pages}{1–14}.
\newblock
\showISBNx{9781450332521}
\href{https://doi.org/10.1145/2670979.2670986}{doi:\nolinkurl{10.1145/2670979.2670986}}


\bibitem[Harding et~al\mbox{.}(2017)]%
        {evaluation-of-cc}
\bibfield{author}{\bibinfo{person}{Rachael Harding}, \bibinfo{person}{Dana Van~Aken}, \bibinfo{person}{Andrew Pavlo}, {and} \bibinfo{person}{Michael Stonebraker}.} \bibinfo{year}{2017}\natexlab{}.
\newblock \showarticletitle{An evaluation of distributed concurrency control}.
\newblock \bibinfo{journal}{\emph{Proc. VLDB Endow.}} \bibinfo{volume}{10}, \bibinfo{number}{5} (\bibinfo{date}{Jan} \bibinfo{year}{2017}), \bibinfo{pages}{553–564}.
\newblock
\showISSN{2150-8097}
\href{https://doi.org/10.14778/3055540.3055548}{doi:\nolinkurl{10.14778/3055540.3055548}}


\bibitem[HashiCorp(2020)]%
        {consul}
\bibfield{author}{\bibinfo{person}{HashiCorp}.} \bibinfo{year}{2020}\natexlab{}.
\newblock \bibinfo{title}{Consul}.
\newblock \bibinfo{howpublished}{\url{https://consul.io}}.
\newblock
\newblock
\shownote{Accessed: 2024-10-16}.


\bibitem[Hawblitzel et~al\mbox{.}(2015)]%
        {ironfleet}
\bibfield{author}{\bibinfo{person}{Chris Hawblitzel}, \bibinfo{person}{Jon Howell}, \bibinfo{person}{Manos Kapritsos}, \bibinfo{person}{Jacob~R. Lorch}, \bibinfo{person}{Bryan Parno}, \bibinfo{person}{Michael~L. Roberts}, \bibinfo{person}{Srinath Setty}, {and} \bibinfo{person}{Brian Zill}.} \bibinfo{year}{2015}\natexlab{}.
\newblock \showarticletitle{IronFleet: Proving Practical Distributed Systems Correct}. In \bibinfo{booktitle}{\emph{Proceedings of the 25th Symposium on Operating Systems Principles}} (Monterey, California) \emph{(\bibinfo{series}{SOSP '15})}. \bibinfo{publisher}{Association for Computing Machinery}, \bibinfo{address}{New York, NY, USA}, \bibinfo{pages}{1–17}.
\newblock
\showISBNx{9781450338349}
\href{https://doi.org/10.1145/2815400.2815428}{doi:\nolinkurl{10.1145/2815400.2815428}}


\bibitem[Helt et~al\mbox{.}(2021)]%
        {regular-sequential-consistency}
\bibfield{author}{\bibinfo{person}{Jeffrey Helt}, \bibinfo{person}{Matthew Burke}, \bibinfo{person}{Amit Levy}, {and} \bibinfo{person}{Wyatt Lloyd}.} \bibinfo{year}{2021}\natexlab{}.
\newblock \showarticletitle{Regular Sequential Serializability and Regular Sequential Consistency}. In \bibinfo{booktitle}{\emph{Proceedings of the ACM SIGOPS 28th Symposium on Operating Systems Principles}} (Virtual Event, Germany) \emph{(\bibinfo{series}{SOSP '21})}. \bibinfo{publisher}{Association for Computing Machinery}, \bibinfo{address}{New York, NY, USA}, \bibinfo{pages}{163–179}.
\newblock
\showISBNx{9781450387095}
\href{https://doi.org/10.1145/3477132.3483566}{doi:\nolinkurl{10.1145/3477132.3483566}}


\bibitem[Herlihy(1987)]%
        {dynamic-quorum}
\bibfield{author}{\bibinfo{person}{Maurice Herlihy}.} \bibinfo{year}{1987}\natexlab{}.
\newblock \showarticletitle{Dynamic quorum adjustment for partitioned data}.
\newblock \bibinfo{journal}{\emph{ACM Trans. Database Syst.}} \bibinfo{volume}{12}, \bibinfo{number}{2} (\bibinfo{date}{Jun} \bibinfo{year}{1987}), \bibinfo{pages}{170–194}.
\newblock
\showISSN{0362-5915}
\href{https://doi.org/10.1145/22952.22953}{doi:\nolinkurl{10.1145/22952.22953}}


\bibitem[Herlihy and Wing(1990)]%
        {linearizability}
\bibfield{author}{\bibinfo{person}{Maurice~P. Herlihy} {and} \bibinfo{person}{Jeannette~M. Wing}.} \bibinfo{year}{1990}\natexlab{}.
\newblock \showarticletitle{Linearizability: A Correctness Condition for Concurrent Objects}.
\newblock \bibinfo{journal}{\emph{ACM Trans. Program. Lang. Syst.}} \bibinfo{volume}{12}, \bibinfo{number}{3} (\bibinfo{date}{jul} \bibinfo{year}{1990}), \bibinfo{pages}{463–492}.
\newblock
\showISSN{0164-0925}
\href{https://doi.org/10.1145/78969.78972}{doi:\nolinkurl{10.1145/78969.78972}}


\bibitem[Hoang~Le et~al\mbox{.}(2019)]%
        {dynastar}
\bibfield{author}{\bibinfo{person}{Long Hoang~Le}, \bibinfo{person}{Enrique Fynn}, \bibinfo{person}{Mojtaba Eslahi-Kelorazi}, \bibinfo{person}{Robert Soulé}, {and} \bibinfo{person}{Fernando Pedone}.} \bibinfo{year}{2019}\natexlab{}.
\newblock \showarticletitle{DynaStar: Optimized Dynamic Partitioning for Scalable State Machine Replication}. In \bibinfo{booktitle}{\emph{2019 IEEE 39th International Conference on Distributed Computing Systems (ICDCS)}}. \bibinfo{pages}{1453--1465}.
\newblock
\href{https://doi.org/10.1109/ICDCS.2019.00145}{doi:\nolinkurl{10.1109/ICDCS.2019.00145}}


\bibitem[Howard et~al\mbox{.}(2016)]%
        {fpaxos}
\bibfield{author}{\bibinfo{person}{Heidi Howard}, \bibinfo{person}{Dahlia Malkhi}, {and} \bibinfo{person}{Alexander Spiegelman}.} \bibinfo{year}{2016}\natexlab{}.
\newblock \bibinfo{title}{Flexible Paxos: Quorum intersection revisited}.
\newblock
\showeprint[arxiv]{1608.06696}~[cs.DC]


\bibitem[Hu et~al\mbox{.}(2024a)]%
        {practical-consistency-summary}
\bibfield{author}{\bibinfo{person}{Guanzhou Hu}, \bibinfo{person}{Andrea Arpaci-Dusseau}, {and} \bibinfo{person}{Remzi Arpaci-Dusseau}.} \bibinfo{year}{2024}\natexlab{a}.
\newblock \bibinfo{title}{A Unified, Practical, and Understandable Summary of Non-transactional Consistency Levels in Distributed Replication}.
\newblock
\showeprint[arxiv]{2409.01576}~[cs.DC]
\urldef\tempurl%
\url{https://arxiv.org/abs/2409.01576}
\showURL{%
\tempurl}


\bibitem[Hu et~al\mbox{.}(2024b)]%
        {aceso-disaggregated-memory}
\bibfield{author}{\bibinfo{person}{Zhisheng Hu}, \bibinfo{person}{Pengfei Zuo}, \bibinfo{person}{Yizou Chen}, \bibinfo{person}{Chao Wang}, \bibinfo{person}{Junliang Hu}, {and} \bibinfo{person}{Ming-Chang Yang}.} \bibinfo{year}{2024}\natexlab{b}.
\newblock \showarticletitle{Aceso: Achieving Efficient Fault Tolerance in Memory-Disaggregated Key-Value Stores}. In \bibinfo{booktitle}{\emph{Proceedings of the ACM SIGOPS 30th Symposium on Operating Systems Principles}} (Austin, TX, USA) \emph{(\bibinfo{series}{SOSP '24})}. \bibinfo{publisher}{Association for Computing Machinery}, \bibinfo{address}{New York, NY, USA}, \bibinfo{pages}{127–143}.
\newblock
\showISBNx{9798400712517}
\href{https://doi.org/10.1145/3694715.3695951}{doi:\nolinkurl{10.1145/3694715.3695951}}


\bibitem[Huang et~al\mbox{.}(2020)]%
        {tidb}
\bibfield{author}{\bibinfo{person}{Dongxu Huang}, \bibinfo{person}{Qi Liu}, \bibinfo{person}{Qiu Cui}, \bibinfo{person}{Zhuhe Fang}, \bibinfo{person}{Xiaoyu Ma}, \bibinfo{person}{Fei Xu}, \bibinfo{person}{Li Shen}, \bibinfo{person}{Liu Tang}, \bibinfo{person}{Yuxing Zhou}, \bibinfo{person}{Menglong Huang}, \bibinfo{person}{Wan Wei}, \bibinfo{person}{Cong Liu}, \bibinfo{person}{Jian Zhang}, \bibinfo{person}{Jianjun Li}, \bibinfo{person}{Xuelian Wu}, \bibinfo{person}{Lingyu Song}, \bibinfo{person}{Ruoxi Sun}, \bibinfo{person}{Shuaipeng Yu}, \bibinfo{person}{Lei Zhao}, \bibinfo{person}{Nicholas Cameron}, \bibinfo{person}{Liquan Pei}, {and} \bibinfo{person}{Xin Tang}.} \bibinfo{year}{2020}\natexlab{}.
\newblock \showarticletitle{TiDB: A Raft-Based HTAP Database}.
\newblock \bibinfo{journal}{\emph{Proc. VLDB Endow.}} \bibinfo{volume}{13}, \bibinfo{number}{12} (\bibinfo{date}{aug} \bibinfo{year}{2020}), \bibinfo{pages}{3072–3084}.
\newblock
\showISSN{2150-8097}
\href{https://doi.org/10.14778/3415478.3415535}{doi:\nolinkurl{10.14778/3415478.3415535}}


\bibitem[Huang et~al\mbox{.}(2017)]%
        {gray-failures}
\bibfield{author}{\bibinfo{person}{Peng Huang}, \bibinfo{person}{Chuanxiong Guo}, \bibinfo{person}{Lidong Zhou}, \bibinfo{person}{Jacob~R. Lorch}, \bibinfo{person}{Yingnong Dang}, \bibinfo{person}{Murali Chintalapati}, {and} \bibinfo{person}{Randolph Yao}.} \bibinfo{year}{2017}\natexlab{}.
\newblock \showarticletitle{Gray Failure: The Achilles' Heel of Cloud-Scale Systems}. In \bibinfo{booktitle}{\emph{Proceedings of the 16th Workshop on Hot Topics in Operating Systems}} (Whistler, BC, Canada) \emph{(\bibinfo{series}{HotOS '17})}. \bibinfo{publisher}{Association for Computing Machinery}, \bibinfo{address}{New York, NY, USA}, \bibinfo{pages}{150–155}.
\newblock
\showISBNx{9781450350686}
\href{https://doi.org/10.1145/3102980.3103005}{doi:\nolinkurl{10.1145/3102980.3103005}}


\bibitem[Hunt et~al\mbox{.}(2010)]%
        {zookeeper}
\bibfield{author}{\bibinfo{person}{Patrick Hunt}, \bibinfo{person}{Mahadev Konar}, \bibinfo{person}{Flavio~P. Junqueira}, {and} \bibinfo{person}{Benjamin Reed}.} \bibinfo{year}{2010}\natexlab{}.
\newblock \showarticletitle{ZooKeeper: Wait-Free Coordination for Internet-Scale Systems}. In \bibinfo{booktitle}{\emph{Proceedings of the 2010 USENIX Conference on USENIX Annual Technical Conference}} (Boston, MA) \emph{(\bibinfo{series}{USENIXATC'10})}. \bibinfo{publisher}{USENIX Association}, \bibinfo{address}{USA}, \bibinfo{pages}{11}.
\newblock


\bibitem[Hunt(2017)]%
        {aws-time-sync-service}
\bibfield{author}{\bibinfo{person}{Randall Hunt}.} \bibinfo{year}{2017}\natexlab{}.
\newblock \bibinfo{title}{Keeping Time With Amazon Time Sync Service}.
\newblock \bibinfo{howpublished}{\url{https://aws.amazon.com/blogs/aws/keeping-time-with-amazon-time-sync-service/}}.
\newblock
\newblock
\shownote{Accessed: 2017-11-29}.


\bibitem[Jha et~al\mbox{.}(2019)]%
        {derecho}
\bibfield{author}{\bibinfo{person}{Sagar Jha}, \bibinfo{person}{Jonathan Behrens}, \bibinfo{person}{Theo Gkountouvas}, \bibinfo{person}{Mae Milano}, \bibinfo{person}{Weijia Song}, \bibinfo{person}{Edward Tremel}, \bibinfo{person}{Robbert~Van Renesse}, \bibinfo{person}{Sydney Zink}, {and} \bibinfo{person}{Kenneth~P. Birman}.} \bibinfo{year}{2019}\natexlab{}.
\newblock \showarticletitle{Derecho: Fast State Machine Replication for Cloud Services}.
\newblock \bibinfo{journal}{\emph{ACM Trans. Comput. Syst.}} \bibinfo{volume}{36}, \bibinfo{number}{2}, Article \bibinfo{articleno}{4} (\bibinfo{date}{apr} \bibinfo{year}{2019}), \bibinfo{numpages}{49}~pages.
\newblock
\showISSN{0734-2071}
\href{https://doi.org/10.1145/3302258}{doi:\nolinkurl{10.1145/3302258}}


\bibitem[Kaldor et~al\mbox{.}(2017)]%
        {canopy}
\bibfield{author}{\bibinfo{person}{Jonathan Kaldor}, \bibinfo{person}{Jonathan Mace}, \bibinfo{person}{Micha\l{} Bejda}, \bibinfo{person}{Edison Gao}, \bibinfo{person}{Wiktor Kuropatwa}, \bibinfo{person}{Joe O'Neill}, \bibinfo{person}{Kian~Win Ong}, \bibinfo{person}{Bill Schaller}, \bibinfo{person}{Pingjia Shan}, \bibinfo{person}{Brendan Viscomi}, \bibinfo{person}{Vinod Venkataraman}, \bibinfo{person}{Kaushik Veeraraghavan}, {and} \bibinfo{person}{Yee~Jiun Song}.} \bibinfo{year}{2017}\natexlab{}.
\newblock \showarticletitle{Canopy: An End-to-End Performance Tracing And Analysis System}. In \bibinfo{booktitle}{\emph{Proceedings of the 26th Symposium on Operating Systems Principles}} (Shanghai, China) \emph{(\bibinfo{series}{SOSP '17})}. \bibinfo{publisher}{Association for Computing Machinery}, \bibinfo{address}{New York, NY, USA}, \bibinfo{pages}{34–50}.
\newblock
\showISBNx{9781450350853}
\href{https://doi.org/10.1145/3132747.3132749}{doi:\nolinkurl{10.1145/3132747.3132749}}


\bibitem[Katsarakis et~al\mbox{.}(2020)]%
        {hermes}
\bibfield{author}{\bibinfo{person}{Antonios Katsarakis}, \bibinfo{person}{Vasilis Gavrielatos}, \bibinfo{person}{M.R.~Siavash Katebzadeh}, \bibinfo{person}{Arpit Joshi}, \bibinfo{person}{Aleksandar Dragojevic}, \bibinfo{person}{Boris Grot}, {and} \bibinfo{person}{Vijay Nagarajan}.} \bibinfo{year}{2020}\natexlab{}.
\newblock \showarticletitle{Hermes: A Fast, Fault-Tolerant and Linearizable Replication Protocol}. In \bibinfo{booktitle}{\emph{Proceedings of the Twenty-Fifth International Conference on Architectural Support for Programming Languages and Operating Systems}} (Lausanne, Switzerland) \emph{(\bibinfo{series}{ASPLOS '20})}. \bibinfo{publisher}{Association for Computing Machinery}, \bibinfo{address}{New York, NY, USA}, \bibinfo{pages}{201–217}.
\newblock
\showISBNx{9781450371025}
\href{https://doi.org/10.1145/3373376.3378496}{doi:\nolinkurl{10.1145/3373376.3378496}}


\bibitem[KRaft(2025)]%
        {kraft}
\bibfield{author}{\bibinfo{person}{KRaft}.} \bibinfo{year}{2025}\natexlab{}.
\newblock \bibinfo{title}{KRaft: Apache Kafka Without ZooKeeper}.
\newblock
\newblock
\shownote{\url{https://developer.confluent.io/learn/kraft/}, Last accessed on 2025-04-12}.


\bibitem[Kung and Robinson(1981)]%
        {occ-methods}
\bibfield{author}{\bibinfo{person}{H.~T. Kung} {and} \bibinfo{person}{John~T. Robinson}.} \bibinfo{year}{1981}\natexlab{}.
\newblock \showarticletitle{On optimistic methods for concurrency control}.
\newblock \bibinfo{journal}{\emph{ACM Trans. Database Syst.}} \bibinfo{volume}{6}, \bibinfo{number}{2} (\bibinfo{date}{jun} \bibinfo{year}{1981}), \bibinfo{pages}{213–226}.
\newblock
\showISSN{0362-5915}
\href{https://doi.org/10.1145/319566.319567}{doi:\nolinkurl{10.1145/319566.319567}}


\bibitem[Lamport(1979)]%
        {sequential-consistency}
\bibfield{author}{\bibinfo{person}{Leslie Lamport}.} \bibinfo{year}{1979}\natexlab{}.
\newblock \showarticletitle{How to Make a Multiprocessor Computer That Correctly Executes Multiprocess Programs}.
\newblock \bibinfo{journal}{\emph{IEEE Transactions on Computers C-28}}  \bibinfo{volume}{9} (\bibinfo{date}{September} \bibinfo{year}{1979}), \bibinfo{pages}{690--691}.
\newblock
\urldef\tempurl%
\url{https://www.microsoft.com/en-us/research/publication/make-multiprocessor-computer-correctly-executes-multiprocess-programs/}
\showURL{%
\tempurl}


\bibitem[Lamport(1998)]%
        {paxos-parliament}
\bibfield{author}{\bibinfo{person}{Leslie Lamport}.} \bibinfo{year}{1998}\natexlab{}.
\newblock \showarticletitle{The Part-Time Parliament}.
\newblock \bibinfo{journal}{\emph{ACM Trans. Comput. Syst.}} \bibinfo{volume}{16}, \bibinfo{number}{2} (\bibinfo{date}{may} \bibinfo{year}{1998}), \bibinfo{pages}{133–169}.
\newblock
\showISSN{0734-2071}
\href{https://doi.org/10.1145/279227.279229}{doi:\nolinkurl{10.1145/279227.279229}}


\bibitem[Lamport(2001)]%
        {paxos-made-simple}
\bibfield{author}{\bibinfo{person}{Leslie Lamport}.} \bibinfo{year}{2001}\natexlab{}.
\newblock \showarticletitle{Paxos Made Simple}.
\newblock \bibinfo{journal}{\emph{ACM SIGACT News (Distributed Computing Column) 32, 4 (Whole Number 121, December 2001)}} (\bibinfo{date}{December} \bibinfo{year}{2001}), \bibinfo{pages}{51--58}.
\newblock
\urldef\tempurl%
\url{https://www.microsoft.com/en-us/research/publication/paxos-made-simple/}
\showURL{%
\tempurl}


\bibitem[Lamport(2005)]%
        {generalized-paxos}
\bibfield{author}{\bibinfo{person}{Leslie Lamport}.} \bibinfo{year}{2005}\natexlab{}.
\newblock \showarticletitle{Generalized consensus and Paxos}.
\newblock \bibinfo{journal}{\emph{Microsoft Research Technical Report}} (\bibinfo{year}{2005}).
\newblock


\bibitem[Lamport(2006)]%
        {fast-paxos}
\bibfield{author}{\bibinfo{person}{Leslie Lamport}.} \bibinfo{year}{2006}\natexlab{}.
\newblock \showarticletitle{Fast Paxos}.
\newblock \bibinfo{journal}{\emph{Distrib. Comput.}} \bibinfo{volume}{19}, \bibinfo{number}{2} (\bibinfo{date}{oct} \bibinfo{year}{2006}), \bibinfo{pages}{79–103}.
\newblock
\showISSN{0178-2770}
\href{https://doi.org/10.1007/s00446-006-0005-x}{doi:\nolinkurl{10.1007/s00446-006-0005-x}}


\bibitem[Lamport et~al\mbox{.}(2009)]%
        {vertical-paxos}
\bibfield{author}{\bibinfo{person}{Leslie Lamport}, \bibinfo{person}{Dahlia Malkhi}, {and} \bibinfo{person}{Lidong Zhou}.} \bibinfo{year}{2009}\natexlab{}.
\newblock \showarticletitle{Vertical paxos and primary-backup replication}. In \bibinfo{booktitle}{\emph{Proceedings of the 28th ACM Symposium on Principles of Distributed Computing}} (Calgary, AB, Canada) \emph{(\bibinfo{series}{PODC '09})}. \bibinfo{publisher}{Association for Computing Machinery}, \bibinfo{address}{New York, NY, USA}, \bibinfo{pages}{312–313}.
\newblock
\showISBNx{9781605583969}
\href{https://doi.org/10.1145/1582716.1582783}{doi:\nolinkurl{10.1145/1582716.1582783}}


\bibitem[Lampson(2001)]%
        {abcd-of-paxos}
\bibfield{author}{\bibinfo{person}{Butler Lampson}.} \bibinfo{year}{2001}\natexlab{}.
\newblock \showarticletitle{The ABCD's of Paxos}. In \bibinfo{booktitle}{\emph{Proceedings of the Twentieth Annual ACM Symposium on Principles of Distributed Computing}} (Newport, RI, USA) \emph{(\bibinfo{series}{PODC '01})}. \bibinfo{publisher}{Association for Computing Machinery}, \bibinfo{address}{New York, NY, USA}, \bibinfo{pages}{13}.
\newblock
\showISBNx{1581133839}
\href{https://doi.org/10.1145/383962.383969}{doi:\nolinkurl{10.1145/383962.383969}}


\bibitem[Lattuada et~al\mbox{.}(2024)]%
        {verus}
\bibfield{author}{\bibinfo{person}{Andrea Lattuada}, \bibinfo{person}{Travis Hance}, \bibinfo{person}{Jay Bosamiya}, \bibinfo{person}{Matthias Brun}, \bibinfo{person}{Chanhee Cho}, \bibinfo{person}{Hayley LeBlanc}, \bibinfo{person}{Pranav Srinivasan}, \bibinfo{person}{Reto Achermann}, \bibinfo{person}{Tej Chajed}, \bibinfo{person}{Chris Hawblitzel}, \bibinfo{person}{Jon Howell}, \bibinfo{person}{Jacob~R. Lorch}, \bibinfo{person}{Oded Padon}, {and} \bibinfo{person}{Bryan Parno}.} \bibinfo{year}{2024}\natexlab{}.
\newblock \showarticletitle{Verus: A Practical Foundation for Systems Verification}. In \bibinfo{booktitle}{\emph{Proceedings of the ACM SIGOPS 30th Symposium on Operating Systems Principles}} (Austin, TX, USA) \emph{(\bibinfo{series}{SOSP '24})}. \bibinfo{publisher}{Association for Computing Machinery}, \bibinfo{address}{New York, NY, USA}, \bibinfo{pages}{438–454}.
\newblock
\showISBNx{9798400712517}
\href{https://doi.org/10.1145/3694715.3695952}{doi:\nolinkurl{10.1145/3694715.3695952}}


\bibitem[Lee et~al\mbox{.}(2015)]%
        {rifl}
\bibfield{author}{\bibinfo{person}{Collin Lee}, \bibinfo{person}{Seo~Jin Park}, \bibinfo{person}{Ankita Kejriwal}, \bibinfo{person}{Satoshi Matsushita}, {and} \bibinfo{person}{John Ousterhout}.} \bibinfo{year}{2015}\natexlab{}.
\newblock \showarticletitle{Implementing Linearizability at Large Scale and Low Latency}. In \bibinfo{booktitle}{\emph{Proceedings of the 25th Symposium on Operating Systems Principles}} (CA) \emph{(\bibinfo{series}{SOSP '15})}. \bibinfo{publisher}{Association for Computing Machinery}, \bibinfo{address}{New York, NY, USA}, \bibinfo{pages}{71–86}.
\newblock
\showISBNx{9781450338349}
\href{https://doi.org/10.1145/2815400.2815416}{doi:\nolinkurl{10.1145/2815400.2815416}}


\bibitem[Lee and Thekkath(1996)]%
        {petal}
\bibfield{author}{\bibinfo{person}{Edward K.~F. Lee} {and} \bibinfo{person}{Chandramohan~A. Thekkath}.} \bibinfo{year}{1996}\natexlab{}.
\newblock \showarticletitle{Petal: distributed virtual disks}. In \bibinfo{booktitle}{\emph{ASPLOS VII}}.
\newblock
\urldef\tempurl%
\url{https://api.semanticscholar.org/CorpusID:314852}
\showURL{%
\tempurl}


\bibitem[Lee et~al\mbox{.}(2016)]%
        {dtp-clock-sync}
\bibfield{author}{\bibinfo{person}{Ki~Suh Lee}, \bibinfo{person}{Han Wang}, \bibinfo{person}{Vishal Shrivastav}, {and} \bibinfo{person}{Hakim Weatherspoon}.} \bibinfo{year}{2016}\natexlab{}.
\newblock \showarticletitle{Globally Synchronized Time via Datacenter Networks}. In \bibinfo{booktitle}{\emph{Proceedings of the 2016 ACM SIGCOMM Conference}} (Florianopolis, Brazil) \emph{(\bibinfo{series}{SIGCOMM '16})}. \bibinfo{publisher}{Association for Computing Machinery}, \bibinfo{address}{New York, NY, USA}, \bibinfo{pages}{454–467}.
\newblock
\showISBNx{9781450341936}
\href{https://doi.org/10.1145/2934872.2934885}{doi:\nolinkurl{10.1145/2934872.2934885}}


\bibitem[Li et~al\mbox{.}(2016)]%
        {nopaxos}
\bibfield{author}{\bibinfo{person}{Jialin Li}, \bibinfo{person}{Ellis Michael}, \bibinfo{person}{Naveen~Kr. Sharma}, \bibinfo{person}{Adriana Szekeres}, {and} \bibinfo{person}{Dan R.~K. Ports}.} \bibinfo{year}{2016}\natexlab{}.
\newblock \showarticletitle{Just Say {NO} to Paxos Overhead: Replacing Consensus with Network Ordering}. In \bibinfo{booktitle}{\emph{12th USENIX Symposium on Operating Systems Design and Implementation (OSDI 16)}}. \bibinfo{publisher}{USENIX Association}, \bibinfo{address}{Savannah, GA}, \bibinfo{pages}{467--483}.
\newblock
\showISBNx{978-1-931971-33-1}
\urldef\tempurl%
\url{https://www.usenix.org/conference/osdi16/technical-sessions/presentation/li}
\showURL{%
\tempurl}


\bibitem[Li et~al\mbox{.}(2020)]%
        {sundial}
\bibfield{author}{\bibinfo{person}{Yuliang Li}, \bibinfo{person}{Gautam Kumar}, \bibinfo{person}{Hema Hariharan}, \bibinfo{person}{Hassan Wassel}, \bibinfo{person}{Peter Hochschild}, \bibinfo{person}{Dave Platt}, \bibinfo{person}{Simon Sabato}, \bibinfo{person}{Minlan Yu}, \bibinfo{person}{Nandita Dukkipati}, \bibinfo{person}{Prashant Chandra}, {and} \bibinfo{person}{Amin Vahdat}.} \bibinfo{year}{2020}\natexlab{}.
\newblock \showarticletitle{Sundial: Fault-tolerant Clock Synchronization for Datacenters}. In \bibinfo{booktitle}{\emph{14th USENIX Symposium on Operating Systems Design and Implementation (OSDI 20)}}. \bibinfo{publisher}{USENIX Association}, \bibinfo{pages}{1171--1186}.
\newblock
\showISBNx{978-1-939133-19-9}
\urldef\tempurl%
\url{https://www.usenix.org/conference/osdi20/presentation/li-yuliang}
\showURL{%
\tempurl}


\bibitem[Lin et~al\mbox{.}(2008)]%
        {pacifica}
\bibfield{author}{\bibinfo{person}{Wei Lin}, \bibinfo{person}{Mao Yang}, \bibinfo{person}{Lintao Zhang}, {and} \bibinfo{person}{Lidong Zhou}.} \bibinfo{year}{2008}\natexlab{}.
\newblock \showarticletitle{PacificA: Replication in Log-Based Distributed Storage Systems}.
\newblock
\urldef\tempurl%
\url{https://api.semanticscholar.org/CorpusID:18304090}
\showURL{%
\tempurl}


\bibitem[{Linux man pages}(2011)]%
        {tc-netem}
\bibfield{author}{\bibinfo{person}{{Linux man pages}}.} \bibinfo{year}{2011}\natexlab{}.
\newblock \bibinfo{title}{tc-netem(8) — Linux manual page}.
\newblock \bibinfo{howpublished}{\url{https://man7.org/linux/man-pages/man8/tc-netem.8.html}}.
\newblock
\newblock
\shownote{[Online; accessed 29-November-2023]}.


\bibitem[Lloyd et~al\mbox{.}(2011)]%
        {cops}
\bibfield{author}{\bibinfo{person}{Wyatt Lloyd}, \bibinfo{person}{Michael~J. Freedman}, \bibinfo{person}{Michael Kaminsky}, {and} \bibinfo{person}{David~G. Andersen}.} \bibinfo{year}{2011}\natexlab{}.
\newblock \showarticletitle{Don't settle for eventual: scalable causal consistency for wide-area storage with COPS}. In \bibinfo{booktitle}{\emph{Proceedings of the Twenty-Third ACM Symposium on Operating Systems Principles}} (Cascais, Portugal) \emph{(\bibinfo{series}{SOSP '11})}. \bibinfo{publisher}{Association for Computing Machinery}, \bibinfo{address}{New York, NY, USA}, \bibinfo{pages}{401–416}.
\newblock
\showISBNx{9781450309776}
\href{https://doi.org/10.1145/2043556.2043593}{doi:\nolinkurl{10.1145/2043556.2043593}}


\bibitem[Lockerman et~al\mbox{.}(2018)]%
        {fuzzylog}
\bibfield{author}{\bibinfo{person}{Joshua Lockerman}, \bibinfo{person}{Jose~M. Faleiro}, \bibinfo{person}{Juno Kim}, \bibinfo{person}{Soham Sankaran}, \bibinfo{person}{Daniel~J. Abadi}, \bibinfo{person}{James Aspnes}, \bibinfo{person}{Siddhartha Sen}, {and} \bibinfo{person}{Mahesh Balakrishnan}.} \bibinfo{year}{2018}\natexlab{}.
\newblock \showarticletitle{The {FuzzyLog}: A Partially Ordered Shared Log}. In \bibinfo{booktitle}{\emph{13th USENIX Symposium on Operating Systems Design and Implementation (OSDI 18)}}. \bibinfo{publisher}{USENIX Association}, \bibinfo{address}{Carlsbad, CA}, \bibinfo{pages}{357--372}.
\newblock
\showISBNx{978-1-939133-08-3}
\urldef\tempurl%
\url{https://www.usenix.org/conference/osdi18/presentation/lockerman}
\showURL{%
\tempurl}


\bibitem[Luo et~al\mbox{.}(2024)]%
        {lazylog}
\bibfield{author}{\bibinfo{person}{Xuhao Luo}, \bibinfo{person}{Shreesha~G. Bhat}, \bibinfo{person}{Jiyu Hu}, \bibinfo{person}{Ramnatthan Alagappan}, {and} \bibinfo{person}{Aishwarya Ganesan}.} \bibinfo{year}{2024}\natexlab{}.
\newblock \showarticletitle{LazyLog: A New Shared Log Abstraction for Low-Latency Applications}. In \bibinfo{booktitle}{\emph{Proceedings of the ACM SIGOPS 30th Symposium on Operating Systems Principles}} (Austin, TX, USA) \emph{(\bibinfo{series}{SOSP '24})}. \bibinfo{publisher}{Association for Computing Machinery}, \bibinfo{address}{New York, NY, USA}, \bibinfo{pages}{296–312}.
\newblock
\showISBNx{9798400712517}
\href{https://doi.org/10.1145/3694715.3695983}{doi:\nolinkurl{10.1145/3694715.3695983}}


\bibitem[Luo et~al\mbox{.}(2022)]%
        {depfast}
\bibfield{author}{\bibinfo{person}{Xuhao Luo}, \bibinfo{person}{Weihai Shen}, \bibinfo{person}{Shuai Mu}, {and} \bibinfo{person}{Tianyin Xu}.} \bibinfo{year}{2022}\natexlab{}.
\newblock \showarticletitle{{DepFast}: Orchestrating Code of Quorum Systems}. In \bibinfo{booktitle}{\emph{2022 USENIX Annual Technical Conference (ATC 22)}}. \bibinfo{publisher}{USENIX Association}, \bibinfo{pages}{557--574}.
\newblock
\showISBNx{978-1-939133-29-14}
\urldef\tempurl%
\url{https://www.usenix.org/conference/atc22/presentation/luo}
\showURL{%
\tempurl}


\bibitem[Ma et~al\mbox{.}(2022)]%
        {sift-veri}
\bibfield{author}{\bibinfo{person}{Haojun Ma}, \bibinfo{person}{Hammad Ahmad}, \bibinfo{person}{Aman Goel}, \bibinfo{person}{Eli Goldweber}, \bibinfo{person}{Jean-Baptiste Jeannin}, \bibinfo{person}{Manos Kapritsos}, {and} \bibinfo{person}{Baris Kasikci}.} \bibinfo{year}{2022}\natexlab{}.
\newblock \showarticletitle{Sift: Using Refinement-guided Automation to Verify Complex Distributed Systems}. In \bibinfo{booktitle}{\emph{2022 USENIX Annual Technical Conference (USENIX ATC 22)}}. \bibinfo{publisher}{USENIX Association}, \bibinfo{address}{Carlsbad, CA}, \bibinfo{pages}{151--166}.
\newblock
\showISBNx{978-1-939133-29-64}
\urldef\tempurl%
\url{https://www.usenix.org/conference/atc22/presentation/ma}
\showURL{%
\tempurl}


\bibitem[Ma et~al\mbox{.}(2019)]%
        {I4}
\bibfield{author}{\bibinfo{person}{Haojun Ma}, \bibinfo{person}{Aman Goel}, \bibinfo{person}{Jean-Baptiste Jeannin}, \bibinfo{person}{Manos Kapritsos}, \bibinfo{person}{Baris Kasikci}, {and} \bibinfo{person}{Karem~A. Sakallah}.} \bibinfo{year}{2019}\natexlab{}.
\newblock \showarticletitle{I4: Incremental Inference of Inductive Invariants for Verification of Distributed Protocols}. In \bibinfo{booktitle}{\emph{Proceedings of the 27th ACM Symposium on Operating Systems Principles}} (Huntsville, Ontario, Canada) \emph{(\bibinfo{series}{SOSP '19})}. \bibinfo{publisher}{Association for Computing Machinery}, \bibinfo{address}{New York, NY, USA}, \bibinfo{pages}{370–384}.
\newblock
\showISBNx{9781450368735}
\href{https://doi.org/10.1145/3341301.3359651}{doi:\nolinkurl{10.1145/3341301.3359651}}


\bibitem[Ma et~al\mbox{.}(2024)]%
        {noctua}
\bibfield{author}{\bibinfo{person}{Kai Ma}, \bibinfo{person}{Cheng Li}, \bibinfo{person}{Enzuo Zhu}, \bibinfo{person}{Ruichuan Chen}, \bibinfo{person}{Feng Yan}, {and} \bibinfo{person}{Kang Chen}.} \bibinfo{year}{2024}\natexlab{}.
\newblock \showarticletitle{Noctua: Towards Automated and Practical Fine-grained Consistency Analysis}. In \bibinfo{booktitle}{\emph{Proceedings of the Nineteenth European Conference on Computer Systems}} (Athens, Greece) \emph{(\bibinfo{series}{EuroSys '24})}. \bibinfo{publisher}{Association for Computing Machinery}, \bibinfo{address}{New York, NY, USA}, \bibinfo{pages}{704–719}.
\newblock
\showISBNx{9798400704376}
\href{https://doi.org/10.1145/3627703.3629570}{doi:\nolinkurl{10.1145/3627703.3629570}}


\bibitem[Mao et~al\mbox{.}(2008)]%
        {mencius}
\bibfield{author}{\bibinfo{person}{Yanhua Mao}, \bibinfo{person}{Flavio~P. Junqueira}, {and} \bibinfo{person}{Keith Marzullo}.} \bibinfo{year}{2008}\natexlab{}.
\newblock \showarticletitle{Mencius: Building Efficient Replicated State Machines for WANs}. In \bibinfo{booktitle}{\emph{Proceedings of the 8th USENIX Conference on Operating Systems Design and Implementation}} (San Diego, California) \emph{(\bibinfo{series}{OSDI'08})}. \bibinfo{publisher}{USENIX Association}, \bibinfo{address}{USA}, \bibinfo{pages}{369–384}.
\newblock


\bibitem[Martin et~al\mbox{.}(2010)]%
        {rfc-ntp}
\bibfield{author}{\bibinfo{person}{Jim Martin}, \bibinfo{person}{Jack Burbank}, \bibinfo{person}{William Kasch}, {and} \bibinfo{person}{Professor David~L. Mills}.} \bibinfo{year}{2010}\natexlab{}.
\newblock \bibinfo{title}{{Network Time Protocol Version 4: Protocol and Algorithms Specification}}.
\newblock \bibinfo{howpublished}{RFC 5905}.
\newblock
\href{https://doi.org/10.17487/RFC5905}{doi:\nolinkurl{10.17487/RFC5905}}


\bibitem[Mehdi et~al\mbox{.}(2017)]%
        {occult}
\bibfield{author}{\bibinfo{person}{Syed~Akbar Mehdi}, \bibinfo{person}{Cody Littley}, \bibinfo{person}{Natacha Crooks}, \bibinfo{person}{Lorenzo Alvisi}, \bibinfo{person}{Nathan Bronson}, {and} \bibinfo{person}{Wyatt Lloyd}.} \bibinfo{year}{2017}\natexlab{}.
\newblock \showarticletitle{I {Can{\textquoteright}t} Believe {It{\textquoteright}s} Not Causal! Scalable Causal Consistency with No Slowdown Cascades}. In \bibinfo{booktitle}{\emph{14th USENIX Symposium on Networked Systems Design and Implementation (NSDI 17)}}. \bibinfo{publisher}{USENIX Association}, \bibinfo{address}{Boston, MA}, \bibinfo{pages}{453--468}.
\newblock
\showISBNx{978-1-931971-37-9}
\urldef\tempurl%
\url{https://www.usenix.org/conference/nsdi17/technical-sessions/presentation/mehdi}
\showURL{%
\tempurl}


\bibitem[Microsoft(2024)]%
        {azure-global-infra}
\bibfield{author}{\bibinfo{person}{Microsoft}.} \bibinfo{year}{2024}\natexlab{}.
\newblock \bibinfo{title}{Azure Global Infrastructure}.
\newblock
\newblock
\shownote{\url{https://datacenters.microsoft.com/globe/explore/}, Last accessed on 2024-11-30}.


\bibitem[Moraru et~al\mbox{.}(2013)]%
        {epaxos}
\bibfield{author}{\bibinfo{person}{Iulian Moraru}, \bibinfo{person}{David~G. Andersen}, {and} \bibinfo{person}{Michael Kaminsky}.} \bibinfo{year}{2013}\natexlab{}.
\newblock \showarticletitle{There is More Consensus in Egalitarian Parliaments}. In \bibinfo{booktitle}{\emph{Proceedings of the 24th ACM Symposium on Operating Systems Principles}} (Farminton, Pennsylvania) \emph{(\bibinfo{series}{SOSP '13})}. \bibinfo{publisher}{Association for Computing Machinery}, \bibinfo{address}{New York, NY, USA}, \bibinfo{pages}{358–372}.
\newblock
\showISBNx{9781450323888}
\href{https://doi.org/10.1145/2517349.2517350}{doi:\nolinkurl{10.1145/2517349.2517350}}


\bibitem[Moraru et~al\mbox{.}(2014a)]%
        {quorum-leases}
\bibfield{author}{\bibinfo{person}{Iulian Moraru}, \bibinfo{person}{David~G. Andersen}, {and} \bibinfo{person}{Michael Kaminsky}.} \bibinfo{year}{2014}\natexlab{a}.
\newblock \showarticletitle{Paxos Quorum Leases: Fast Reads Without Sacrificing Writes}. In \bibinfo{booktitle}{\emph{Proceedings of the ACM Symposium on Cloud Computing}} (Seattle, WA, USA) \emph{(\bibinfo{series}{SOCC '14})}. \bibinfo{publisher}{Association for Computing Machinery}, \bibinfo{address}{New York, NY, USA}, \bibinfo{pages}{1–13}.
\newblock
\showISBNx{9781450332521}
\href{https://doi.org/10.1145/2670979.2671001}{doi:\nolinkurl{10.1145/2670979.2671001}}


\bibitem[Moraru et~al\mbox{.}(2014b)]%
        {quorum-leases-report}
\bibfield{author}{\bibinfo{person}{Iulian Moraru}, \bibinfo{person}{David~G. Andersen}, {and} \bibinfo{person}{Michael Kaminsky}.} \bibinfo{year}{2014}\natexlab{b}.
\newblock \showarticletitle{Paxos Quorum Leases: Fast Reads Without Sacrificing Writes}.
\newblock \bibinfo{journal}{\emph{Carnegie Mellon University PDL Technical Report}} (\bibinfo{year}{2014}).
\newblock


\bibitem[Mostefaoui et~al\mbox{.}(2024)]%
        {randomized-consensus-coins}
\bibfield{author}{\bibinfo{person}{Achour Mostefaoui}, \bibinfo{person}{Matthieu Perrin}, {and} \bibinfo{person}{Julien Weibel}.} \bibinfo{year}{2024}\natexlab{}.
\newblock \showarticletitle{Brief Announcement: Randomized Consensus: Common Coins Are not the Holy Grail!}. In \bibinfo{booktitle}{\emph{Proceedings of the 43rd ACM Symposium on Principles of Distributed Computing}} (Nantes, France) \emph{(\bibinfo{series}{PODC '24})}. \bibinfo{publisher}{Association for Computing Machinery}, \bibinfo{address}{New York, NY, USA}, \bibinfo{pages}{36–39}.
\newblock
\showISBNx{9798400706684}
\href{https://doi.org/10.1145/3662158.3662824}{doi:\nolinkurl{10.1145/3662158.3662824}}


\bibitem[Murat et~al\mbox{.}(2024)]%
        {swarm-disaggregated-memory}
\bibfield{author}{\bibinfo{person}{Antoine Murat}, \bibinfo{person}{Clément Burgelin}, \bibinfo{person}{Athanasios Xygkis}, \bibinfo{person}{Igor Zablotchi}, \bibinfo{person}{Marcos~Kawazoe Aguilera}, {and} \bibinfo{person}{Rachid Guerraoui}.} \bibinfo{year}{2024}\natexlab{}.
\newblock \showarticletitle{SWARM: Replicating Shared Disaggregated-Memory Data in No Time}. In \bibinfo{booktitle}{\emph{Proceedings of the ACM SIGOPS 30th Symposium on Operating Systems Principles}} \emph{(\bibinfo{series}{SOSP ’24})}. \bibinfo{publisher}{ACM}, \bibinfo{pages}{24–45}.
\newblock
\href{https://doi.org/10.1145/3694715.3695945}{doi:\nolinkurl{10.1145/3694715.3695945}}


\bibitem[Nawab et~al\mbox{.}(2018)]%
        {dpaxos}
\bibfield{author}{\bibinfo{person}{Faisal Nawab}, \bibinfo{person}{Divyakant Agrawal}, {and} \bibinfo{person}{Amr El~Abbadi}.} \bibinfo{year}{2018}\natexlab{}.
\newblock \showarticletitle{DPaxos: Managing Data Closer to Users for Low-Latency and Mobile Applications}. In \bibinfo{booktitle}{\emph{Proceedings of the 2018 International Conference on Management of Data}} (Houston, TX, USA) \emph{(\bibinfo{series}{SIGMOD '18})}. \bibinfo{publisher}{Association for Computing Machinery}, \bibinfo{address}{New York, NY, USA}, \bibinfo{pages}{1221–1236}.
\newblock
\showISBNx{9781450347037}
\href{https://doi.org/10.1145/3183713.3196928}{doi:\nolinkurl{10.1145/3183713.3196928}}


\bibitem[Ngo et~al\mbox{.}(2020)]%
        {copilots}
\bibfield{author}{\bibinfo{person}{Khiem Ngo}, \bibinfo{person}{Siddhartha Sen}, {and} \bibinfo{person}{Wyatt Lloyd}.} \bibinfo{year}{2020}\natexlab{}.
\newblock \showarticletitle{Tolerating Slowdowns in Replicated State Machines using Copilots}. In \bibinfo{booktitle}{\emph{14th USENIX Symposium on Operating Systems Design and Implementation (OSDI 20)}}. \bibinfo{publisher}{USENIX Association}, \bibinfo{pages}{583--598}.
\newblock
\showISBNx{978-1-939133-19-9}
\urldef\tempurl%
\url{https://www.usenix.org/conference/osdi20/presentation/ngo}
\showURL{%
\tempurl}


\bibitem[Oki and Liskov(1988)]%
        {viewstamped-replication}
\bibfield{author}{\bibinfo{person}{Brian~M. Oki} {and} \bibinfo{person}{Barbara~H. Liskov}.} \bibinfo{year}{1988}\natexlab{}.
\newblock \showarticletitle{Viewstamped Replication: A New Primary Copy Method to Support Highly-Available Distributed Systems}. In \bibinfo{booktitle}{\emph{Proceedings of the Seventh Annual ACM Symposium on Principles of Distributed Computing}} (Toronto, Ontario, Canada) \emph{(\bibinfo{series}{PODC '88})}. \bibinfo{publisher}{Association for Computing Machinery}, \bibinfo{address}{New York, NY, USA}, \bibinfo{pages}{8–17}.
\newblock
\showISBNx{0897912772}
\href{https://doi.org/10.1145/62546.62549}{doi:\nolinkurl{10.1145/62546.62549}}


\bibitem[Ongaro(2014)]%
        {raft-thesis}
\bibfield{author}{\bibinfo{person}{Diego Ongaro}.} \bibinfo{year}{2014}\natexlab{}.
\newblock \emph{\bibinfo{title}{Consensus: Bridging Theory and Practice}}.
\newblock \bibinfo{thesistype}{Ph.\,D. Dissertation}. \bibinfo{address}{Stanford, CA, USA}.
\newblock Advisor(s) K., Ousterhout, John and David, Mazi\`{e}res, and Mendel, Rosenblum,.
\newblock
\showISBNx{9798662514218}
\newblock
\shownote{AAI28121474}.


\bibitem[Ongaro and Ousterhout(2014)]%
        {raft}
\bibfield{author}{\bibinfo{person}{Diego Ongaro} {and} \bibinfo{person}{John Ousterhout}.} \bibinfo{year}{2014}\natexlab{}.
\newblock \showarticletitle{In Search of an Understandable Consensus Algorithm}. In \bibinfo{booktitle}{\emph{Proceedings of the 2014 USENIX Conference on USENIX Annual Technical Conference}} (Philadelphia, PA) \emph{(\bibinfo{series}{USENIX ATC'14})}. \bibinfo{publisher}{USENIX Association}, \bibinfo{address}{USA}, \bibinfo{pages}{305–320}.
\newblock
\showISBNx{9781931971102}


\bibitem[Optics(2021)]%
        {live-optics-rw-ratio}
\bibfield{author}{\bibinfo{person}{Team~Live Optics}.} \bibinfo{year}{2021}\natexlab{}.
\newblock \bibinfo{title}{Live Optics Basics: Read / Write Ratio}.
\newblock \bibinfo{howpublished}{\url{https://support.liveoptics.com/hc/en-us/articles/229590547-Live-Optics-Basics-Read-Write-Ratio}}.
\newblock
\newblock
\shownote{Accessed: 2024-12-01}.


\bibitem[Pan et~al\mbox{.}(2021)]%
        {rabia}
\bibfield{author}{\bibinfo{person}{Haochen Pan}, \bibinfo{person}{Jesse Tuglu}, \bibinfo{person}{Neo Zhou}, \bibinfo{person}{Tianshu Wang}, \bibinfo{person}{Yicheng Shen}, \bibinfo{person}{Xiong Zheng}, \bibinfo{person}{Joseph Tassarotti}, \bibinfo{person}{Lewis Tseng}, {and} \bibinfo{person}{Roberto Palmieri}.} \bibinfo{year}{2021}\natexlab{}.
\newblock \showarticletitle{Rabia: Simplifying State-Machine Replication Through Randomization}. In \bibinfo{booktitle}{\emph{Proceedings of the ACM SIGOPS 28th Symposium on Operating Systems Principles}} (Virtual Event, Germany) \emph{(\bibinfo{series}{SOSP '21})}. \bibinfo{publisher}{Association for Computing Machinery}, \bibinfo{address}{New York, NY, USA}, \bibinfo{pages}{472–487}.
\newblock
\showISBNx{9781450387095}
\href{https://doi.org/10.1145/3477132.3483582}{doi:\nolinkurl{10.1145/3477132.3483582}}


\bibitem[Park and Ousterhout(2019)]%
        {curp-commutativity}
\bibfield{author}{\bibinfo{person}{Seo~Jin Park} {and} \bibinfo{person}{John Ousterhout}.} \bibinfo{year}{2019}\natexlab{}.
\newblock \showarticletitle{Exploiting Commutativity For Practical Fast Replication}. In \bibinfo{booktitle}{\emph{16th USENIX Symposium on Networked Systems Design and Implementation (NSDI 19)}}. \bibinfo{publisher}{USENIX Association}, \bibinfo{address}{Boston, MA}, \bibinfo{pages}{47--64}.
\newblock
\showISBNx{978-1-931971-49-2}
\urldef\tempurl%
\url{https://www.usenix.org/conference/nsdi19/presentation/park}
\showURL{%
\tempurl}


\bibitem[Pasuparthy and Agarwal(2023)]%
        {benchmarking-spanner}
\bibfield{author}{\bibinfo{person}{Suraj Pasuparthy} {and} \bibinfo{person}{Lokesh Agarwal}.} \bibinfo{year}{2023}\natexlab{}.
\newblock \bibinfo{title}{Benchmarking Spanner’s price-performance for key-value workloads}.
\newblock \bibinfo{howpublished}{\url{https://cloud.google.com/blog/products/databases/benchmarking-spanner-for-key-value-workloads/}}.
\newblock
\newblock
\shownote{Accessed: 2024-12-01}.


\bibitem[Petersen et~al\mbox{.}(1997)]%
        {bayou}
\bibfield{author}{\bibinfo{person}{Karin Petersen}, \bibinfo{person}{Mike~J. Spreitzer}, \bibinfo{person}{Douglas~B. Terry}, \bibinfo{person}{Marvin~M. Theimer}, {and} \bibinfo{person}{Alan~J. Demers}.} \bibinfo{year}{1997}\natexlab{}.
\newblock \showarticletitle{Flexible update propagation for weakly consistent replication}. In \bibinfo{booktitle}{\emph{Proceedings of the Sixteenth ACM Symposium on Operating Systems Principles}} (Saint Malo, France) \emph{(\bibinfo{series}{SOSP '97})}. \bibinfo{publisher}{Association for Computing Machinery}, \bibinfo{address}{New York, NY, USA}, \bibinfo{pages}{288–301}.
\newblock
\showISBNx{0897919165}
\href{https://doi.org/10.1145/268998.266711}{doi:\nolinkurl{10.1145/268998.266711}}


\bibitem[Ports et~al\mbox{.}(2015)]%
        {speculative-paxos}
\bibfield{author}{\bibinfo{person}{Dan R.~K. Ports}, \bibinfo{person}{Jialin Li}, \bibinfo{person}{Vincent Liu}, \bibinfo{person}{Naveen~Kr. Sharma}, {and} \bibinfo{person}{Arvind Krishnamurthy}.} \bibinfo{year}{2015}\natexlab{}.
\newblock \showarticletitle{Designing Distributed Systems Using Approximate Synchrony in Data Center Networks}. In \bibinfo{booktitle}{\emph{12th USENIX Symposium on Networked Systems Design and Implementation (NSDI 15)}}. \bibinfo{publisher}{USENIX Association}, \bibinfo{address}{Oakland, CA}, \bibinfo{pages}{43--57}.
\newblock
\showISBNx{978-1-931971-218}
\urldef\tempurl%
\url{https://www.usenix.org/conference/nsdi15/technical-sessions/presentation/ports}
\showURL{%
\tempurl}


\bibitem[Qi et~al\mbox{.}(2021)]%
        {bidl-blockchain}
\bibfield{author}{\bibinfo{person}{Ji Qi}, \bibinfo{person}{Xusheng Chen}, \bibinfo{person}{Yunpeng Jiang}, \bibinfo{person}{Jianyu Jiang}, \bibinfo{person}{Tianxiang Shen}, \bibinfo{person}{Shixiong Zhao}, \bibinfo{person}{Sen Wang}, \bibinfo{person}{Gong Zhang}, \bibinfo{person}{Li Chen}, \bibinfo{person}{Man~Ho Au}, {and} \bibinfo{person}{Heming Cui}.} \bibinfo{year}{2021}\natexlab{}.
\newblock \showarticletitle{Bidl: A High-throughput, Low-latency Permissioned Blockchain Framework for Datacenter Networks}. In \bibinfo{booktitle}{\emph{Proceedings of the ACM SIGOPS 28th Symposium on Operating Systems Principles}} (Virtual Event, Germany) \emph{(\bibinfo{series}{SOSP '21})}. \bibinfo{publisher}{Association for Computing Machinery}, \bibinfo{address}{New York, NY, USA}, \bibinfo{pages}{18–34}.
\newblock
\showISBNx{9781450387095}
\href{https://doi.org/10.1145/3477132.3483574}{doi:\nolinkurl{10.1145/3477132.3483574}}


\bibitem[RabbitMQ(2025)]%
        {rabbitmq}
\bibfield{author}{\bibinfo{person}{RabbitMQ}.} \bibinfo{year}{2025}\natexlab{}.
\newblock \bibinfo{title}{RabbitMQ: One broker to queue them all}.
\newblock
\newblock
\shownote{\url{https://www.rabbitmq.com/}, Last accessed on 2025-04-08}.


\bibitem[Redpanda(2024)]%
        {redpanda}
\bibfield{author}{\bibinfo{person}{Redpanda}.} \bibinfo{year}{2024}\natexlab{}.
\newblock \bibinfo{title}{Redpanda: The Unified Streaming Data Platform}.
\newblock
\newblock
\shownote{\url{https://www.redpanda.com/}, Last accessed on 2024-09-05}.


\bibitem[Renesse and Schneider(2004)]%
        {chain-replication}
\bibfield{author}{\bibinfo{person}{Robbert~Van Renesse} {and} \bibinfo{person}{Fred~B. Schneider}.} \bibinfo{year}{2004}\natexlab{}.
\newblock \showarticletitle{Chain Replication for Supporting High Throughput and Availability}. In \bibinfo{booktitle}{\emph{6th Symposium on Operating Systems Design \& Implementation (OSDI 04)}}. \bibinfo{publisher}{USENIX Association}, \bibinfo{address}{CA}.
\newblock
\urldef\tempurl%
\url{https://www.usenix.org/conference/osdi-04/chain-replication-supporting-high-throughput-and-availability}
\showURL{%
\tempurl}


\bibitem[Rensin(2015)]%
        {kubernetes}
\bibfield{author}{\bibinfo{person}{David~K. Rensin}.} \bibinfo{year}{2015}\natexlab{}.
\newblock \bibinfo{booktitle}{\emph{Kubernetes - Scheduling the Future at Cloud Scale}}.
\newblock \bibinfo{publisher}{O'Reilly and Associates}, \bibinfo{address}{1005 Gravenstein Highway North Sebastopol, CA 95472}. All pages.
\newblock
\urldef\tempurl%
\url{http://www.oreilly.com/webops-perf/free/kubernetes.csp}
\showURL{%
\tempurl}


\bibitem[Rizvi et~al\mbox{.}(2017)]%
        {canopus}
\bibfield{author}{\bibinfo{person}{Sajjad Rizvi}, \bibinfo{person}{Bernard Wong}, {and} \bibinfo{person}{Srinivasan Keshav}.} \bibinfo{year}{2017}\natexlab{}.
\newblock \showarticletitle{Canopus: A Scalable and Massively Parallel Consensus Protocol}. In \bibinfo{booktitle}{\emph{Proceedings of the 13th International Conference on Emerging Networking EXperiments and Technologies}} (Incheon, Republic of Korea) \emph{(\bibinfo{series}{CoNEXT '17})}. \bibinfo{publisher}{Association for Computing Machinery}, \bibinfo{address}{New York, NY, USA}, \bibinfo{pages}{426–438}.
\newblock
\showISBNx{9781450354226}
\href{https://doi.org/10.1145/3143361.3143394}{doi:\nolinkurl{10.1145/3143361.3143394}}


\bibitem[Ryabinin et~al\mbox{.}(2024)]%
        {swiftpaxos}
\bibfield{author}{\bibinfo{person}{Fedor Ryabinin}, \bibinfo{person}{Alexey Gotsman}, {and} \bibinfo{person}{Pierre Sutra}.} \bibinfo{year}{2024}\natexlab{}.
\newblock \showarticletitle{{SwiftPaxos}: Fast {Geo-Replicated} State Machines}. In \bibinfo{booktitle}{\emph{21st USENIX Symposium on Networked Systems Design and Implementation (NSDI 24)}}. \bibinfo{publisher}{USENIX Association}, \bibinfo{address}{Santa Clara, CA}, \bibinfo{pages}{345--369}.
\newblock
\showISBNx{978-1-939133-39-7}
\urldef\tempurl%
\url{https://www.usenix.org/conference/nsdi24/presentation/ryabinin}
\showURL{%
\tempurl}


\bibitem[Schneider(1990)]%
        {smr-approach}
\bibfield{author}{\bibinfo{person}{Fred~B. Schneider}.} \bibinfo{year}{1990}\natexlab{}.
\newblock \showarticletitle{Implementing Fault-Tolerant Services Using the State Machine Approach: A Tutorial}.
\newblock \bibinfo{journal}{\emph{ACM Comput. Surv.}} \bibinfo{volume}{22}, \bibinfo{number}{4} (\bibinfo{date}{dec} \bibinfo{year}{1990}), \bibinfo{pages}{299–319}.
\newblock
\showISSN{0360-0300}
\href{https://doi.org/10.1145/98163.98167}{doi:\nolinkurl{10.1145/98163.98167}}


\bibitem[ScyllaDB(2023)]%
        {scylladb}
\bibfield{author}{\bibinfo{person}{ScyllaDB}.} \bibinfo{year}{2023}\natexlab{}.
\newblock \bibinfo{title}{Beyond Legacy NoSQL: 7 Design Principles Behind ScyllaDB}.
\newblock
\newblock
\shownote{\url{https://lp.scylladb.com/real-time-big-data-database-principles-thanks.html}, Last accessed on 2023-11-13}.


\bibitem[Stathakopoulou et~al\mbox{.}(2022)]%
        {insanely-scalable-smr}
\bibfield{author}{\bibinfo{person}{Chrysoula Stathakopoulou}, \bibinfo{person}{Matej Pavlovic}, {and} \bibinfo{person}{Marko Vukoli\'{c}}.} \bibinfo{year}{2022}\natexlab{}.
\newblock \showarticletitle{State Machine Replication Scalability Made Simple}. In \bibinfo{booktitle}{\emph{Proceedings of the Seventeenth European Conference on Computer Systems}} (Rennes, France) \emph{(\bibinfo{series}{EuroSys '22})}. \bibinfo{publisher}{Association for Computing Machinery}, \bibinfo{address}{New York, NY, USA}, \bibinfo{pages}{17–33}.
\newblock
\showISBNx{9781450391627}
\href{https://doi.org/10.1145/3492321.3519579}{doi:\nolinkurl{10.1145/3492321.3519579}}


\bibitem[Sun et~al\mbox{.}(2024)]%
        {anvil}
\bibfield{author}{\bibinfo{person}{Xudong Sun}, \bibinfo{person}{Wenjie Ma}, \bibinfo{person}{Jiawei~Tyler Gu}, \bibinfo{person}{Zicheng Ma}, \bibinfo{person}{Tej Chajed}, \bibinfo{person}{Jon Howell}, \bibinfo{person}{Andrea Lattuada}, \bibinfo{person}{Oded Padon}, \bibinfo{person}{Lalith Suresh}, \bibinfo{person}{Adriana Szekeres}, {and} \bibinfo{person}{Tianyin Xu}.} \bibinfo{year}{2024}\natexlab{}.
\newblock \showarticletitle{Anvil: Verifying Liveness of Cluster Management Controllers}. In \bibinfo{booktitle}{\emph{18th USENIX Symposium on Operating Systems Design and Implementation (OSDI 24)}}. \bibinfo{publisher}{USENIX Association}, \bibinfo{address}{Santa Clara, CA}, \bibinfo{pages}{649--666}.
\newblock
\showISBNx{978-1-939133-40-3}
\urldef\tempurl%
\url{https://www.usenix.org/conference/osdi24/presentation/sun-xudong}
\showURL{%
\tempurl}


\bibitem[Suri-Payer et~al\mbox{.}(2021)]%
        {basil}
\bibfield{author}{\bibinfo{person}{Florian Suri-Payer}, \bibinfo{person}{Matthew Burke}, \bibinfo{person}{Zheng Wang}, \bibinfo{person}{Yunhao Zhang}, \bibinfo{person}{Lorenzo Alvisi}, {and} \bibinfo{person}{Natacha Crooks}.} \bibinfo{year}{2021}\natexlab{}.
\newblock \showarticletitle{Basil: Breaking up BFT with ACID (Transactions)}. In \bibinfo{booktitle}{\emph{Proceedings of the ACM SIGOPS 28th Symposium on Operating Systems Principles}} (Virtual Event, Germany) \emph{(\bibinfo{series}{SOSP '21})}. \bibinfo{publisher}{Association for Computing Machinery}, \bibinfo{address}{New York, NY, USA}, \bibinfo{pages}{1–17}.
\newblock
\showISBNx{9781450387095}
\href{https://doi.org/10.1145/3477132.3483552}{doi:\nolinkurl{10.1145/3477132.3483552}}


\bibitem[Takruri et~al\mbox{.}(2020)]%
        {flair}
\bibfield{author}{\bibinfo{person}{Hatem Takruri}, \bibinfo{person}{Ibrahim Kettaneh}, \bibinfo{person}{Ahmed Alquraan}, {and} \bibinfo{person}{Samer Al-Kiswany}.} \bibinfo{year}{2020}\natexlab{}.
\newblock \showarticletitle{{FLAIR}: Accelerating Reads with {Consistency-Aware} Network Routing}. In \bibinfo{booktitle}{\emph{17th USENIX Symposium on Networked Systems Design \& Implementation (NSDI 20)}}. \bibinfo{publisher}{USENIX Association}, \bibinfo{address}{CA}, \bibinfo{pages}{723--737}.
\newblock
\showISBNx{978-1-939133-13-7}
\urldef\tempurl%
\url{https://usenix.org/conference/nsdi20/presentation/takruri}
\showURL{%
\tempurl}


\bibitem[Terrace and Freedman(2009)]%
        {craq}
\bibfield{author}{\bibinfo{person}{Jeff Terrace} {and} \bibinfo{person}{Michael~J. Freedman}.} \bibinfo{year}{2009}\natexlab{}.
\newblock \showarticletitle{Object Storage on {CRAQ}: {High-Throughput} Chain Replication for {Read-Mostly} Workloads}. In \bibinfo{booktitle}{\emph{2009 USENIX Annual Technical Conference (USENIX ATC 09)}}. \bibinfo{publisher}{USENIX Association}, \bibinfo{address}{San Diego, CA}.
\newblock
\urldef\tempurl%
\url{https://www.usenix.org/conference/usenix-09/object-storage-craq-high-throughput-chain-replication-read-mostly-workloads}
\showURL{%
\tempurl}


\bibitem[Terry et~al\mbox{.}(1994)]%
        {session-guarantees}
\bibfield{author}{\bibinfo{person}{Douglas~B. Terry}, \bibinfo{person}{Alan~J. Demers}, \bibinfo{person}{Karin Petersen}, \bibinfo{person}{Mike~J. Spreitzer}, \bibinfo{person}{Marvin~M. Theimer}, {and} \bibinfo{person}{Brent~B. Welch}.} \bibinfo{year}{1994}\natexlab{}.
\newblock \showarticletitle{Session Guarantees for Weakly Consistent Replicated Data}. In \bibinfo{booktitle}{\emph{Proceedings of the Third International Conference on on Parallel and Distributed Information Systems}} (Autin, Texas, USA) \emph{(\bibinfo{series}{PDIS '94})}. \bibinfo{publisher}{IEEE Computer Society Press}, \bibinfo{address}{Washington, DC, USA}, \bibinfo{pages}{140–150}.
\newblock
\showISBNx{0818664010}


\bibitem[Thiessen et~al\mbox{.}(2024)]%
        {reconfigurable-linearizable-reads}
\bibfield{author}{\bibinfo{person}{Myles Thiessen}, \bibinfo{person}{Aleksey Panas}, \bibinfo{person}{Guy Khazma}, {and} \bibinfo{person}{Eyal de Lara}.} \bibinfo{year}{2024}\natexlab{}.
\newblock \bibinfo{title}{Towards Reconfigurable Linearizable Reads}.
\newblock
\showeprint[arxiv]{2404.05470}~[cs.DC]
\urldef\tempurl%
\url{https://arxiv.org/abs/2404.05470}
\showURL{%
\tempurl}


\bibitem[TigerBeetle(2024)]%
        {tigerbeetle}
\bibfield{author}{\bibinfo{person}{TigerBeetle}.} \bibinfo{year}{2024}\natexlab{}.
\newblock \bibinfo{title}{TigerBeetle: The Financial Transactions Database}.
\newblock
\newblock
\shownote{\url{https://tigerbeetle.com/}, Last accessed on 2024-11-12}.


\bibitem[Tollman et~al\mbox{.}(2021)]%
        {epaxos-revisited}
\bibfield{author}{\bibinfo{person}{Sarah Tollman}, \bibinfo{person}{Seo~Jin Park}, {and} \bibinfo{person}{John Ousterhout}.} \bibinfo{year}{2021}\natexlab{}.
\newblock \showarticletitle{{EPaxos} Revisited}. In \bibinfo{booktitle}{\emph{18th USENIX Symposium on Networked Systems Design and Implementation (NSDI 21)}}. \bibinfo{publisher}{USENIX Association}, \bibinfo{pages}{613--632}.
\newblock
\showISBNx{978-1-939133-21-2}
\urldef\tempurl%
\url{https://www.usenix.org/conference/nsdi21/presentation/tollman}
\showURL{%
\tempurl}


\bibitem[Trach et~al\mbox{.}(2020)]%
        {t-lease}
\bibfield{author}{\bibinfo{person}{Bohdan Trach}, \bibinfo{person}{Rasha Faqeh}, \bibinfo{person}{Oleksii Oleksenko}, \bibinfo{person}{Wojciech Ozga}, \bibinfo{person}{Pramod Bhatotia}, {and} \bibinfo{person}{Christof Fetzer}.} \bibinfo{year}{2020}\natexlab{}.
\newblock \showarticletitle{T-Lease: a trusted lease primitive for distributed systems}. In \bibinfo{booktitle}{\emph{Proceedings of the 11th ACM Symposium on Cloud Computing}} (Virtual Event, USA) \emph{(\bibinfo{series}{SoCC '20})}. \bibinfo{publisher}{Association for Computing Machinery}, \bibinfo{address}{New York, NY, USA}, \bibinfo{pages}{387–400}.
\newblock
\showISBNx{9781450381376}
\href{https://doi.org/10.1145/3419111.3421273}{doi:\nolinkurl{10.1145/3419111.3421273}}


\bibitem[Uluyol et~al\mbox{.}(2020)]%
        {tradeoffs-geo-distributed}
\bibfield{author}{\bibinfo{person}{Muhammed Uluyol}, \bibinfo{person}{Anthony Huang}, \bibinfo{person}{Ayush Goel}, \bibinfo{person}{Mosharaf Chowdhury}, {and} \bibinfo{person}{Harsha~V. Madhyastha}.} \bibinfo{year}{2020}\natexlab{}.
\newblock \showarticletitle{{Near-Optimal} Latency Versus Cost Tradeoffs in {Geo-Distributed} Storage}. In \bibinfo{booktitle}{\emph{17th USENIX Symposium on Networked Systems Design and Implementation (NSDI 20)}}. \bibinfo{publisher}{USENIX Association}, \bibinfo{address}{Santa Clara, CA}, \bibinfo{pages}{157--180}.
\newblock
\showISBNx{978-1-939133-13-7}
\urldef\tempurl%
\url{https://www.usenix.org/conference/nsdi20/presentation/uluyol}
\showURL{%
\tempurl}


\bibitem[VanBenschoten et~al\mbox{.}(2022)]%
        {cockroachdb}
\bibfield{author}{\bibinfo{person}{Nathan VanBenschoten}, \bibinfo{person}{Arul Ajmani}, \bibinfo{person}{Marcus Gartner}, \bibinfo{person}{Andrei Matei}, \bibinfo{person}{Aayush Shah}, \bibinfo{person}{Irfan Sharif}, \bibinfo{person}{Alexander Shraer}, \bibinfo{person}{Adam Storm}, \bibinfo{person}{Rebecca Taft}, \bibinfo{person}{Oliver Tan}, \bibinfo{person}{Andy Woods}, {and} \bibinfo{person}{Peyton Walters}.} \bibinfo{year}{2022}\natexlab{}.
\newblock \showarticletitle{Enabling the Next Generation of Multi-Region Applications with CockroachDB}. In \bibinfo{booktitle}{\emph{Proceedings of the 2022 International Conference on Management of Data}} (Philadelphia, PA, USA) \emph{(\bibinfo{series}{SIGMOD '22})}. \bibinfo{publisher}{Association for Computing Machinery}, \bibinfo{address}{New York, NY, USA}, \bibinfo{pages}{2312–2325}.
\newblock
\showISBNx{9781450392495}
\href{https://doi.org/10.1145/3514221.3526053}{doi:\nolinkurl{10.1145/3514221.3526053}}


\bibitem[Veeraraghavan et~al\mbox{.}(2018)]%
        {maelstrom}
\bibfield{author}{\bibinfo{person}{Kaushik Veeraraghavan}, \bibinfo{person}{Justin Meza}, \bibinfo{person}{Scott Michelson}, \bibinfo{person}{Sankaralingam Panneerselvam}, \bibinfo{person}{Alex Gyori}, \bibinfo{person}{David Chou}, \bibinfo{person}{Sonia Margulis}, \bibinfo{person}{Daniel Obenshain}, \bibinfo{person}{Shruti Padmanabha}, \bibinfo{person}{Ashish Shah}, \bibinfo{person}{Yee~Jiun Song}, {and} \bibinfo{person}{Tianyin Xu}.} \bibinfo{year}{2018}\natexlab{}.
\newblock \showarticletitle{Maelstrom: Mitigating Datacenter-level Disasters by Draining Interdependent Traffic Safely and Efficiently}. In \bibinfo{booktitle}{\emph{13th USENIX Symposium on Operating Systems Design and Implementation (OSDI 18)}}. \bibinfo{publisher}{USENIX Association}, \bibinfo{address}{Carlsbad, CA}, \bibinfo{pages}{373--389}.
\newblock
\showISBNx{978-1-939133-08-3}
\urldef\tempurl%
\url{https://www.usenix.org/conference/osdi18/presentation/veeraraghavan}
\showURL{%
\tempurl}


\bibitem[Vogels(2008)]%
        {eventual-consistency}
\bibfield{author}{\bibinfo{person}{Werner Vogels}.} \bibinfo{year}{2008}\natexlab{}.
\newblock \showarticletitle{Eventually Consistent: Building Reliable Distributed Systems at a Worldwide Scale Demands Trade-Offs Between Consistency and Availability.}
\newblock \bibinfo{journal}{\emph{Queue}} \bibinfo{volume}{6}, \bibinfo{number}{6} (\bibinfo{date}{oct} \bibinfo{year}{2008}), \bibinfo{pages}{14–19}.
\newblock
\showISSN{1542-7730}
\href{https://doi.org/10.1145/1466443.1466448}{doi:\nolinkurl{10.1145/1466443.1466448}}


\bibitem[Wang et~al\mbox{.}(2017)]%
        {apus-rdma-paxos}
\bibfield{author}{\bibinfo{person}{Cheng Wang}, \bibinfo{person}{Jianyu Jiang}, \bibinfo{person}{Xusheng Chen}, \bibinfo{person}{Ning Yi}, {and} \bibinfo{person}{Heming Cui}.} \bibinfo{year}{2017}\natexlab{}.
\newblock \showarticletitle{APUS: fast and scalable paxos on RDMA}. In \bibinfo{booktitle}{\emph{Proceedings of the 2017 Symposium on Cloud Computing}} (Santa Clara, California) \emph{(\bibinfo{series}{SoCC '17})}. \bibinfo{publisher}{Association for Computing Machinery}, \bibinfo{address}{New York, NY, USA}, \bibinfo{pages}{94–107}.
\newblock
\showISBNx{9781450350280}
\href{https://doi.org/10.1145/3127479.3128609}{doi:\nolinkurl{10.1145/3127479.3128609}}


\bibitem[Wang et~al\mbox{.}(2019)]%
        {parallel-paxos-raft}
\bibfield{author}{\bibinfo{person}{Zhaoguo Wang}, \bibinfo{person}{Changgeng Zhao}, \bibinfo{person}{Shuai Mu}, \bibinfo{person}{Haibo Chen}, {and} \bibinfo{person}{Jinyang Li}.} \bibinfo{year}{2019}\natexlab{}.
\newblock \showarticletitle{On the Parallels between Paxos and Raft, and how to Port Optimizations}. In \bibinfo{booktitle}{\emph{Proceedings of the 2019 ACM Symposium on Principles of Distributed Computing}} (Toronto ON, Canada) \emph{(\bibinfo{series}{PODC '19})}. \bibinfo{publisher}{Association for Computing Machinery}, \bibinfo{address}{New York, NY, USA}, \bibinfo{pages}{445–454}.
\newblock
\showISBNx{9781450362177}
\href{https://doi.org/10.1145/3293611.3331595}{doi:\nolinkurl{10.1145/3293611.3331595}}


\bibitem[Wei et~al\mbox{.}(2023)]%
        {off-path-smartnic}
\bibfield{author}{\bibinfo{person}{Xingda Wei}, \bibinfo{person}{Rongxin Cheng}, \bibinfo{person}{Yuhan Yang}, \bibinfo{person}{Rong Chen}, {and} \bibinfo{person}{Haibo Chen}.} \bibinfo{year}{2023}\natexlab{}.
\newblock \showarticletitle{Characterizing Off-path {SmartNIC} for Accelerating Distributed Systems}. In \bibinfo{booktitle}{\emph{17th USENIX Symposium on Operating Systems Design and Implementation (OSDI 23)}}. \bibinfo{publisher}{USENIX Association}, \bibinfo{address}{Boston, MA}, \bibinfo{pages}{987--1004}.
\newblock
\showISBNx{978-1-939133-34-2}
\urldef\tempurl%
\url{https://www.usenix.org/conference/osdi23/presentation/wei-smartnic}
\showURL{%
\tempurl}


\bibitem[Whittaker et~al\mbox{.}(2021a)]%
        {compartmentalization}
\bibfield{author}{\bibinfo{person}{Michael Whittaker}, \bibinfo{person}{Ailidani Ailijiang}, \bibinfo{person}{Aleksey Charapko}, \bibinfo{person}{Murat Demirbas}, \bibinfo{person}{Neil Giridharan}, \bibinfo{person}{Joseph~M. Hellerstein}, \bibinfo{person}{Heidi Howard}, \bibinfo{person}{Ion Stoica}, {and} \bibinfo{person}{Adriana Szekeres}.} \bibinfo{year}{2021}\natexlab{a}.
\newblock \showarticletitle{Scaling Replicated State Machines with Compartmentalization}.
\newblock \bibinfo{journal}{\emph{Proc. VLDB Endow.}} \bibinfo{volume}{14}, \bibinfo{number}{11} (\bibinfo{date}{jul} \bibinfo{year}{2021}), \bibinfo{pages}{2203–2215}.
\newblock
\showISSN{2150-8097}
\href{https://doi.org/10.14778/3476249.3476273}{doi:\nolinkurl{10.14778/3476249.3476273}}


\bibitem[Whittaker et~al\mbox{.}(2021b)]%
        {quorums-made-practical}
\bibfield{author}{\bibinfo{person}{Michael Whittaker}, \bibinfo{person}{Aleksey Charapko}, \bibinfo{person}{Joseph~M. Hellerstein}, \bibinfo{person}{Heidi Howard}, {and} \bibinfo{person}{Ion Stoica}.} \bibinfo{year}{2021}\natexlab{b}.
\newblock \showarticletitle{Read-Write Quorum Systems Made Practical}. In \bibinfo{booktitle}{\emph{Proceedings of the 8th Workshop on Principles and Practice of Consistency for Distributed Data}} (Online, United Kingdom) \emph{(\bibinfo{series}{PaPoC '21})}. \bibinfo{publisher}{Association for Computing Machinery}, \bibinfo{address}{New York, NY, USA}, Article \bibinfo{articleno}{7}, \bibinfo{numpages}{8}~pages.
\newblock
\showISBNx{9781450383387}
\href{https://doi.org/10.1145/3447865.3457962}{doi:\nolinkurl{10.1145/3447865.3457962}}


\bibitem[Yi et~al\mbox{.}(2023)]%
        {glean-consensus}
\bibfield{author}{\bibinfo{person}{Jian Yi}, \bibinfo{person}{Qing Li}, \bibinfo{person}{Bin Zhang}, \bibinfo{person}{Yong Jiang}, \bibinfo{person}{Dan Zhao}, \bibinfo{person}{Yuan Yang}, {and} \bibinfo{person}{Zhenhui Yuan}.} \bibinfo{year}{2023}\natexlab{}.
\newblock \showarticletitle{Gleaning the Consensus for Linearizable and Conflict-Free Per-Replica Local Reads}. In \bibinfo{booktitle}{\emph{Proceedings of the 7th Asia-Pacific Workshop on Networking}} (Hong Kong, China) \emph{(\bibinfo{series}{APNet '23})}. \bibinfo{publisher}{Association for Computing Machinery}, \bibinfo{address}{New York, NY, USA}, \bibinfo{pages}{143–149}.
\newblock
\showISBNx{9798400707827}
\href{https://doi.org/10.1145/3600061.3603175}{doi:\nolinkurl{10.1145/3600061.3603175}}


\bibitem[Yin et~al\mbox{.}(2019)]%
        {hotstuff}
\bibfield{author}{\bibinfo{person}{Maofan Yin}, \bibinfo{person}{Dahlia Malkhi}, \bibinfo{person}{Michael~K. Reiter}, \bibinfo{person}{Guy~Golan Gueta}, {and} \bibinfo{person}{Ittai Abraham}.} \bibinfo{year}{2019}\natexlab{}.
\newblock \showarticletitle{HotStuff: BFT Consensus with Linearity and Responsiveness}. In \bibinfo{booktitle}{\emph{Proceedings of the 2019 ACM Symposium on Principles of Distributed Computing}} (Toronto ON, Canada) \emph{(\bibinfo{series}{PODC '19})}. \bibinfo{publisher}{Association for Computing Machinery}, \bibinfo{address}{New York, NY, USA}, \bibinfo{pages}{347–356}.
\newblock
\showISBNx{9781450362177}
\href{https://doi.org/10.1145/3293611.3331591}{doi:\nolinkurl{10.1145/3293611.3331591}}


\bibitem[Zhang et~al\mbox{.}(2024)]%
        {fisslock}
\bibfield{author}{\bibinfo{person}{Hanze Zhang}, \bibinfo{person}{Ke Cheng}, \bibinfo{person}{Rong Chen}, {and} \bibinfo{person}{Haibo Chen}.} \bibinfo{year}{2024}\natexlab{}.
\newblock \showarticletitle{Fast and Scalable In-network Lock Management Using Lock Fission}. In \bibinfo{booktitle}{\emph{18th USENIX Symposium on Operating Systems Design and Implementation (OSDI 24)}}. \bibinfo{publisher}{USENIX Association}, \bibinfo{address}{Santa Clara, CA}, \bibinfo{pages}{251--268}.
\newblock
\showISBNx{978-1-939133-40-3}
\urldef\tempurl%
\url{https://www.usenix.org/conference/osdi24/presentation/zhang-hanze}
\showURL{%
\tempurl}


\bibitem[Zhang et~al\mbox{.}(2020)]%
        {byzantine-ordered-consensus}
\bibfield{author}{\bibinfo{person}{Yunhao Zhang}, \bibinfo{person}{Srinath Setty}, \bibinfo{person}{Qi Chen}, \bibinfo{person}{Lidong Zhou}, {and} \bibinfo{person}{Lorenzo Alvisi}.} \bibinfo{year}{2020}\natexlab{}.
\newblock \showarticletitle{Byzantine Ordered Consensus without Byzantine Oligarchy}. In \bibinfo{booktitle}{\emph{14th USENIX Symposium on Operating Systems Design and Implementation (OSDI 20)}}. \bibinfo{publisher}{USENIX Association}, \bibinfo{pages}{633--649}.
\newblock
\showISBNx{978-1-939133-19-9}
\urldef\tempurl%
\url{https://www.usenix.org/conference/osdi20/presentation/zhang-yunhao}
\showURL{%
\tempurl}


\bibitem[Zhao et~al\mbox{.}(2018)]%
        {sdpaxos}
\bibfield{author}{\bibinfo{person}{Hanyu Zhao}, \bibinfo{person}{Quanlu Zhang}, \bibinfo{person}{Zhi Yang}, \bibinfo{person}{Ming Wu}, {and} \bibinfo{person}{Yafei Dai}.} \bibinfo{year}{2018}\natexlab{}.
\newblock \showarticletitle{SDPaxos: Building Efficient Semi-Decentralized Geo-replicated State Machines}. In \bibinfo{booktitle}{\emph{Proceedings of the ACM Symposium on Cloud Computing}} (Carlsbad, CA, USA) \emph{(\bibinfo{series}{SoCC '18})}. \bibinfo{publisher}{Association for Computing Machinery}, \bibinfo{address}{New York, NY, USA}, \bibinfo{pages}{68–81}.
\newblock
\showISBNx{9781450360111}
\href{https://doi.org/10.1145/3267809.3267837}{doi:\nolinkurl{10.1145/3267809.3267837}}


\bibitem[Zhou and Mu(2021)]%
        {pull-based-consensus-mongodb}
\bibfield{author}{\bibinfo{person}{Siyuan Zhou} {and} \bibinfo{person}{Shuai Mu}.} \bibinfo{year}{2021}\natexlab{}.
\newblock \showarticletitle{{Fault-Tolerant} Replication with {Pull-Based} Consensus in {MongoDB}}. In \bibinfo{booktitle}{\emph{18th USENIX Symposium on Networked Systems Design and Implementation (NSDI 21)}}. \bibinfo{publisher}{USENIX Association}, \bibinfo{pages}{687--703}.
\newblock
\showISBNx{978-1-939133-21-2}
\urldef\tempurl%
\url{https://www.usenix.org/conference/nsdi21/presentation/zhou}
\showURL{%
\tempurl}


\bibitem[Zhou et~al\mbox{.}(2023)]%
        {electrode}
\bibfield{author}{\bibinfo{person}{Yang Zhou}, \bibinfo{person}{Zezhou Wang}, \bibinfo{person}{Sowmya Dharanipragada}, {and} \bibinfo{person}{Minlan Yu}.} \bibinfo{year}{2023}\natexlab{}.
\newblock \showarticletitle{Electrode: Accelerating Distributed Protocols with {eBPF}}. In \bibinfo{booktitle}{\emph{20th USENIX Symposium on Networked Systems Design and Implementation (NSDI 23)}}. \bibinfo{publisher}{USENIX Association}, \bibinfo{address}{Boston, MA}, \bibinfo{pages}{1391--1407}.
\newblock
\showISBNx{978-1-939133-33-5}
\urldef\tempurl%
\url{https://www.usenix.org/conference/nsdi23/presentation/zhou}
\showURL{%
\tempurl}


\end{thebibliography}

\appendix
\section{\revise{Complete Presentation of the \bodega Protocol as an Implementation Guide}}
\label{sec:appendix-full-algo}

\revise{Figure~\ref{fig:algo-blocks-full} presents a full summary of the \bodega protocol, covering the complete design described in \S\ref{sec:design-roster}-\ref{sec:design-roster-leases}. It offers a rigorous description of the protocol, and can be used as a directly followable reference for implementing \bodega on top of classic consensus.}

\section{Formal \tlaplus Specification of \bodega}
\label{sec:appendix-tla-spec}

We also present a formal specification of \bodega written as a PlusCal algorithm that can be auto-translated into \tlaplus. It has been model-checked for both \textit{linearizability} and \textit{fault-tolerance} properties on symbolic inputs of 3 nodes (thus 1 allowed failure), 3 ballot numbers, 2 distinct write requests plus 2 distinct read requests, and all possible choices of responders. These inputs should be large enough to explore the distinct, interesting execution paths of the protocol. Please refer to the inlined comments for details.

\clearpage
\pagenumbering{gobble}

\begin{figure*}[!ht]
    \setstretch{1.05}
    \begin{minipage}{0.12\textwidth}
        \begin{mdframed}[%
                linewidth=0pt,
                innertopmargin=12.5pt,
                innerbottommargin=12.5pt,
                backgroundcolor=gray!20,
                roundcorner=0pt,
                font=\small\bfseries,
                fontcolor=black!85]
            \begin{center}
                Relevant \\
                states \\
                on node $r$:
            \end{center}
        \end{mdframed}
    \end{minipage}\hspace{-5px}
    \setstretch{1.12}
    \begin{minipage}{0.872\textwidth}
        \begin{mdframed}[%
                linewidth=0pt,
                leftmargin=-4pt,
                rightmargin=4pt,
                backgroundcolor=gray!8,
                roundcorner=0pt,
                font=\small,
                fontcolor=black!75]
            Normal SMR log of \avar{\color{black}\bfseries slot}s \hspace{3.2em}
            Current ballot\# \& \revise{roster}: \avar{\color{black}\bfseries bal} \ \ \avar{\color{black}\bfseries ros} \hspace{3.2em}
            Heartbeat per-peer $p$ timers: \atimer{\color{black}\bfseries heartbeat, $p$} \\
            Grantor-side per-peer $p$ timers: \atimer{\color{black}\bfseries guarding, $p$} \ \ \atimer{\color{black}\bfseries endowing, $p$} \hspace{3.2em}
            Grantor-side status sets: \aset{\color{black}\bfseries guarding} \ \ \aset{\color{black}\bfseries endowing} \\
            Grantee-side per-peer $p$ timers: \atimer{\color{black}\bfseries guarded, $p$} \ \ \ \atimer{\color{black}\bfseries endowed, $p$} \hspace{3.6em}
            Grantee-side status sets: \aset{\color{black}\bfseries guarded} \ \ \ \aset{\color{black}\bfseries endowed} \\
            Grantee-side per-peer $p$ local-read safety slot\#: \avar{\color{black}\bfseries thresh$_p$} \hspace{3.1em}
            Per-slot early \texttt{AcceptNote} markers: \aset{\color{black}\bfseries accnote, \avar{\color{black}\bfseries slot}}
        \end{mdframed}
    \end{minipage}%

    \vspace*{8pt}

    \setstretch{1.0}
    \begin{minipage}[t]{0.323\textwidth}
        \begin{algoblock}[%
            frametitle={Proactive \revise{roster} change to \avpm{ros}:},
            frametitlebackgroundcolor=\acviolet]
            \begin{algolist}[series=algo]
                \item let \avar{bal} be the current ballot
                \item do \afun{announce\_\revise{roster}}{\avar{bal}, \avpm{ros}}
            \end{algolist}
        \end{algoblock}

        \begin{center}
            \vspace{-6.5pt}
            \begin{tikzpicture}[y=0.25cm]
                \coordinate (A) at (0, 0);
                \coordinate (B) at (0, -1);
                \draw[-, line width=1.5pt, draw=Orchid!75] (A) -- (B);
            \end{tikzpicture}
            \hspace{35pt}
            \vspace{-3pt}
        \end{center}
        
        \begin{algoblock}[%
            frametitle={Upon \atimer{heartbeat, $p$} timeout for peer $p$:},
            frametitlebackgroundcolor=\acviolet]
            \begin{algolist}[resume*=algo]
                \item let \avar{bal}, \avar{ros} be the current ballot \& \revise{roster}
                \item if $p$ has a special role in \avar{ros}:
                    \begin{algolist}
                        \item create new \revise{roster} \avpm{ros} copying \avar{ros}
                        \item unmark $p$ in any responder set of \avpm{ros}
                        \item if $p$ is the leader of \avar{ros}: set the leader of \avpm{ros} to me (or any responsive node)
                        \item do \afun{announce\_\revise{roster}}{\avar{bal}, \avpm{ros}}
                    \end{algolist}
            \end{algolist}
        \end{algoblock}

        \begin{center}
            \vspace{-4pt}
            \begin{tikzpicture}[y=0.35cm]
                \coordinate (A) at (0, 0);
                \coordinate (B) at (0, -1);
                \draw[->, line width=1.5pt, draw=Orchid!75] (A) -- (B);
            \end{tikzpicture}
            \hspace{23pt}
            \begin{tikzpicture}[y=0.35cm]
                \coordinate (A) at (0, 0);
                \coordinate (B) at (0, -1);
                \draw[->, line width=1.5pt, draw=Orchid!75] (A) -- (B);
            \end{tikzpicture}
            \vspace{-3pt}
        \end{center}
        
        \begin{algoblock}[%
            frametitle={Do\,\,\,\afun{announce\_\revise{roster}}{\avar{bal}, \avpm{ros}}:},
            frametitlebackgroundcolor=\acviolet]
            \begin{algolist}[resume*=algo]
                \item compose a unique higher ballot\# \avpm{bal} $=$ $(b+1).r$ID from old \avar{bal} $= b.$\_
                \item broadcast \amsg{Heartbeat}{\avpm{bal}, \avpm{ros}} to all
            \end{algolist}
        \end{algoblock}

        \begin{center}
            \vspace{-4pt}
            \begin{tikzpicture}[y=0.4cm]
                \coordinate (A) at (0, 0);
                \coordinate (B) at (0, -1);
                \draw[->, line width=1.5pt, draw=Orchid!75] (A) -- (B);
            \end{tikzpicture}
            \vspace{-3pt}
        \end{center}

        \begin{algoblock}[%
            frametitle={Recv \amsg{Heartbeat}{\avpm{bal}, \avpm{ros}} $\leftarrow$ $p$:},
            frametitlebackgroundcolor=\acorange]
            \begin{algolist}[resume*=algo]
                \item if \avpm{bal} $>$ current ballot \avar{bal}:
                    \begin{algolist}
                        \item do \afun{revoke\_leases}{\avar{bal}}
                        \item update my ballot \& \revise{roster} to \avpm{bal}, \avpm{ros}
                        \item do \afun{initiate\_leases}{\avpm{bal}}
                        \item if I am the leader of \avpm{ros}: restart all in-progress slots from \texttt{Prepare} phase
                    \end{algolist}
                \vspace{6pt}
                \hrule
                \vspace{2pt}
                \item refresh heartbeat timer \atimer{heartbeat, $p$}
            \end{algolist}
        \end{algoblock}

        \begin{center}
            \vspace{-4pt}
            \begin{tikzpicture}[y=0.4cm]
                \coordinate (A) at (0, 0);
                \coordinate (B) at (0, -1);
                \draw[->, line width=1.5pt, draw=YellowOrange!80] (B) -- (A);
            \end{tikzpicture}
            \vspace{-3pt}
        \end{center}

        \begin{algoblock}[%
            frametitle={Repeat every heartbeat interval:},
            frametitlebackgroundcolor=\acorange]
            \begin{algolist}[resume*=algo]
                \item let \avar{bal}, \avar{ros} be the current ballot \& \revise{roster}
                \item broadcast \amsg{Heartbeat}{\avar{bal}, \avar{ros}} to all
                \item for every $p \in$ \aset{endowing}:
                    \begin{algolist}
                        \item extend \atimer{endowing, $p$} by \atmdur{lease}$+$\atmdur{$\Delta$}
                        \item piggyback lease \amsg{Renew}{\avar{bal}}
                    \end{algolist}
                \item for every $p \in$ \aset{endowed} from whom there are unreplied \amsg{Renew}{\avar{bal}}s:
                    \begin{algolist}
                        \item also piggyback \amsg{RenewReply}{\avar{bal}}
                    \end{algolist}
            \end{algolist}
        \end{algoblock}

        \begin{flushright}
            \vspace{-4pt}
            \begin{tikzpicture}[x=1.4cm,y=0.37cm]
                \coordinate (A) at (0, 0);
                \coordinate (B) at (0, -1);
                \coordinate (C) at (1, 0);
                \coordinate (D) at (1, -0.5);
                \coordinate (E) at (1, -0.42);
                \coordinate (F) at (2.4, -0.42);
                \draw[->, line width=1.5pt, draw=YellowOrange!80] (A) -- (B);
                \draw[-, line width=1.5pt, draw=YellowOrange!80] (C) -- (D);
                \draw[-, line width=1.5pt, draw=YellowOrange!80] (E) -- (F);
            \end{tikzpicture}
            \vspace{-3pt}
            \hspace*{-4pt}
        \end{flushright}
        
        \begin{algoblock}[%
            frametitle={Recv lease \amsg{RenewReply}{\avar{bal}} $\leftarrow$ $p$:},
            frametitlebackgroundcolor=\acorange]
            \begin{algolist}[resume*=algo]
                \item if \avar{bal} $\ne$ current ballot: ignore
                \item if $p$ in \aset{endowing}:
                    \begin{algolist}
                        \item kickoff \atimer{endowing, $p$} by \atmdur{lease}$+$\atmdur{$\Delta$}
                    \end{algolist}
            \end{algolist}
        \end{algoblock}
    \end{minipage}%
    \begin{minipage}[t]{0.017\textwidth}
        \begin{center}
            \begin{tikzpicture}[x=0.166cm,y=0.42cm,baseline=(current bounding box.north)]
                \path[use as bounding box] (0,0) -- (2,-33);
                \coordinate (A) at (0,   -20.6);
                \coordinate (B) at (0.9, -20.6);
                \coordinate (C) at (0.8, -20.6);
                \coordinate (D) at (0.8, -0.44);
                \coordinate (E) at (0.7, -0.44);
                \coordinate (F) at (2,   -0.44);
                \draw[-,  line width=1.5pt, draw=Maroon!70] (A) -- (B);
                \draw[-,  line width=1.5pt, draw=Maroon!70] (C) -- (D);
                \draw[->, line width=1.5pt, draw=Maroon!70] (E) -- (F);
                \coordinate (G) at (2,   -32.3);
                \coordinate (H) at (1.1, -32.3);
                \coordinate (I) at (1.2, -32.3);
                \coordinate (J) at (1.2, -23.1);
                \coordinate (K) at (1.3, -23.1);
                \coordinate (L) at (0,   -23.1);
                \draw[-,  line width=1.5pt, draw=Maroon!70] (G) -- (H);
                \draw[-,  line width=1.5pt, draw=Maroon!70] (I) -- (J);
                \draw[->, line width=1.5pt, draw=Maroon!70] (K) -- (L);
                \coordinate (M) at (0, -36.94);
                \coordinate (N) at (2, -36.94);
                \draw[->, line width=1.5pt, draw=YellowOrange!80] (M) -- (N);
            \end{tikzpicture}
        \end{center}
    \end{minipage}%
    \begin{minipage}[t]{0.327\textwidth}
        \begin{algoblock}[%
            frametitle={Do\,\,\,\afun{revoke\_leases}{\avar{bal}}:},
            frametitlebackgroundcolor=\acred]
            \begin{algolist}[resume*=algo]
                \item clear \aset{guarding} to empty set
                \item broadcast lease \amsg{Revoke}{\avar{bal}} to all
                \item for every $p \in$ \aset{endowing}, wait until:
                    \begin{algolist}
                        \item either \amsg{RevokeReply}{\avar{bal}} is received from $p$ successfully, or
                        \item $p$ removed from \aset{endowing} by timeout
                    \end{algolist}
            \end{algolist}
        \end{algoblock}

        \begin{center}
            \vspace{-4.5pt}
            \begin{tikzpicture}[y=0.35cm]
                \coordinate (A) at (0, 0);
                \coordinate (B) at (0, -1);
                \draw[->, line width=1.5pt, draw=Maroon!70] (A) -- (B);
            \end{tikzpicture}
            \vspace{-3pt}
        \end{center}
        
        \begin{algoblock}[%
            frametitle={Recv lease \amsg{Revoke}{\avar{bal}} $\leftarrow$ $p$:},
            frametitlebackgroundcolor=\acred]
            \begin{algolist}[resume*=algo]
                \item if \avar{bal} $\geqslant$ current ballot:
                    \begin{algolist}
                        \item remove $p$ from \aset{guarded} and \aset{endowed}
                    \end{algolist}
                \item reply \amsg{RevokeReply}{\avar{bal}} to $p$
            \end{algolist}
        \end{algoblock}

        \begin{center}
            \vspace{-4.5pt}
            \begin{tikzpicture}[y=0.35cm]
                \coordinate (A) at (0, 0);
                \coordinate (B) at (0, -1);
                \draw[->, line width=1.5pt, draw=Maroon!70] (A) -- (B);
            \end{tikzpicture}
            \vspace{-3pt}
        \end{center}
        
        \begin{algoblock}[%
            frametitle={Do\,\,\,\afun{initiate\_leases}{\avar{bal}}:},
            frametitlebackgroundcolor=\acred]
            \begin{algolist}[resume*=algo]
                \item for every node $p$:
                    \begin{algolist}
                        \item kickoff \atimer{guarding, $p$} by \atmdur{guard}$+$\atmdur{$\Delta$}
                        \item add $p$ to \aset{guarding}
                    \end{algolist}
                \item broadcast lease \amsg{Guard}{\avar{bal}, \avar{thresh}}, where \avar{thresh} is the highest slot\# I ever accepted
                \item for every $p \in$ \aset{guarding}, {\color{black!50} asynchronously}:
                    \begin{algolist}
                        \item wait until recv \amsg{GuardReply}{\avar{bal}}
                        \item if $p$ has been removed from \aset{guarding} by timeout, retry this step for $p$
                        \item move $p$ from \aset{guarding} to \aset{endowing}
                        \item kickoff \atimer{endowing, $p$} by \atmdur{guard}$+$\atmdur{lease}$+$\atmdur{$\Delta$}
                    \end{algolist}
            \end{algolist}
        \end{algoblock}

        \begin{center}
            \vspace{-4.5pt}
            \begin{tikzpicture}[y=0.35cm]
                \coordinate (A) at (0, 0);
                \coordinate (B) at (0, -1);
                \draw[->, line width=1.5pt, draw=Maroon!70] (A) -- (B);
            \end{tikzpicture}
            \vspace{-3pt}
        \end{center}
        
        \begin{algoblock}[%
            frametitle={Recv lease \amsg{Guard}{\avar{bal}, \avar{thresh}} $\leftarrow$ $p$:},
            frametitlebackgroundcolor=\acred]
            \begin{algolist}[resume*=algo]
                \item if \avar{bal} $\ne$ current ballot: ignore
                \item if $p \notin$ \aset{guarded} and $\notin$ \aset{endowed}:
                    \begin{algolist}
                        \item add $p$ to \aset{guarded}, and record \avar{thresh}$_p$ for later use in local read checks
                        \item kickoff \atimer{guarded, $p$} by \atmdur{guard}$-$\atmdur{$\Delta$}
                        \item reply \amsg{GuardReply}{\avar{bal}} to $p$
                    \end{algolist}
            \end{algolist}
        \end{algoblock}

        \vspace{11.5pt}

        \begin{algoblock}[%
            frametitle={Recv lease \amsg{Renew}{\avar{bal}} $\leftarrow$ $p$:},
            frametitlebackgroundcolor=\acorange]
            \begin{algolist}[resume*=algo]
                \item if \avar{bal} $\ne$ current ballot: ignore
                \item {\color{black!50} (also ignore duplicate \texttt{Renew}s; can be implemented with \textit{seq}\# or unique msg IDs)}
                \item if $p$ in \aset{guarded}:
                    \begin{algolist}
                        \item move $p$ from \aset{guarded} to \aset{endowed}
                    \end{algolist}
                \item if $p$ in \aset{endowed}:
                    \begin{algolist}
                        \item kickoff \atimer{endowed, $p$} by \atmdur{lease}$-$\atmdur{$\Delta$}
                    \end{algolist}
            \end{algolist}
        \end{algoblock}
    \end{minipage}
    \hfill
    \begin{minipage}[t]{0.323\textwidth}
        \begin{algoblock}[%
            frametitle={Upon any lease timer \atimer{{\color{darkgray!75}\normalfont intent}, $p$} timeout:},
            frametitlebackgroundcolor=\acgold]
            \begin{algolist}[resume*=algo]
                \item remove $p$ from the corresponding \aset{\color{darkgray!75} intent}
            \end{algolist}
        \end{algoblock}
        
        \vspace{9pt}
        
        \begin{algoblock}[%
            frametitle={Handle \amsg{Write}{$k$, $v$} req $\leftarrow$ client:},
            frametitlebackgroundcolor=\acblue]
            \begin{algolist}[resume*=algo]
                \item let \avar{bal}, \avar{ros} be the current ballot \& \revise{roster}
                \item if I am a leader: proceed to \texttt{Accept} phase at next available \avar{slot}; this is just normal consensus except with two changes:
                    \begin{algolist}
                        \item leader decides commit only after all responders of $k$ in \avar{ros} have replied
                        \item all nodes broadcast \amsg{AcceptNote}{\avar{bal}, \avar{slot}} to responders of $k$, besides replying \amsg{AcceptReply}{\avar{bal}, \avar{slot}} to leader
                    \end{algolist}
                \item else: redirect client to leader node
            \end{algolist}
        \end{algoblock}

        \begin{center}
            \vspace{-5pt}
            \begin{tikzpicture}[y=0.3cm]
                \coordinate (A) at (0, 0);
                \coordinate (B) at (0, -1);
                \draw[->, line width=1.5pt, draw=RoyalBlue!50] (A) -- (B);
            \end{tikzpicture}
            \vspace{-3pt}
        \end{center}
        
        \begin{algoblock}[%
            frametitle={Recv \amsg{AcceptNote}{\avar{bal}, \avar{slot}} $\leftarrow$ $p$:},
            frametitlebackgroundcolor=\acblue]
            \begin{algolist}[resume*=algo]
                \item if \avar{bal} $=$ current ballot prepared at \avar{slot}:
                    \begin{algolist}
                        \item add $p$ to \aset{accnote, \avar{slot}}
                    \end{algolist}
            \end{algolist}
        \end{algoblock}

        \vspace{8.5pt}
        
        \begin{algoblock}[%
            frametitle={Handle \amsg{Read}{$k$} req $\leftarrow$ client:},
            frametitlebackgroundcolor=\acgreen]
            \begin{algolist}[resume*=algo]
                \item if $|$\aset{endowed}$|$ $< m$, or if I am not a responder of $k$ in current \revise{roster} \avar{ros}:
                    \begin{algolist}
                        \item if I am {\color{black!50} (unstable)} leader: proceed to \texttt{Accept} phase as normal write cmd
                        \item else: redirect client to leader node
                    \end{algolist}
                \item else if there does not exist a size-$m$ subset of \aset{endowed} where, for all $p$ in the subset, I have committed all slots up to \avar{thresh}$_p$:
                    \begin{algolist}
                        \item {\color{black!50} I probably just joined current roster,} hold until true or redirect to leader
                    \end{algolist}
                \item else if I am the {\color{black!50} (stable)} leader of \avar{ros}:
                    \begin{algolist}
                        \item reply with the last committed val of $k$
                    \end{algolist}
                \item else{\color{black!50}, (I am a non-leader responder in \avar{ros})}:
                    \begin{algolist}
                        \item do \afun{responder\_read}{$k$}
                    \end{algolist}
            \end{algolist}
        \end{algoblock}

        \begin{center}
            \vspace{-5pt}
            \begin{tikzpicture}[y=0.3cm]
                \coordinate (A) at (0, 0);
                \coordinate (B) at (0, -1);
                \draw[->, line width=1.5pt, draw=ForestGreen!60] (A) -- (B);
            \end{tikzpicture}
            \vspace{-3pt}
        \end{center}
        
        \begin{algoblock}[%
            frametitle={Do\,\,\,\afun{responder\_read}{$k$}:},
            frametitlebackgroundcolor=\acgreen]
            \begin{algolist}[resume*=algo]
                \item find highest \avar{slot}$_k$ containing write to $k$
                \item if \avar{slot}$_k$ has committed, or is accepted at current \avar{bal} and $|$\aset{accnote, \avar{slot}$_k$}$|$ $\geqslant m$:
                    \begin{algolist}
                        \item reply with the value in \avar{slot}$_k$
                    \end{algolist}
                \item else{\color{black!50}, (not sure if last write commits)}:
                    \begin{algolist}
                        \item hold until the above becomes true
                        \item promptly redirect if \atmdur{unhold} timeout
                    \end{algolist}
            \end{algolist}
        \end{algoblock}
    \end{minipage}
    
    \captionsetup{justification=raggedright}
    \vspace{-3pt}
    \floatcapoffig{Complete Summary of the \bodega algorithm}{Lists all actions a node $r$ would take upon certain conditions, grouped by purposes for clarity: \acbox{\acviolet} triggers for a new \revise{roster}, \acbox{\acred} granting procedure of new \revise{roster} leases, \acbox{\acorange} heartbeats and lease renewals, \acbox{\acblue} handling client write requests, \acbox{\acgreen} handling client read requests. The description is based on a regular key-value API. Nodes implicitly retransmit non-acked messages. Broadcast msg receivers include the sender itself. Clock drift between nodes is assumed to be bounded by $t_{\Delta}$, as is required by any distributed lease algorithm; clock skews are irrelevant thanks to Guards. \revise{The arrows annotate a natural reading order that follows the usual flow of the protocol.}}
    \label{fig:algo-blocks-full}
\end{figure*}

\clearpage



\section*{Supplementary material}

---------------- MODULE *Bodega* ----------------

EXTENDS *FiniteSets*, *Sequences*, *Integers*, *TLC*

Model inputs & assumptions.

CONSTANT *Replicas*,   symmetric set of server nodes
         *Writes*,     symmetric set of write commands (each w/ unique value)
         *Reads*,      symmetric set of read commands
         *MaxBallot*,  maximum ballot pickable for leader preemption
         *NodeFailuresOn*   if true, turn on node failures injection

$ReplicasAssumption \triangleq \land IsFiniteSet(Replicas)$
$\qquad\qquad\qquad\qquad\;\; \land Cardinality(Replicas) \geq 1$
$\qquad\qquad\qquad\qquad\;\; \land \text{``none''} \notin Replicas$

$Population \triangleq Cardinality(Replicas)$

$MajorityNum \triangleq (Population \div 2) + 1$

$WritesAssumption \triangleq \land IsFiniteSet(Writes)$
$\qquad\qquad\qquad\qquad\; \land Cardinality(Writes) \geq 1$
$\qquad\qquad\qquad\qquad\; \land \text{``nil''} \notin Writes$

a write command model value serves as both the
ID of the command and the value to be written

$ReadsAssumption \triangleq \land IsFiniteSet(Reads)$
$\qquad\qquad\qquad\qquad \land Cardinality(Reads) \geq 0$
$\qquad\qquad\qquad\qquad \land \text{``nil''} \notin Writes$

$MaxBallotAssumption \triangleq \land MaxBallot \in Nat$
$\qquad\qquad\qquad\qquad\qquad \land MaxBallot \geq 2$

$NodeFailuresOnAssumption \triangleq NodeFailuresOn \in \text{BOOLEAN}$

ASSUME $\land ReplicasAssumption$
$\qquad\;\; \land WritesAssumption$
$\qquad\;\; \land ReadsAssumption$
$\qquad\;\; \land MaxBallotAssumption$
$\qquad\;\; \land NodeFailuresOnAssumption$

---

Useful constants & typedefs.

$Commands \triangleq Writes \cup Reads$

$NumWrites \triangleq Cardinality(Writes)$

$NumReads \triangleq Cardinality(Reads)$

$NumCommands \triangleq Cardinality(Commands)$

$Range(seq) \triangleq \{seq[i] : i \in 1 \,..\, Len(seq)\}$

Client observable events.

$ClientEvents \triangleq \quad [type : \{\text{``Req''}\}, cmd : Commands]$
$\qquad\qquad\qquad \cup \quad [type : \{\text{``Ack''}\}, cmd : Commands,$
$\qquad\qquad\qquad\qquad\qquad\qquad\; val \;\: : \{\text{``nil''}\} \cup Writes,$
$\qquad\qquad\qquad\qquad\qquad\qquad\; by : Replicas]$

$ReqEvent(c) \triangleq [type \mapsto \text{``Req''}, cmd \mapsto c]$

$AckEvent(c, v, n) \triangleq [type \mapsto \text{``Ack''}, cmd \mapsto c, val \mapsto v, by \mapsto n]$

  *val* is the old value for a write command

$InitPending \triangleq$ (CHOOSE $ws \in [1 \mathinner{\ldotp\ldotp} Cardinality(Writes) \to Writes]$
                        $: Range(ws) = Writes)$
            $\circ$ (CHOOSE $rs \in [1 \mathinner{\ldotp\ldotp} Cardinality(Reads) \to Reads]$
                        $: Range(rs) = Reads)$

  *W.L.O.G.*, choose any sequence concatenating writes
  commands and read commands as the sequence of reqs;
  all other cases are either symmetric or less useful
  than this one

Server-side constants & states.

$Ballots \triangleq 1 \mathinner{\ldotp\ldotp} MaxBallot$

$Slots \triangleq 1 \mathinner{\ldotp\ldotp} NumWrites$

$Statuses \triangleq \{\text{``Preparing''}, \text{``Accepting''}, \text{``Committed''}\}$

$InstStates \triangleq [status : \{\text{``Empty''}\} \cup Statuses,$
              $write : \{\text{``nil''}\} \cup Writes,$
              $voted : [bal : \{0\} \cup Ballots,$
                      $write : \{\text{``nil''}\} \cup Writes]]$

$NullInst \triangleq [status \mapsto \text{``Empty''},$
            $write \mapsto \text{``nil''},$
            $voted \mapsto [bal \mapsto 0, write \mapsto \text{``nil''}]]$

$NodeStates \triangleq [leader : \{\text{``none''}\} \cup Replicas,$
              $commitUpTo : \{0\} \cup Slots,$
              $commitPrev\ : \{0\} \cup Slots \cup \{NumWrites + 1\},$
              $balPrepared\quad : \{0\} \cup Ballots,$
              $balMaxKnown : \{0\} \cup Ballots,$
              $insts : [Slots \to InstStates]]$

$NullNode \triangleq [leader \mapsto \text{``none''},$
            $commitUpTo \mapsto 0,$
            $commitPrev\ \mapsto 0,$
            $balPrepared\quad \mapsto 0,$
            $balMaxKnown \mapsto 0,$
            $insts \mapsto [s \in Slots \mapsto NullInst]]$

  *commitPrev* is the last slot which might have been
  committed by an old leader; a newly prepared leader
  can safely serve reads locally only after its log has
  been committed up to this slot. The time before this
  condition becomes satisfied may be considered the
  "recovery" or "ballot transfer" time

$FirstEmptySlot(insts) \triangleq$
   IF $\forall s \in Slots : insts[s].status \neq \text{``Empty''}$
      THEN $NumWrites + 1$
      ELSE CHOOSE $s \in Slots :$
              $\wedge insts[s].status = \text{``Empty''}$
              $\wedge \forall t \in 1 \mathinner{\ldotp\ldotp} (s-1) : insts[t].status \neq \text{``Empty''}$

$LastNonEmptySlot(insts) \triangleq$
   IF $\forall s \in Slots : insts[s].status = \text{``Empty''}$

```
                        then 0
                        else  choose s ∈ Slots :
                                ∧ insts[s].status ≠ "Empty"
                                ∧ ∀ t ∈ (s + 1) .. NumWrites : insts[t].status = "Empty"
                              note that this is not the same as FirstEmptySlot − 1
                              due to possible existence of holes
```

Service-internal messages.

$PrepareMsgs \triangleq [type : \{\text{"Prepare"}\}, src : Replicas,$
$\qquad\qquad\qquad\qquad\qquad bal : Ballots]$

$PrepareMsg(r, b) \triangleq [type \mapsto \text{"Prepare"}, src \mapsto r,$
$\qquad\qquad\qquad\qquad\qquad bal \mapsto b]$

$InstsVotes \triangleq [Slots \rightarrow [bal : \{0\} \cup Ballots,$
$\qquad\qquad\qquad\qquad write : \{\text{"nil"}\} \cup Writes]]$

$VotesByNode(n) \triangleq [s \in Slots \mapsto n.insts[s].voted]$

$PrepareReplyMsgs \triangleq [type : \{\text{"PrepareReply"}\}, src : Replicas,$
$\qquad\qquad\qquad\qquad\qquad bal : Ballots,$
$\qquad\qquad\qquad\qquad\qquad votes : InstsVotes]$

$PrepareReplyMsg(r, b, iv) \triangleq [type \mapsto \text{"PrepareReply"}, src \mapsto r,$
$\qquad\qquad\qquad\qquad\qquad bal \mapsto b,$
$\qquad\qquad\qquad\qquad\qquad votes \mapsto iv]$

$PeakVotedWrite(prs, s) \triangleq$
$\quad$ if $\forall pr \in prs : pr.votes[s].bal = 0$
$\qquad$ then "nil"
$\qquad$ else  let $ppr \triangleq$
$\qquad\qquad\qquad$ choose $ppr \in prs :$
$\qquad\qquad\qquad\qquad \forall pr \in prs : pr.votes[s].bal \leq ppr.votes[s].bal$
$\qquad\qquad$ in $\ ppr.votes[s].write$

$LastTouchedSlot(prs) \triangleq$
$\quad$ if $\forall s \in Slots : PeakVotedWrite(prs, s) = \text{"nil"}$
$\qquad$ then 0
$\qquad$ else  choose $s \in Slots :$
$\qquad\qquad\qquad \land PeakVotedWrite(prs, s) \neq \text{"nil"}$
$\qquad\qquad\qquad \land \forall t \in (s + 1) .. NumWrites : PeakVotedWrite(prs, t) = \text{"nil"}$

$PrepareNoticeMsgs \triangleq [type : \{\text{"PrepareNotice"}\}, src : Replicas,$
$\qquad\qquad\qquad\qquad\qquad bal : Ballots,$
$\qquad\qquad\qquad\qquad\qquad commit\_prev : \{0\} \cup Slots]$

$PrepareNoticeMsg(r, b, cp) \triangleq [type \mapsto \text{"PrepareNotice"}, src \mapsto r,$
$\qquad\qquad\qquad\qquad\qquad bal \mapsto b,$
$\qquad\qquad\qquad\qquad\qquad commit\_prev \mapsto cp]$
$\qquad\qquad\qquad\qquad$ this messasge is added to allow
$\qquad\qquad\qquad\qquad$ followers to learn about $commitPrev$

$AcceptMsgs \triangleq [type : \{\text{"Accept"}\}, src : Replicas,$
$\qquad\qquad\qquad\qquad bal : Ballots,$
$\qquad\qquad\qquad\qquad slot : Slots,$
$\qquad\qquad\qquad\qquad write : Writes]$

$AcceptMsg(r, b, s, c) \triangleq [type \mapsto \text{"Accept"}, src \mapsto r,$

$$\quad bal \mapsto b,$$
$$slot \mapsto s,$$
$$write \mapsto c]$$

$AcceptReplyMsgs \triangleq [type : \{\text{"AcceptReply"}\}, src : Replicas,$
$\qquad\qquad\qquad\qquad\qquad\qquad bal : Ballots,$
$\qquad\qquad\qquad\qquad\qquad\qquad slot : Slots]$

$AcceptReplyMsg(r, b, s) \triangleq [type \mapsto \text{"AcceptReply"}, src \mapsto r,$
$\qquad\qquad\qquad\qquad\qquad\qquad\qquad bal \mapsto b,$
$\qquad\qquad\qquad\qquad\qquad\qquad\qquad slot \mapsto s]$

> no need to carry command *ID* in *AcceptReply* because ballot and slot uniquely identifies the write

$CommitNoticeMsgs \triangleq [type : \{\text{"CommitNotice"}\}, upto : Slots]$

$CommitNoticeMsg(u) \triangleq [type \mapsto \text{"CommitNotice"}, upto \mapsto u]$

$Messages \triangleq \quad PrepareMsgs$
$\qquad\qquad \cup \quad PrepareReplyMsgs$
$\qquad\qquad \cup \quad PrepareNoticeMsgs$
$\qquad\qquad \cup \quad AcceptMsgs$
$\qquad\qquad \cup \quad AcceptReplyMsgs$
$\qquad\qquad \cup \quad CommitNoticeMsgs$

*Roster* lease related typedefs.

$Rosters \triangleq \{ros \in [bal : Ballots, leader : Replicas, responders : \text{SUBSET } Replicas] :$
$\qquad\qquad ros.leader \notin ros.responders\}$

$Roster(b, l, resps) \triangleq [bal \mapsto b, leader \mapsto l, responders \mapsto resps]$

> each new ballot number maps to a new *roster*; this includes the change of leader (as in classic *MultiPaxos*) and/or the change of who're *responders*

$LeaseGrants \triangleq [from : Replicas, roster : Rosters]$

$LeaseGrant(f, ros) \triangleq [from \mapsto f, roster \mapsto ros]$

> this is the only type of message that may be "removed" from the global set of messages to make a "cheated" model of leasing: if a *LeaseGrant* message is removed, it means that promise has expired and the grantor did not refresh, probably making way for switching to a different *roster*

---

Main algorithm in *PlusCal*.

**algorithm** *Bodega*

**variable** $msgs = \{\},$               messages in the network
$\qquad\quad grants = \{\},$              lease *msgs* in the network
$\qquad\quad node = [r \in Replicas \mapsto NullNode],$    replica node state
$\qquad\quad pending = InitPending,$        sequence of pending reqs
$\qquad\quad observed = \langle\rangle,$             client observed events
$\qquad\quad crashed = [r \in Replicas \mapsto \text{FALSE}];$    replica crashed flag

**define**

$$CurrentRoster \triangleq$$
$$\text{LET } leased(b) \triangleq Cardinality(\{g \in grants :$$
$$g.roster.bal = b\}) \geq MajorityNum$$
$$\text{IN } \text{IF } \neg \exists b \in Ballots : leased(b)$$
$$\text{THEN } Roster(0, \text{"none"}, 0)$$
$$\text{ELSE } (\text{CHOOSE } g \in grants : leased(g.roster.bal)).roster$$

*the leasing mechanism ensures that at any time, there's at most one leader*

$$ThinkAmLeader(r) \triangleq \land node[r].leader = r$$
$$\land node[r].balPrepared = node[r].balMaxKnown$$
$$\land CurrentRoster.bal > 0$$
$$\land CurrentRoster.bal = node[r].balMaxKnown$$
$$\land CurrentRoster.leader = r$$

$$ThinkAmFollower(r) \triangleq \land node[r].leader \neq r$$
$$\land CurrentRoster.bal > 0$$
$$\land CurrentRoster.bal = node[r].balMaxKnown$$
$$\land CurrentRoster.leader \neq r$$

$$ThinkAmResponder(r) \triangleq \land ThinkAmFollower(r)$$
$$\land r \in CurrentRoster.responders$$

$$BallotTransfered(r) \triangleq node[r].commitUpTo \geq node[r].commitPrev$$

$$WriteCommittable(ars) \triangleq$$
$$\land Cardinality(\{ar.src : ar \in ars\}) \geq MajorityNum$$
$$\land CurrentRoster.responders \subseteq \{ar.src : ar \in ars\}$$

$$reqsMade \triangleq \{e.cmd : e \in \{e \in Range(observed) : e.type = \text{"Req"}\}\}$$

$$acksRecv \triangleq \{e.cmd : e \in \{e \in Range(observed) : e.type = \text{"Ack"}\}\}$$

$$AppendObserved(seq) \triangleq$$
$$\text{LET } filter(e) \triangleq \text{IF } e.type = \text{"Req"} \text{ THEN } e.cmd \notin reqsMade$$
$$\text{ELSE } e.cmd \notin acksRecv$$
$$\text{IN } observed \circ SelectSeq(seq, filter)$$

$$UnseenPending(r) \triangleq$$
$$\text{LET } filter(c) \triangleq \forall s \in Slots : node[r].insts[s].write \neq c$$
$$\text{IN } SelectSeq(pending, filter)$$

$$RemovePending(cmd) \triangleq$$
$$\text{LET } filter(c) \triangleq c \neq cmd$$
$$\text{IN } SelectSeq(pending, filter)$$

$$terminated \triangleq \land Len(pending) = 0$$
$$\land Cardinality(reqsMade) = NumCommands$$
$$\land Cardinality(acksRecv) = NumCommands$$

$$numCrashed \triangleq Cardinality(\{r \in Replicas : crashed[r]\})$$

**end define ;**

*Send a set of messages helper.*
**macro** $Send(set)$ **begin**
$\quad msgs := msgs \cup set ;$
**end macro ;**

*Expire existing lease grant from $f$, and make a new repeatedly refreshed*

lease grant to new *roster ros*.
**macro** $Lease(f, ros)$ **begin**
  $grants := \{g \in grants : g.from \neq f\} \cup \{LeaseGrant(f, ros)\}$;
**end macro** ;

Observe client events helper.
**macro** $Observe(seq)$ **begin**
  $observed := AppendObserved(seq)$;
**end macro** ;

Resolve a pending command helper.
**macro** $Resolve(c)$ **begin**
  $pending := RemovePending(c)$;
**end macro** ;

Someone steps up as leader and sends *Prepare* message to followers.
To simplify this spec *W.L.O.G.*, we change the *responders roster* only when
a new leader steps up; in practice, a separate and independent type of
trigger will be used to change the *roster*.
**macro** $BecomeLeader(r)$ **begin**
  if I'm not a current leader
  **await** $node[r].leader \neq r$ ;
  pick a greater ballot number and a *roster*
  **with** $b \in Ballots$,
     $resps \in \text{SUBSET } \{f \in Replicas : f \neq r\}$,
  **do**
    **await** $\land\ b > node[r].balMaxKnown$
       $\land\ \neg \exists\, m \in msgs : (m.type = \text{"Prepare"}) \land (m.bal = b)$ ;
       *W.L.O.G.*, using this clause to model that ballot
       numbers from different proposers be unique
    update states and restart *Prepare* phase for in-progress instances
    $node[r].leader := r\ ||$
    $node[r].commitPrev := NumWrites + 1\ ||$
    $node[r].balPrepared\quad := 0\ ||$
    $node[r].balMaxKnown := b\ ||$
    $node[r].insts :=$
      $[s \in Slots \mapsto$
       $[node[r].insts[s]$
         $\text{EXCEPT } !.status = \text{IF } @ = \text{"Accepting"}$
                $\text{THEN "Preparing"}$
                $\text{ELSE}\quad @ ]]$ ;
    broadcast *Prepare* and reply to myself instantly
    $Send(\{PrepareMsg(r, b),$
       $PrepareReplyMsg(r, b, VotesByNode(node[r]))\})$ ;
    expire my old lease grant if any and grant to myself
    $Lease(r, Roster(b, r, resps))$ ;
  **end with** ;
**end macro** ;

Replica replies to a *Prepare* message.
**macro** $HandlePrepare(r)$ **begin**
  if receiving a *Prepare* message with larger ballot than ever seen
  **with** $m \in msgs$ **do**
    **await** $\land\ m.type = \text{"Prepare"}$
       $\land\ m.bal > node[r].balMaxKnown$ ;

update states and reset statuses
$node[r].leader := m.src \,\|$
$node[r].commitPrev := NumWrites + 1 \,\|$
$node[r].balMaxKnown := m.bal \,\|$
$node[r].insts :=$
$\quad [s \in Slots \mapsto$
$\quad\quad [node[r].insts[s]$
$\quad\quad\quad \text{EXCEPT } !.status = \text{IF } @ = \text{``Accepting''}$
$\quad\quad\quad\quad\quad\quad\quad\quad\quad \text{THEN ``Preparing''}$
$\quad\quad\quad\quad\quad\quad\quad\quad\quad \text{ELSE } @]];$

send back *PrepareReply* with my voted list
$Send(\{PrepareReplyMsg(r, m.bal, VotesByNode(node[r]))\});$

expire my old lease grant if any and grant to new leader
remember that we simplify this spec by merging *responders*
*roster* change into leader change Prepares
$Lease(r, (\text{CHOOSE } g \in grants : g.from = m.src).roster);$
**end with** ;
**end macro** ;

Leader gathers *PrepareReply* messages until condition met, then marks
the corresponding ballot as prepared and saves highest voted commands.
**macro** $HandlePrepareReplies(r)$ **begin**
if I'm waiting for *PrepareReplies*
**await** $\land node[r].leader = r$
$\quad\quad\quad \land node[r].balPrepared = 0;$
when there are enough number of *PrepareReplies* of desired ballot
**with** $prs = \{m \in msgs : \land m.type = \text{``PrepareReply''}$
$\quad\quad\quad\quad\quad\quad\quad\quad \land m.bal = node[r].balMaxKnown\}$
**do**
$\quad$ **await** $Cardinality(\{pr.src : pr \in prs\}) \geq MajorityNum;$
marks this ballot as prepared and saves highest voted command
in each slot if any
$\quad node[r].balPrepared := node[r].balMaxKnown \,\|$
$\quad node[r].insts :=$
$\quad\quad [s \in Slots \mapsto$
$\quad\quad\quad [node[r].insts[s]$
$\quad\quad\quad\quad \text{EXCEPT } !.status = \text{IF } \lor @ = \text{``Preparing''}$
$\quad\quad\quad\quad\quad\quad\quad\quad\quad\quad\quad \lor \land @ = \text{``Empty''}$
$\quad\quad\quad\quad\quad\quad\quad\quad\quad\quad\quad\quad\quad \land PeakVotedWrite(prs, s) \neq \text{``nil''}$
$\quad\quad\quad\quad\quad\quad\quad\quad\quad\quad\quad \text{THEN ``Accepting''}$
$\quad\quad\quad\quad\quad\quad\quad\quad\quad\quad\quad \text{ELSE } @,$
$\quad\quad\quad\quad\quad\quad\quad !.write = PeakVotedWrite(prs, s)]] \,\|$
$\quad node[r].commitPrev := LastTouchedSlot(prs);$
send *Accept* messages for in-progress instances and reply to myself
instantly; send *PrepareNotices* as well
$\quad Send(\text{UNION}$
$\quad\quad \{\{AcceptMsg(r, node[r].balPrepared, s, node[r].insts[s].write),$
$\quad\quad\quad AcceptReplyMsg(r, node[r].balPrepared, s)\} :$
$\quad\quad\quad s \in \{s \in Slots : node[r].insts[s].status = \text{``Accepting''}\}\}$
$\quad \cup \quad \{PrepareNoticeMsg(r, node[r].balPrepared, LastTouchedSlot(prs))\});$
**end with** ;
**end macro** ;

Follower receives *PrepareNotice* from a prepared and recovered leader, and
updates its *commitPrev* accordingly.

**macro** $HandlePrepareNotice(r)$ **begin**
    if I'm a follower waiting on *PrepareNotice*
    **await** $\wedge\, ThinkAmFollower(r)$
           $\wedge\, node[r].commitPrev = NumWrites + 1$ ;
    when there's a *PrepareNotice* message in effect
    **with** $m \in msgs$ **do**
        **await** $\wedge\, m.type = $ "PrepareNotice"
              $\wedge\, m.bal = node[r].balMaxKnown$ ;
        update my *commitPrev*
        $node[r].commitPrev := m.commit\_prev$ ;
    **end with** ;
**end macro** ;

A prepared leader takes a new write request into the next empty slot.
**macro** $TakeNewWriteRequest(r)$ **begin**
    if I'm a prepared leader and there's pending write request
    **await** $\wedge\, ThinkAmLeader(r)$
           $\wedge\, \exists s \in Slots : node[r].insts[s].status = $ "Empty"
           $\wedge\, Len(UnseenPending(r)) > 0$
           $\wedge\, Head(UnseenPending(r)) \in Writes$ ;
    find the next empty slot and pick a pending request
    **with** $s = FirstEmptySlot(node[r].insts)$,
          $c = Head(UnseenPending(r))$
                   $W.L.O.G.$, only pick a command not seen in current
                   prepared log to have smaller state space; in practice,
                   duplicated client requests should be treated by some
                   idempotency mechanism such as using request *IDs*
    **do**
        update slot status and voted
        $node[r].insts[s].status := $ "Accepting" $\|$
        $node[r].insts[s].write := c \,\|$
        $node[r].insts[s].voted.bal := node[r].balPrepared \,\|$
        $node[r].insts[s].voted.write := c$ ;
        broadcast *Accept* and reply to myself instantly
        $Send(\{AcceptMsg(r, node[r].balPrepared, s, c),$
                   $AcceptReplyMsg(r, node[r].balPrepared, s)\})$ ;
        append to observed events sequence if haven't yet
        $Observe(\langle ReqEvent(c) \rangle)$ ;
    **end with** ;
**end macro** ;

Replica replies to an *Accept* message.
**macro** $HandleAccept(r)$ **begin**
    if I'm a follower
    **await** $ThinkAmFollower(r)$ ;
    if receiving an unreplied *Accept* message with valid ballot
    **with** $m \in msgs$ **do**
        **await** $\wedge\, m.type = $ "Accept"
              $\wedge\, m.bal \geq node[r].balMaxKnown$
              $\wedge\, m.bal \geq node[r].insts[m.slot].voted.bal$ ;
        update node states and corresponding instance's states
        $node[r].leader := m.src \,\|$
        $node[r].balMaxKnown := m.bal \,\|$
        $node[r].insts[m.slot].status := $ "Accepting" $\|$
        $node[r].insts[m.slot].write := m.write \,\|$

$node[r].insts[m.slot].voted.bal := m.bal \parallel$
$node[r].insts[m.slot].voted.write := m.write$;
send back *AcceptReply*
$Send(\{AcceptReplyMsg(r, m.bal, m.slot)\})$;
**end with** ;
**end macro** ;

Leader gathers *AcceptReply* messages for a slot until condition met,
then marks the slot as committed and acknowledges the client.
**macro** $HandleAcceptReplies(r)$ **begin**
if I'm a prepared leader
**await** $\land ThinkAmLeader(r)$
$\land node[r].commitUpTo < NumWrites$
$\land node[r].insts[node[r].commitUpTo+1].status =$ "Accepting";
*W.L.O.G.*, only enabling the next slot after *commitUpTo*
here to make the body of this macro simpler; in practice,
messages are received proactively and there should be a
separate "Executed" status
for this slot, when there is a good set of *AcceptReplies* that is at
least a majority number and that covers all *responders*
**with** $s = node[r].commitUpTo+1$,
$c = node[r].insts[s].write$,
$ls = s-1$,
$v =$ IF $ls = 0$ THEN "nil" ELSE $node[r].insts[ls].write$,
$ars = \{m \in msgs : \land m.type =$ "AcceptReply"
$\land m.slot = s$
$\land m.bal = node[r].balPrepared\}$
**do**
**await** $WriteCommittable(ars)$;
marks this slot as committed and apply command
$node[r].insts[s].status :=$ "Committed" $\parallel$
$node[r].commitUpTo := s$;
append to observed events sequence if haven't yet, and remove
the command from pending
$Observe(\langle AckEvent(c, v, r)\rangle)$;
$Resolve(c)$;
broadcast *CommitNotice* to followers
$Send(\{CommitNoticeMsg(s)\})$;
**end with** ;
**end macro** ;

Replica receives new commit notification.
**macro** $HandleCommitNotice(r)$ **begin**
if I'm a follower waiting on *CommitNotice*
**await** $\land ThinkAmFollower(r)$
$\land node[r].commitUpTo < NumWrites$
$\land node[r].insts[node[r].commitUpTo+1].status =$ "Accepting";
*W.L.O.G.*, only enabling the next slot after *commitUpTo*
here to make the body of this macro simpler
for this slot, when there's a *CommitNotice* message
**with** $s = node[r].commitUpTo+1$,
$c = node[r].insts[s].write$,
$m \in msgs$
**do**
**await** $\land m.type =$ "CommitNotice"

$\qquad \wedge m.upto = s$ ;
        marks this slot as committed and apply command
        $node[r].insts[s].status :=$ "Committed" $\|$
        $node[r].commitUpTo := s$ ;
    **end with** ;
**end macro** ;

A prepared leader or a responder follower takes a new read request and serves it locally.
**macro** $TakeNewReadRequest(r)$ **begin**
    if I'm a caught-up leader or responder follower
    **await** $\wedge \vee ThinkAmLeader(r)$
              $\vee ThinkAmResponder(r)$
         $\wedge BallotTransfered(r)$
         $\wedge Len(UnseenPending(r)) > 0$
         $\wedge Head(UnseenPending(r)) \in Reads$ ;
    pick a pending request; examine my log and find the last non-empty slot, check its status
    **with** $s = LastNonEmptySlot(node[r].insts)$,
         $v =$ IF $s = 0$ THEN "nil" ELSE $node[r].insts[s].write$,
         $c = Head(UnseenPending(r))$
                $W.L.O.G.$, only pick a command not seen in current
                prepared log to have smaller state space; in practice,
                duplicated client requests should be treated by some
                idempotency mechanism such as using request $IDs$
    **do**
        if the latest value is in $Committed$ status, can directly reply;
        otherwise, should hold until I've received enough broadcasted
        $AcceptReplies$ indicating that the write is surely to be committed
        **await** $\vee s = 0$
                $\vee s > 0 \wedge node[r].insts[s].status =$ "Committed"
                $\vee$ LET $ars \triangleq \{m \in msgs : \wedge m.type =$ "AcceptReply"
                                    $\wedge m.slot = s$
                                    $\wedge m.bal = node[r].balMaxKnown\}$
            IN $WriteCommittable(ars)$ ;
        acknowledge client with the latest value, and remove the command from pending
        $Observe(\langle ReqEvent(c), AckEvent(c, v, r)\rangle)$ ;
        $Resolve(c)$ ;
    **end with** ;
**end macro** ;

Replica node crashes itself under promised conditions.
**macro** $ReplicaCrashes(r)$ **begin**
    if less than $(N -$ majority$)$ number of replicas have failed
    **await** $\wedge MajorityNum + numCrashed < Cardinality(Replicas)$
          $\wedge \neg crashed[r]$
          $\wedge node[r].balMaxKnown < MaxBallot$ ;
            this clause is needed only because we have an upper
            bound ballot number for modeling checking; in practice
            someone else could always come up with a higher ballot
    mark myself as crashed
    $crashed[r] :=$ TRUE ;
**end macro** ;

Replica server node main loop.
**process** $Replica \in Replicas$
**begin**
    $rloop$: **while** $(\neg terminated) \land (\neg crashed[self])$ **do**
        **either**
            $BecomeLeader(self)$;
        **or**
            $HandlePrepare(self)$;
        **or**
            $HandlePrepareReplies(self)$;
        **or**
            $HandlePrepareNotice(self)$;
        **or**
            $TakeNewWriteRequest(self)$;
        **or**
            $HandleAccept(self)$;
        **or**
            $HandleAcceptReplies(self)$;
        **or**
            $HandleCommitNotice(self)$;
        **or**
            $TakeNewReadRequest(self)$;
        **or**
            **if** $NodeFailuresOn$ **then**
                $ReplicaCrashes(self)$;
            **end if**;
        **end either**;
    **end while**;
**end process**;

**end algorithm**;

───────────── MODULE $Bodega\_MC$ ─────────────

EXTENDS $Bodega$

$TLC$ roster-related defs.

$ConditionalPerm(set) \triangleq$ IF $Cardinality(set) > 1$
                                  THEN $Permutations(set)$
                                  ELSE $\{\}$

$SymmetricPerms \triangleq \quad ConditionalPerm(Replicas)$
                        $\cup \quad ConditionalPerm(Writes)$
                        $\cup \quad ConditionalPerm(Reads)$

$ConstMaxBallot \triangleq 3$

Type check invariant.

$TypeOK \triangleq \land \forall m \in msgs : m \in Messages$
                $\land \forall g \in grants : g \in LeaseGrants$
                $\land Cardinality(\{g.from : g \in grants\}) = Cardinality(grants)$
                $\land \forall r \in Replicas : node[r] \in NodeStates$
                $\land Len(pending) \leq NumCommands$
                $\land Cardinality(Range(pending)) = Len(pending)$

$\land \forall c \in Range(pending) : c \in Commands$
$\land Len(observed) \leq 2 * NumCommands$
$\land Cardinality(Range(observed)) = Len(observed)$
$\land Cardinality(reqsMade) \geq Cardinality(acksRecv)$
$\land \forall e \in Range(observed) : e \in ClientEvents$
$\land \forall r \in Replicas : crashed[r] \in \text{BOOLEAN}$

THEOREM $Spec \Rightarrow \Box TypeOK$

---

*Linearizability* constraint.

$ReqPosOfCmd(c) \triangleq \text{CHOOSE } i \in 1 \ldots Len(observed):$
$\qquad\qquad\qquad\quad \land observed[i].type = \text{``Req''}$
$\qquad\qquad\qquad\quad \land observed[i].cmd = c$

$AckPosOfCmd(c) \triangleq \text{CHOOSE } i \in 1 \ldots Len(observed):$
$\qquad\qquad\qquad\quad \land observed[i].type = \text{``Ack''}$
$\qquad\qquad\qquad\quad \land observed[i].cmd = c$

$ResultOfCmd(c) \triangleq observed[AckPosOfCmd(c)].val$

$OrderIdxOfCmd(order, c) \triangleq \text{CHOOSE } j \in 1 \ldots Len(order) : order[j] = c$

$LastWriteBefore(order, j) \triangleq$
$\quad \text{LET } k \triangleq \text{CHOOSE } k \in 0 \ldots (j-1):$
$\qquad\qquad\qquad\quad \land (k = 0 \lor order[k] \in Writes)$
$\qquad\qquad\qquad\quad \land \forall l \in (k+1) \ldots (j-1) : order[l] \in Reads$
$\quad \text{IN} \quad \text{IF } k = 0 \text{ THEN } \text{``nil''} \text{ ELSE } order[k]$

$IsLinearOrder(order) \triangleq$
$\quad \land \{order[j] : j \in 1 \ldots Len(order)\} = Commands$
$\quad \land \forall j \in 1 \ldots Len(order):$
$\qquad ResultOfCmd(order[j]) = LastWriteBefore(order, j)$

$ObeysRealTime(order) \triangleq$
$\quad \forall c1, c2 \in Commands:$
$\qquad (AckPosOfCmd(c1) < ReqPosOfCmd(c2))$
$\qquad\quad \Rightarrow (OrderIdxOfCmd(order, c1) < OrderIdxOfCmd(order, c2))$

$Linearizability \triangleq$
$\quad terminated \Rightarrow$
$\qquad \exists order \in [1 \ldots NumCommands \to Commands]:$
$\qquad\quad \land IsLinearOrder(order)$
$\qquad\quad \land ObeysRealTime(order)$

THEOREM $Spec \Rightarrow Linearizability$

---

───── Bodega_MC.cfg ─────

```
SPECIFICATION Spec

CONSTANTS
    Replicas = {s1, s2, s3}
    Writes = {w1, w2}
    Reads = {r1, r2}
```

```
            MaxBallot <- ConstMaxBallot
            NodeFailuresOn <- TRUE

    SYMMETRY SymmetricPerms

    INVARIANTS
        TypeOK
        Linearizability

    CHECK_DEADLOCK TRUE
```



\end{document}